\documentclass[screen]{acmart}

\PassOptionsToPackage{hyphens}{url}
\usepackage{hyperref}
\usepackage{graphicx}
\usepackage{multirow}
\usepackage{tabularx}
\usepackage{array}
\usepackage{changepage}
\usepackage[inline]{enumitem}
\usepackage{booktabs}
\usepackage{natbib}
\bibpunct{\nolinebreak{}[}{]}{,}{n}{}{,}
\usepackage{changepage} \usepackage{array}
\usepackage{varwidth}
\usepackage{adjustbox}
\usepackage{rotating}
\usepackage{soul}
\usepackage{xcolor}
\usepackage{amsmath}
\usepackage{caption}
\usepackage{makecell}
\usepackage{wrapfig}
\usepackage{fontawesome}
\usepackage{longtable}
\usepackage{tikz}
\usetikzlibrary{positioning}
\usepackage{tikz-network}
\usepackage{subcaption}

\newcommand{\publicationnote}{\parbox{\textwidth}{\centering\footnotesize
    This is the preprint version of the article published in ACM Computing Surveys. Please cite the published version: \\
    \textit{ACM Computing Surveys}, 58(16), Article 401 (2026).\\
    \href{https://doi.org/10.1145/3839225}
         {https://doi.org/10.1145/3839225}}}

\definecolor{color1}{HTML}{80B1D3}
\definecolor{color2}{HTML}{FB8072}
\definecolor{color3}{HTML}{7fc97f}
\definecolor{color4}{HTML}{beaed4}
\definecolor{color5}{HTML}{fdc086}

\newcolumntype{L}[1]{>{\raggedright\let\newline\\\arraybackslash\hspace{0pt}}m{#1}}
\newcolumntype{C}[1]{>{\centering\let\newline\\\arraybackslash\hspace{0pt}}m{#1}}
\newcolumntype{R}[1]{>{\raggedleft\let\newline\\\arraybackslash\hspace{0pt}}m{#1}}

\newcommand*\emptydot[1][1ex]{\tikz\draw (0,0) circle (#1);} 

\newcommand*\fulldot[1][1ex]{\tikz\fill (0,0) circle (#1);}

\newcounter{definitioncounter}[section] \renewcommand{\thedefinitioncounter}{\thesection.\arabic{definitioncounter}}

\newenvironment{definition}[2][]
{
  \refstepcounter{definitioncounter}
  \addvspace{0.5em} \begin{adjustwidth}{1em}{1em} \small \textbf{Definition \thedefinitioncounter. #1} (#2):
    \begin{itshape}
    \small
}
{
    \end{itshape}
  \end{adjustwidth}
  \addvspace{0.5em} }

\newenvironment{ourdefinition}[1][]
{
  \refstepcounter{definitioncounter}
  \addvspace{0.5em} \begin{adjustwidth}{1em}{1em} \small
    \textbf{Definition \thedefinitioncounter. #1}:
    \begin{itshape}
    \small
}
{
    \end{itshape}
  \end{adjustwidth}
  \addvspace{0.5em} }

\AtBeginDocument{}

\setcopyright{cc}

\begin{document}

\title{Detection and Characterization of Coordinated Online Behavior: A Survey}

\author{Lorenzo Mannocci}
\correspondingauthor
\affiliation{\institution{University of Pisa}
  \city{Pisa}
  \country{Italy}
}
\affiliation{\institution{
Institute for Informatics and Telematics (IIT-CNR)}
\city{Pisa}
  \country{Italy}
}
\email{lorenzo.mannocci@di.unipi.it}
\orcid{0000-0002-5556-3746}

\author{Michele Mazza}
\affiliation{\institution{
Institute for Informatics and Telematics (IIT-CNR)}
\city{Pisa}
  \country{Italy}
}
\email{michele.mazza@iit.cnr.it}
\orcid{0000-0003-1874-3753}

\author{Anna Monreale}
\affiliation{
 \institution{University of Pisa}
 \city{Pisa}
 \country{Italy}
}
\email{anna.monreale@unipi.it}
\orcid{0000-0001-8541-0284}

\author{Maurizio Tesconi}
\affiliation{\institution{
Institute for Informatics and Telematics (IIT-CNR)}
\city{Pisa}
  \country{Italy}
}
\email{maurizio.tesconi@iit.cnr.it}
\orcid{0000-0001-8228-7807}

\author{Stefano Cresci}
\affiliation{\institution{
Institute for Informatics and Telematics, National Research Council (IIT-CNR)}
\city{Pisa}
  \country{Italy}
}
\email{stefano.cresci@iit.cnr.it}
\orcid{0000-0003-0170-2445}

\authorsaddresses{Authors' Contact Information:
L. Mannocci (\href{mailto:lorenzo.mannocci@di.unipi.it}{\nolinkurl{lorenzo.mannocci@di.unipi.it}});
M. Mazza (\href{mailto:michele.mazza@iit.cnr.it}{\nolinkurl{michele.mazza@iit.cnr.it}});
A. Monreale (\href{mailto:anna.monreale@unipi.it}{\nolinkurl{anna.monreale@unipi.it}});
M. Tesconi (\href{mailto:maurizio.tesconi@iit.cnr.it}{\nolinkurl{maurizio.tesconi@iit.cnr.it}});
S. Cresci (\href{mailto:stefano.cresci@iit.cnr.it}{\nolinkurl{stefano.cresci@iit.cnr.it}}).
}

\renewcommand{\shortauthors}{Mannocci et al.}
\newcommand{\new}[1]{\textcolor{red}{#1}}
\newcommand{\rev}[1]{{\color{blue}#1}}

\DeclareRobustCommand{\hlcyan}[1]{{\sethlcolor{cyan}\hl{#1}}}

\begin{abstract}
Coordination is a fundamental aspect of life. The advent of social media has made it integral also to online human interactions, such as those that characterize thriving online communities and social movements. At the same time, coordination is also core to effective disinformation, manipulation, and hate campaigns. This survey collects, categorizes, and critically discusses the body of work produced as a result of the growing interest on coordinated online behavior. We reconcile industry and academic definitions, propose a comprehensive framework to study coordinated online behavior, and review and critically discuss the existing detection and characterization methods. Our analysis identifies open challenges and promising directions of research, serving as a guide for scholars, practitioners, and policymakers in understanding and addressing the complexities inherent to online coordination. We also provide an interactive companion website for exploring the surveyed literature.
\end{abstract}

\begin{CCSXML}
<ccs2012>
   <concept>
       <concept_id>10002951.10003227.10003233.10010519</concept_id>
       <concept_desc>Information systems~Social networking sites</concept_desc>
       <concept_significance>500</concept_significance>
       </concept>
   <concept>
       <concept_id>10010147.10010257</concept_id>
       <concept_desc>Computing methodologies~Machine learning</concept_desc>
       <concept_significance>500</concept_significance>
       </concept>
   <concept>
       <concept_id>10003120.10003130.10003131</concept_id>
       <concept_desc>Human-centered computing~Collaborative and social computing theory, concepts and paradigms</concept_desc>
       <concept_significance>500</concept_significance>
       </concept>
   <concept>
       <concept_id>10010147.10010341.10010346.10010348</concept_id>
       <concept_desc>Computing methodologies~Network science</concept_desc>
       <concept_significance>500</concept_significance>
       </concept>
 </ccs2012>
\end{CCSXML}

\ccsdesc[500]{Information systems~Social networking sites}
\ccsdesc[500]{Computing methodologies~Machine learning}
\ccsdesc[500]{Human-centered computing~Collaborative and social computing theory, concepts and paradigms}
\ccsdesc[500]{Computing methodologies~Network science}

\keywords{coordinated behavior, coordination, synchronization, online social networks, network science}

\maketitle

\makeatletter
\apptocmd{\ps@firstpagestyle}{\fancyfoot{}\fancyfoot[C]{\publicationnote}}{}{}

\apptocmd{\ps@standardpagestyle}{\fancyfoot{}\fancyfoot[C]{\publicationnote}}{}{}
\makeatother

\pagestyle{standardpagestyle}
\thispagestyle{firstpagestyle}

\section{Introduction}
\label{sec:introduction}

Coordination, the process in which multiple connected actors are involved to pursue goals~\cite{malone1988modeling}, is a fundamental aspect in the existence of various life forms, including human beings. From flocks of birds engaging in synchronized flight to insects working together in colonies, coordination enhances efficiency, safety, and resource utilization~\cite{malone1990coordination}. The ability to coordinate actions boosts the chances to overcome environmental challenges, fostering not only individual survival but also the resilience and success of entire communities. For these reasons, human coordination has been extensively scrutinized in multiple scientific disciplines interested in the dynamics of our offline interactions~\cite{turvey1990coordination,sibertin2005coordination}.

\begin{figure}
  \begin{minipage}{0.58\textwidth}
    \centering
    \includegraphics[width=\linewidth]{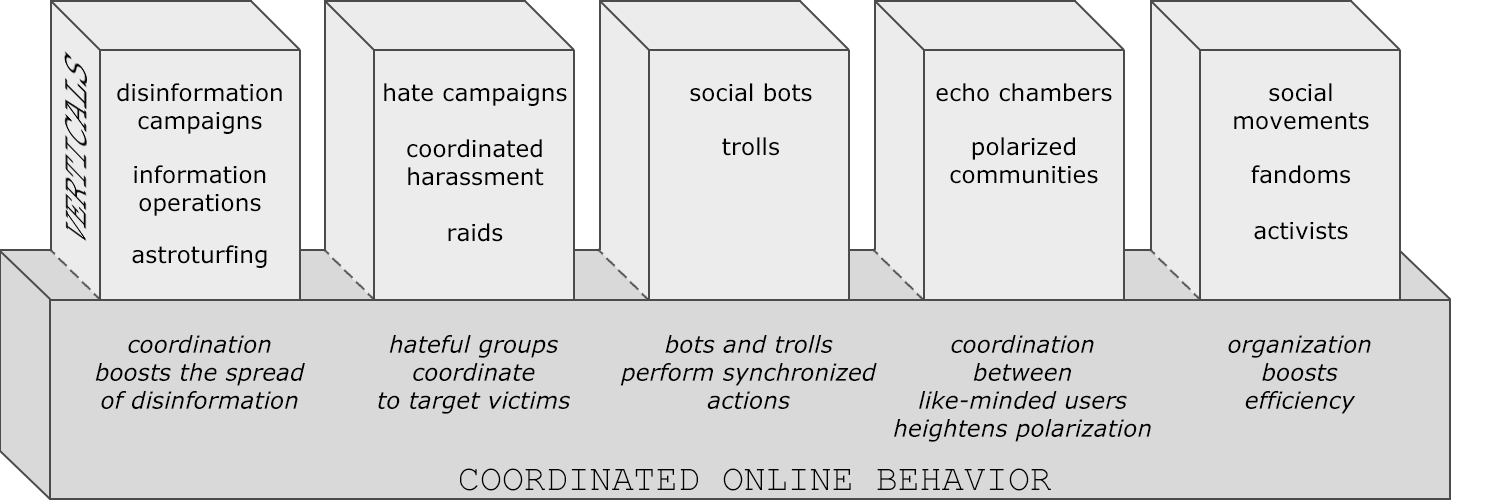}
    \caption{Coordination is a fundamental aspect of online human interactions and the study of coordinated online behavior can complement the analyses of many other online phenomena.}
    \label{fig:cbphenomena}
  \end{minipage}
  \quad
  \begin{minipage}{0.38\textwidth}
    \centering
    \includegraphics[width=0.8\linewidth]{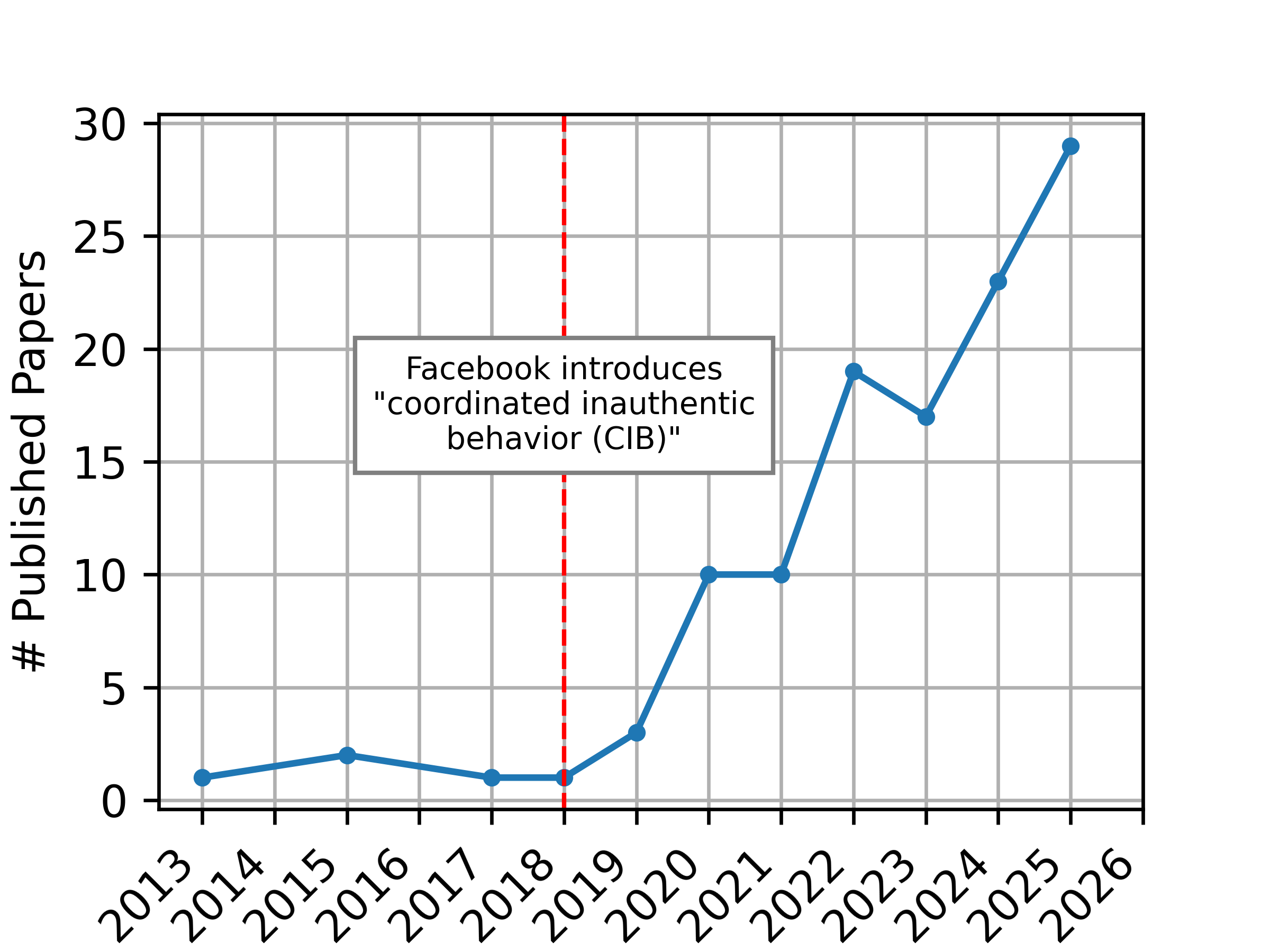}
    \caption{Number of articles published yearly on coordinated online behavior. } 
    \label{fig:publishedpapers}
  \end{minipage}\end{figure}

With the advent of social media platforms, coordination has also become a fundamental component of \textit{online} interactions. Social media users are now provided with a broad array of tools to coordinate with each other, such as hashtags that enable them to collectively discuss specific topics~\cite{tardelli2024temporal}. Online platforms have become a suitable environment for organizing social and political movements, giving rise to phenomena such as online activism~\cite{ng2022combined,nizzoli2021coordinated}, boycotts~\cite{lucchini2022reddit}, protests~\cite{magelinski2022synchronized,steinert2015online}, and coordinated crisis communication by democratic institutions during emergencies~\cite{yoo2024speak}. The 2011 Arab Springs are a notable example, being largely organized through social media~\cite{steinert2015online}. 
At the same time however, scholars found evidence of online coordination being exploited by nefarious actors for all sorts of malicious purposes.
For instance, disinformation campaigns often leverage actors that coordinate their actions to maximize the outreach of their false narratives~\cite{starbird2019disinformation, keller2020political,vargas2020detection}. Similarly, coordination is employed within information operations~\cite{cima2024coordinated,vargas2020detection,ng2022coordinated}, coordinated social media manipulation~\cite{milzner2025just,thiele2025attributing}, and astroturfing, which involves creating the false appearance of grassroots support for a target cause, product, or person~\cite{keller2020political}. Also, social bots (automated accounts~\cite{ng2025global}) and trolls (human-driven accounts engaging in collective disruptive or deceptive behavior~\cite{zannettou2019who}) exploit coordination to amplify messages, manipulate trends, or spread disinformation~\cite{mannocci2022mulbot,luceri2020detecting,zannettou2019who}.
Coordination both results from and contributes to echo chambers and online polarization~\cite{vasconcelos2021segregation}. Figure~\ref{fig:cbphenomena} illustrates its multifaceted nature, underlying diverse and interconnected phenomena in online social media.

Recognizing the profound impact of online coordination on social media as well as its consequences on the offline world, both researchers~\cite{vargas2020detection, pacheco2020unveiling, weber2021amplifying, nizzoli2021coordinated} and industry stakeholders~\cite{gleicher2018coordinated} devoted a great deal of efforts to study its dynamics and to develop effective strategies to detect and mitigate its malicious instances. Research sped up significantly after 2018, when Facebook introduced the concept of \textit{coordinated inauthentic behavior} (CIB)~\cite{gleicher2018coordinated}, marking a milestone in the development of the field. As shown in Figure \ref{fig:publishedpapers}, subsequent years saw a surge of interest on the subject, testified by the steadily growing number of published papers, with only a minor fluctuation in 2023. This survey is motivated by this thriving interest on online coordination, which resulted in the availability of a large body of work. However, while Facebook's interest towards online coordination was constrained to inauthentic behaviors as a response to the threat of orchestrated campaigns~\cite{gleicher2018coordinated}, here we embrace a more holistic and unbiased view by focusing on the broader concept of \textit{coordinated behavior}. This inclusive approach allows for the analysis of a broader spectrum of works, including those focused on legitimate collective actions, offering a more comprehensive understanding of the coordination dynamics that shape digital spaces and fostering nuanced perspectives that go beyond mere threat detection.  
In spite of the many efforts, the existing literature still reflects the complexities and ambiguities surrounding this phenomenon. Among them is the limited agreement on a shared definition, which also hinders operationalization. Complexity also emerges from the diversity of methods proposed for detecting and characterizing coordinated online behavior, which impairs comparisons between different works and limits the generalizability of findings. Finally, the use of the same coordination technique by actors with disparate motivations poses challenges to estimating the impact and effects of online coordination.

This survey offers an extensive overview and critical analysis on coordinated behavior, starting from reconciling existing definitions from industry and academia, and proposing a general and comprehensive conceptual framework. We systematically analyze and categorize existing approaches for detecting and characterizing online coordination, elucidating open challenges and delineating promising directions for future research. Our work provides a roadmap for scholars, practitioners, and policymakers navigating the evolving complexities of coordinated online behavior.\vspace{0.1cm}

\subsection{Interactive companion website}
\begin{figure}
\includegraphics[width=1\textwidth]{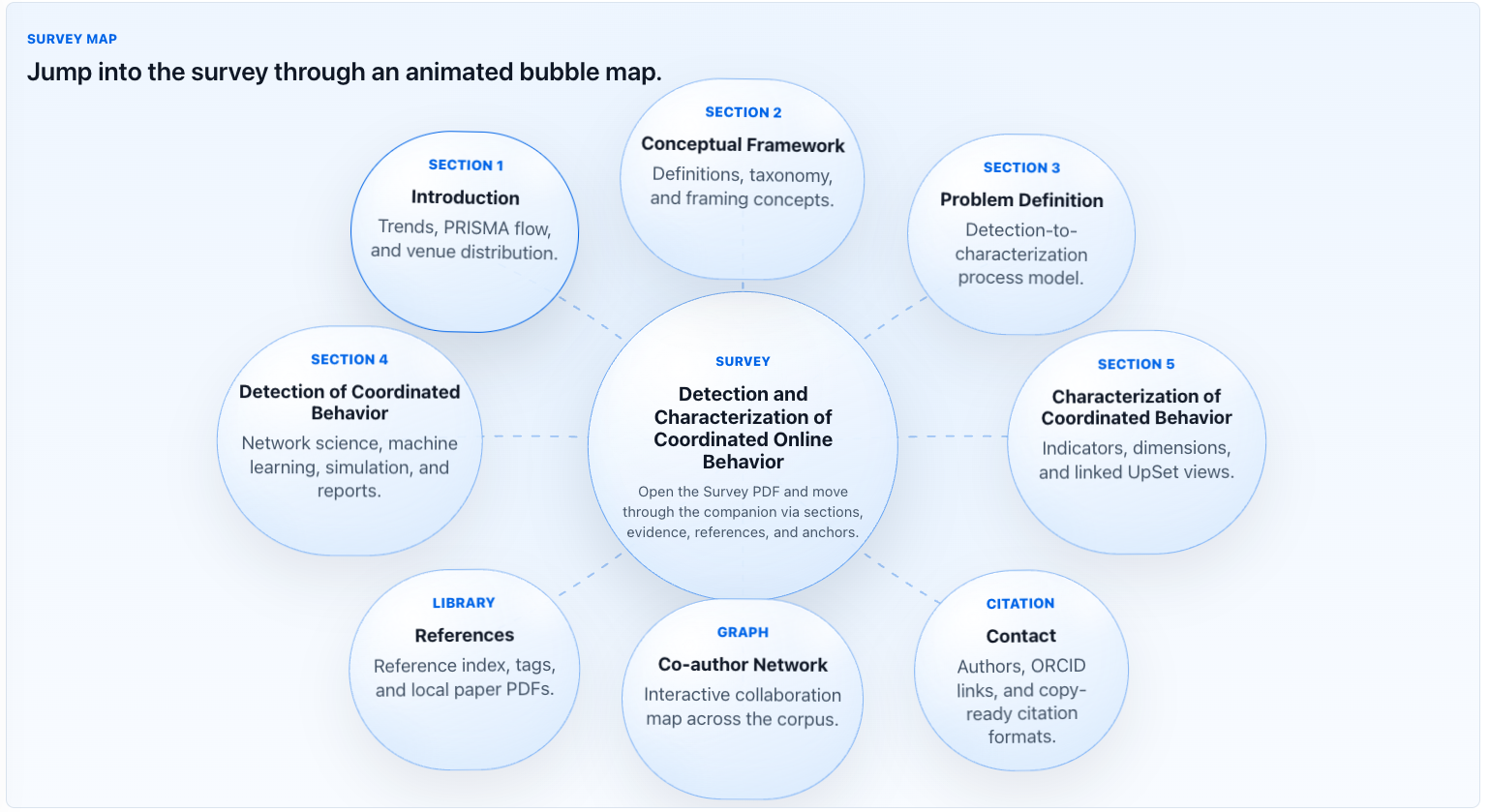}
    \centering
    \caption{Interactive companion website, available at \url{https://www.lorenzomannocci.it/coordinated-behavior/}, to facilitate the exploration of the reviewed literature.}
    \label{fig:website}
\end{figure}

In addition to this survey, we developed an interactive companion website\footnote{\url{https://www.lorenzomannocci.it/coordinated-behavior/}}, to facilitate the exploration of the reviewed literature. The website complements the static survey by allowing readers to search and filter papers, navigate the proposed taxonomy, and interactively explore the comparison tables and other resources presented throughout this work. By providing an accessible and structured interface to the surveyed literature, it supports researchers and practitioners in efficiently navigating this rapidly evolving research area. Figure~\ref{fig:website} shows a screenshot of the homepage of the website.

\textit{Significance.} Comprehensively modeling coordinated online behavior has far-reaching implications. On a theoretical level, it reconciles diverse definitions and provides a foundational framework for future research. On a technical level, it critically evaluates and categorizes existing detection and characterization methods, informing the development of the next generation of robust and adaptive tools for studying both malicious and neutral instances of coordinated behavior. By shedding light on online coordination---a fundamental dynamic of computer-mediated human behavior---this survey contributes to safeguarding online integrity and to fostering positive interactions in digital spaces. It offers valuable insights for shaping future methodologies, platforms, and policies, and it also contributes to enriching the interdisciplinary research occurring at the intersection between computer science and social dynamics.\vspace{0.1cm}

\textit{Scope.}
Coordinated online behavior is orthogonal to many of the research topics tackled in fields such as Web and social media analysis, online social networks security, as well as social computing and computational social science, as shown in Figure~\ref{fig:cbphenomena}. Therefore, a large number of works from these scientific communities implicitly or explicitly deal with online coordination. While coordinated behavior can manifest across a wide range of online environments (e.g., e-commerce platforms, review systems, or online marketplaces), this survey focuses primarily on online social media platforms, where coordination has been most extensively studied and where interactions are more readily observable at scale. Accordingly, this survey is constrained to those papers that address online coordination explicitly within this context and that provide relevant contributions to its detection, characterization, or understanding.\vspace{0.1cm}

\begin{figure}
  \begin{minipage}{0.54\textwidth}
        \centering
        \resizebox{0.9\linewidth}{!}{\begin{tikzpicture}[
        prismabox/.style={
            draw,
            rectangle,
            minimum width=6.5cm,
            minimum height=1.5cm,
            align=left,
            rounded corners=4pt,
            font=\Large
        },
        exclusionbox/.style={
            draw,
            rectangle,
            minimum width=6.5cm,
            minimum height=1.8cm,
            align=left,
            dashed,
            rounded corners=4pt,
            font=\Large
        },
        headerbox/.style={
            draw,
            rectangle,
            minimum width=6.5cm,
            text width=6cm,
            minimum height=1cm,
            align=center,
            fill=gray!20,
            rounded corners=4pt,
            font=\bfseries
        },
        sidebox/.style={
            draw,
            rectangle,
            minimum width=1cm,
            minimum height=2.5cm,
            align=center,
            rounded corners=4pt,
            fill=gray!20,
            font=\bfseries
        },
        arrow/.style={->, thick}
    ]
    
\node[sidebox] (id) at (0,0) {\rotatebox{90}{Identification}};
    \node[sidebox] (scr) at (0,-4) {\rotatebox{90}{Screening}};
    \node[sidebox] (inc) at (0,-8) {\rotatebox{90}{Included}};
    
\node[prismabox, right=0.6cm of id] (box1) {
    Records identified from:\\
    \hspace{0.4cm}Scopus ($n=353$)\\
    \hspace{0.4cm}Google Scholar ($n=1000$)
    };
    
\node[prismabox, right=2cm of box1] (snowball) {
    Additional studies identified through\\
    backward reference searching\\
    \hspace{0.4cm}($n=42$)
    };
    
\node[prismabox, right=0.6cm of scr] (box2) {
    Records screened ($n = 1206$)
    };
    
\node[prismabox, right=0.6cm of inc] (box3) {
    Studies included in review ($n = 125$)
    };
    
\draw[arrow] (box1.south) -- coordinate[midway] (mid12) (box2.north);
    \draw[arrow] (box2.south) -- coordinate[midway] (mid23) (box3.north);
    
\node[exclusionbox, right=1.5cm of mid12] (box_excl1) {
    Records removed before screening:\\
    \hspace{0.4cm} Duplicate records removed ($n=147$)
    };
    
\draw[arrow] (snowball.south) |- (box3.east);
    
\node[exclusionbox, right=1.5cm of mid23] (box_excl2) {
    Full-text articles excluded ($n=1123$):\\
    \hspace{0.4cm}Not in English ($n = 4$)\\
    \hspace{0.4cm}Out of scope ($n = 1118$)
    };
    
\draw[arrow] (mid12) -- (box_excl1.west);
    \draw[arrow] (mid23) -- (box_excl2.west);
    
\node[headerbox, above=0.5cm of box1] (head1) {
    Identification of studies via databases and registers
    };
    
    \node[headerbox, above=0.5cm of snowball] (head2) {
    Identification of studies\\via other methods
    };
\end{tikzpicture} }
    \caption{PRISMA diagram summarizing the stages used to construct the final corpus of studies considered in this survey.}
    \label{fig:prisma}
  \end{minipage}
  \hspace{0.2pt}
  \begin{minipage}{0.42\textwidth}
    \centering
    \includegraphics[width=0.9\linewidth]{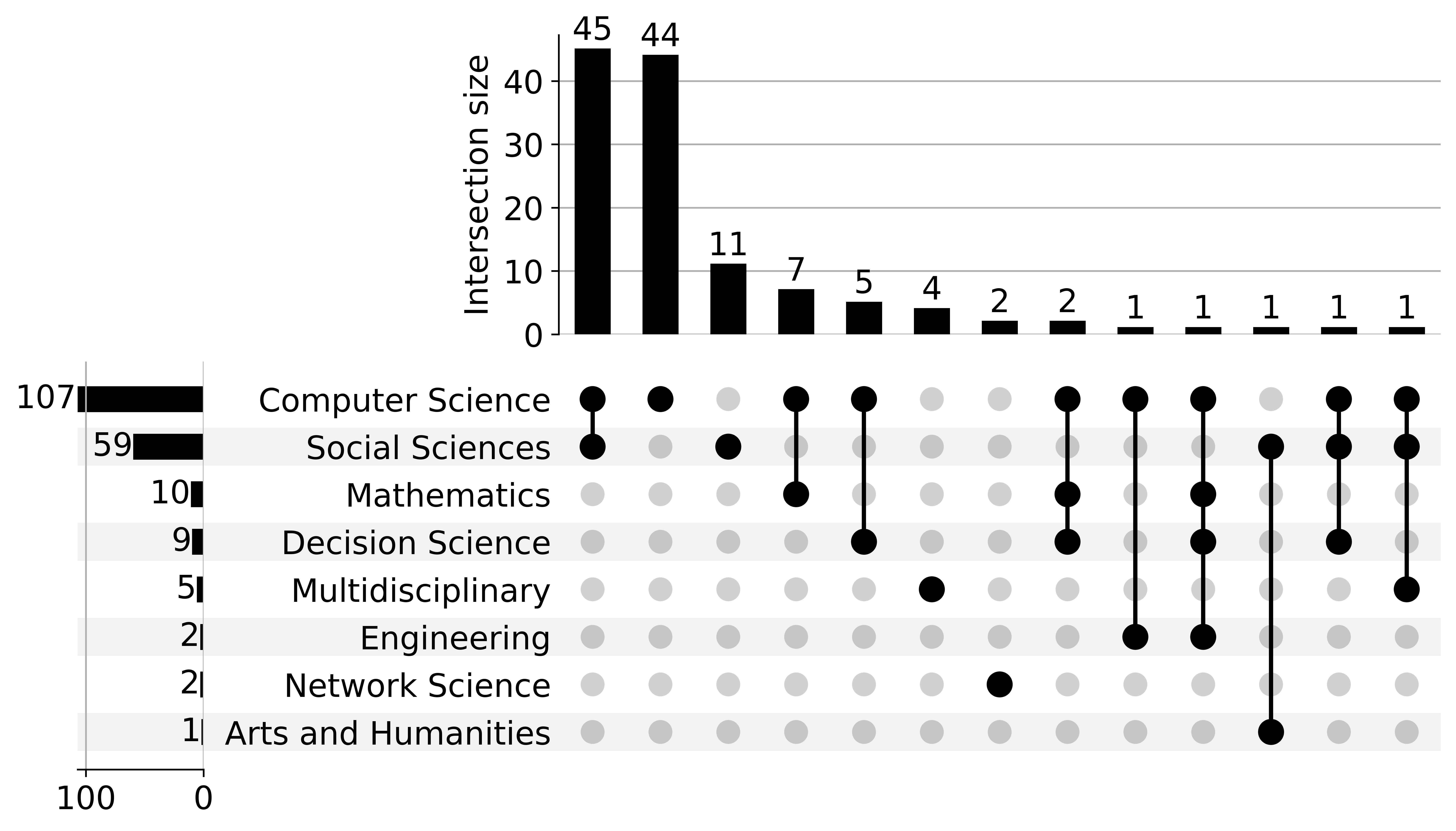}
    \caption{Distribution of papers in the dataset according to the disciplinary area of the publication venue, highlighting the predominance of Computer Science venues alongside contributions from Social Sciences and interdisciplinary outlets.}
    \label{fig:upset_venue}
  \end{minipage}\end{figure} 
\subsection{Literature search and selection.}
A systematic literature search was conducted to identify studies addressing coordinated behavior on social media platforms using two databases: Scopus and Google Scholar. The Scopus query targeted papers containing the following search strings:

\begin{quote}\small
(``coordinated behavior'' OR ``coordinated behaviour'' OR ``coordinated activity'' OR
``coordinated campaign'' OR ``coordinated inauthentic behavior'' OR
``coordinated inauthentic behaviour'' OR ``inauthentic behavior'' OR
``inauthentic behaviour'' OR ``information operation*'')
AND
(``social media'' OR online OR twitter OR facebook OR instagram OR
reddit OR telegram OR tiktok OR youtube)
\end{quote}

The search was restricted to publications between 2014 and 2026, written in English, and within relevant subject areas (Computer Science, Social Sciences, Decision Sciences, Engineering, and Multidisciplinary). This query returned 353 records. To broaden coverage, a complementary search was performed on Google Scholar using similar search strings, collecting the first 1000 results returned by the platform. This choice aims to maximize coverage while reducing potential bias introduced by the platform’s ranking mechanism. After merging the two sources, 147 duplicate records were removed, leaving 1206 records for screening. Titles and abstracts were manually reviewed, leading to the exclusion of 4 non-English papers and 1118 records that were out of scope (e.g., biological coordination, general sociological coordination, or studies on bot detection or information operations not explicitly addressing coordinated behavior). Most of these discarded records derived from the search on Google Scholar, likely due to the choice of selecting the first 1000 results to maximize the coverage. After the screening phase, 83 papers were retained. A backward reference search of these studies identified an additional 42 relevant papers, resulting in a final corpus of 125 papers included in this survey. The full selection workflow is illustrated in the PRISMA diagram in Figure~\ref{fig:prisma}.

\subsection{Subject area distribution.} To analyze the disciplinary distribution of the collected studies, each paper was assigned a subject area based on the official classification associated with the publication outlet. For journal articles, the subject area was determined using the Scimago Journal Rank classification system\footnote{\url{https://www.scimagojr.com}}. For conference papers, the classification provided by the CORE conference ranking portal was used\footnote{\url{https://portal.core.edu.au/conf-ranks/}}. For preprints available on arXiv, the category assigned by arXiv itself was adopted. The resulting distribution of papers across subject areas is shown in Figure~\ref{fig:upset_venue}.

\textit{Organization.} This survey is structured as follows. Section~\ref{sec:conceptual-framework} presents the theoretical foundations of coordinated online behavior, proposing a new general definition and laying out a comprehensive conceptual framework. Section~\ref{sec:problem-definition} bridges the theoretical and methodological parts by defining the detection and characterization tasks. Section~\ref{sec:detection} discusses the existing literature on the detection task, while Section~\ref{sec:characterization} focuses on the characterization task. Section~\ref{sec:discussion} summarizes the main outstanding challenges and suggests promising directions of research. Finally, Section~\ref{sec:conclusions} concludes the survey.
 \section{Conceptual framework}
\label{sec:conceptual-framework}
The study of online coordination has its roots in the earlier studies of offline coordination. Recently, this study was advanced both by commercial platforms and academia, with a large array of different proposals. This section examines previously proposed definitions of coordination and discusses their advantages and limitations. Based on the results of this analysis, we then propose a general definition and a comprehensive conceptual framework.

\subsection{Offline coordination}
\label{sec:defining_coordination}
Coordination has already been extensively studied well before the emergence of social media across disciplines such as computer science, organization theory, management science, economics, and psychology~\cite{sibertin2005coordination,turvey1990coordination,malone1988modeling}. Although the meaning of coordination is intuitive, researchers suggested many definitions to frame the concept. A concise and precise definition was given in~\cite{malone1988what}:

\begin{definition}[Coordination]{1988}
\label{def:coordination}
  The additional information processing performed when multiple, connected actors pursue goals that a single actor pursuing the same goals would not perform~\cite{malone1988what}.
\end{definition}
Definition~\ref{def:coordination} denotes coordination as the organizational overhead that multiple actors incur into when pursuing goals together. 
We note that this and similar definitions~\cite{malone1987modeling, malone1988modeling,baligh1981describing,baligh1986decision} implicitly leverage the fundamental components of coordination, which we explicitly define as follows:
\begin{ourdefinition}[Coordination components]
\label{def:coordination-components}
    A set of two or more \underline{actors} who perform \underline{activities} in order to achieve \underline{goals}.
\end{ourdefinition}
Definition~\ref{def:coordination-components} introduces the fundamental components of coordination: \textit{actors}, \textit{activities}, and \textit{goals}. Being coordination a nuanced concept, the theoretical modeling of these components can have major implications on downstream analyses and results. For example, in the case of communities or groups of users, each user in the group can be treated as a standalone actor, or alternatively the entire group may be considered as a single actor. Similar choices must be made when modeling the activities that allow actors to coordinate. Each actor typically performs multiple activities during any given time frame, and each of these activities might contribute differently to the overall coordination. Therefore, the choice of activities to model during an analysis directly impacts the resulting observed coordination~\cite{malone1990coordination}. 
Finally, the analyst is often interested in knowing the goal that the actors pursue when performing the activities. However, the actors may not all have the same goal, or even have any explicit goal at all~\cite{malone1988what}.
These reflections on the components of coordination surface some of the challenges that early scholars faced since the 80s when studying offline coordination. Interestingly, many of these challenges carry over to the study of online coordination, informing the development of a new comprehensive definition and conceptual framework. 

\subsection{Concepts and definitions by online platforms}
\label{sec:smp_definitions}
Beginning around 2016, mounting societal pressure impelled major social media platforms to confront pervasive challenges such as the organized dissemination of mis- and disinformation~\cite{wardle2017information}. In consequence of this pressure, each platform adopted disclosure practices to communicate their results at exposing orchestrated deceptive activities perpetrated by organized actors.
Given the importance of coordination for the success of large-scale disinformation campaigns~\cite{nizzoli2021coordinated,pacheco2021uncovering}, within these public disclosure initiatives each platform addressed some instances of malicious online coordination. This section explores the concepts and definitions introduced by major social media platforms that are related to online coordination, discussing both their merits and limitations.

\subsubsection{Meta (Facebook, Instagram, Threads)}
After the public disclosure that the Internet Research Agency (IRA) had strategically exploited the platform to influence the 2016 US presidential election~\cite{zhang2021vigdet,weber2021amplifying,sharma2021identifying}, Facebook began publishing reports detailing how their services were abused and the actions taken in response. In unveiling further actions against the IRA, in July 2018 Facebook introduced the concept of \textit{coordinated inauthentic behavior} (CIB)~\cite{gleicher2018coordinated}. A few months later, they supplied it with a first definition, and in October 2019 with a second one. These definitions, originally introduced by Facebook, also applied to Instagram due to their shared ownership. Then they have been extended across Meta’s platforms, including Facebook, Instagram, and Threads.

\begin{definition}[Coordinated inauthentic behavior]{2018}
\label{def:cib}
Groups of pages or people working together to mislead others on who they are or what they are doing~\cite{gleicher2018coordinated}.
\end{definition}

\begin{definition}[Coordinated inauthentic behavior]{2019}
\label{def:cib2}
The use of multiple Facebook or Instagram assets (accounts, pages, groups, or events), working in concert to engage in inauthentic behavior, i.e., to mislead people or Facebook, where the use of fake accounts is central to the operation~\cite{policies2018inauthenticbehavior}.
\end{definition}
Definition~\ref{def:cib} underscores the collaborative nature of disinformation campaigns~\cite{starbird2019disinformation}, emphasizing the objective of misleading others about the purported identity of the involved actors. To this end, it introduces the concept of \textit{inauthenticity} of the actors, which is ever since often used in conjunction with the notion of coordination. Definition~\ref{def:cib2} explicitly articulates the concept of inauthenticity and elucidates that the act of deceiving others involves the extensive use of fake accounts~\cite{gleicher2019howwe}. A widespread critique of these initial definitions is that they exclusively address CIB, overlooking other types of malicious and possibly harmful coordination, let alone the neutral or benign ones~\cite{nizzoli2021coordinated,cinelli2022coordinated,giglietto2020takes,gruzd2022coordinated}. In September 2021, Facebook provided additional definitions focusing on harmfulness rather than inauthenticity.

\begin{definition}[Coordinated social harm]{2021}
\label{def:coordinated-social-harm}
  Networks of primarily authentic users who organize to systematically violate policies to cause harm on or off the platform~\cite{gleicher2021removing}.
\end{definition}
\begin{definition}[Coordinated mass harassment]{2021}
\label{def:coordinated-mass-harassment}
    Coordinated efforts of mass harassment that target individuals at heightened risk of offline harm~\cite{davis2021advancing}.
\end{definition}
Definitions~\ref{def:coordinated-social-harm} and~\ref{def:coordinated-mass-harassment} adopt the concept of harmfulness in place of inauthenticity, thereby broadening the scope to also encompass authentic yet coordinated actors. These, in fact, hold the potential to cause negative consequences both on and off the platforms, as underscored in Definition~\ref{def:coordinated-social-harm}.

\subsubsection{Twitter/X}
In October 2018, the platform released a public archive containing data about identified \textit{information operations} (IOs). Although not explicitly stated in Twitter's definition at the time, a certain degree of coordination is necessary for the success of an IO~\cite{vargas2020detection,cima2024coordinated}. 
However, it was not until January 2021 that Twitter adopted a similar approach to Facebook and released a definition that explicitly addresses instances where coordination is leveraged to cause harm both online and offline. 

\begin{definition}[Information operation]{2018}
\label{def:io}
    People directly involved in manipulation that can be reliably attributed to a government or state-linked actor~\cite{gadde2018enabling}.
\end{definition}
\begin{definition}[Coordinated harmful activity]{2021}
\label{def:coordinated-harmful-activity}
   Groups, movements, or campaigns that are engaged in coordinated activity resulting in harm on and off of Twitter~\cite{twitter2021coordinated}.
\end{definition}

\subsubsection{YouTube/Google}
In its reports, primarily concerning abuses that occurred on YouTube, Google makes reference to \textit{coordinated influence operations} (CIO)~\cite{huntley2020updates}. Even though Google did not provide a definition for CIOs, they nonetheless highlighted the importance of coordination in these online manipulations. 

\subsubsection{Reddit}
In contrast to other platforms, Reddit embraced the broad concept of \textit{content manipulation} to characterize the campaigns that violate its rules~\cite{reddit2022transparency}. The platform shared only a small number of such campaigns, leaving it unclear whether these represent the entirety of the identified cases, or only a selection of them. Despite the absence of an explicit reference to coordination, large-scale content manipulations gain advantage from coordinated activities, similarly to what we discussed earlier about IOs.

\begin{definition}[Content manipulation]{2022}
\label{def:content-manipulation}
   Things like spam, community interference, vote manipulation, and other attempts to artificially promote content~\cite{reddit2022transparency}.
\end{definition}

\subsubsection{Tiktok}
In 2023, TikTok revisited Facebook’s definition of CIB, explicitly linking coordination with inauthenticity. In doing so, it emphasizes coordinated activity aimed at misleading users or influencing public discourse. Among the cited examples, TikTok highlights coordination in the context of information operations, such as accounts covertly promoting political candidates or issues, or posting content on behalf of foreign entities without proper disclosure.
\begin{definition}[Covert Influence Operations]{2023}
\label{def:tiktok}
   Coordinated, inauthentic behaviors where networks of accounts work together to mislead people or our systems and try to strategically influence public discussion. This may include attempting to undermine the results of an election, influencing parts of an armed conflict, or shaping public discussion of social issues.~\cite{tiktok2025integrity}.
\end{definition}

\subsubsection{Ambiguities and limitations}
This brief survey of the main concepts and definitions proposed by social media shows that online platforms are at the forefront in the analysis of coordinated online behavior. However, their conceptualizations are driven primarily by pressing practical regulation needs and by immediate contingencies, rather than by methodological rigor and theoretical soundness~\cite{gleicher2018coordinated}. Faced with specific instances of malicious coordination, online platforms adopt different and at times contrasting definitions, adding confusion and ambiguity to the already challenging task of defining an inherently nuanced and complex phenomenon~\cite{murero2023coordinated}.

The current tangled landscape of platform definitions means that certain coordinated efforts may be categorized as such by some platforms but not by others, contingent upon the concepts they adhere to. What may be identified as coordinated behavior on one platform could be overlooked or dismissed on another, leading to inconsistencies in detection and mitigation. As a notable example, in June 2020 many teenagers organized on TikTok to reserve tickets for a Donald Trump rally to be held in Tulsa, OK (US). By mass-reserving and later cancelling their participation, they prevented others from making reservations and artificially inflated expected attendance numbers, ultimately causing an embarrassing number of empty seats at the rally~\cite{douek2020does}. Facebook’s head of security commented that, while tactical and sophisticated, they would have not acted upon this campaign as an instance of CIB (as per Definition~\ref{def:cib2}), since it did not make use of fake accounts nor aimed to mislead Facebook users~\cite{douek2020does}. A similar case is the Chinese Spamouflage campaign, a cross-platform propaganda effort against the Hong Kong pro-democracy movement.\footnote{\url{https://graphika.com/reports/spamouflage-breakout} (accessed: 28/07/2026)} Various platforms reported takedowns of Spamouflage instances, with Google and Twitter labeling it as CIO and IO (as per Definition~\ref{def:io}), respectively. However, Reddit interpreted Spamouflage differently, recognizing its low-quality and one-sided content, but refraining from considering these activities as rule violations.\footnote{\url{https://www.reddit.com/r/redditsecurity/comments/dp9nbg/reddit_security_report_october_30_2019/} (accessed: 28/07/2026)}

Evidently, platforms address instances of coordinated online behavior disparately, and the criteria to establish the legitimacy of user behaviors lack objectivity. These discrepancies underscore the need for a unified understanding and standardized approach to defining and addressing online coordination, one that transcends platform-specific idiosyncrasies and fosters a more cohesive response to this challenge.

\begin{table}[t]
    \caption{Examples of operational definitions used in recent academic literature. No definition is general enough to comprehensively describe coordinated online behavior. However, each definition grasps one or more relevant properties (highlighted in bold) that we leverage in our framework.}
    \label{tab:notions}
    \centering
    \small
    \setlength{\tabcolsep}{6pt}
    \scalebox{0.75}{
    \begin{tabular}{L{3.5cm}L{13cm}}
        \toprule
        \textbf{reference} & \textbf{definition}\\
        \midrule
        \citet{nizzoli2021coordinated} & Unexpected, suspicious, or exceptional \textbf{similarity} between a number of users \\ [0.5ex]
        \citet{cinelli2022coordinated} & The number of times two accounts \textbf{behaved similarly}, such as when they \textbf{repeatedly} retweet the same post \\ [0.5ex]
        \citet{giglietto2020takes} & The act of making people and/or things be involved in an \textbf{organized cooperation} \\ [0.5ex]
\citet{magelinski2020detecting} & \textbf{Many instances} of a tweet-behavior, i.e. tweeted hashtag [...] within a \textbf{small predetermined time window} \\ [0.5ex]
        \citet{weber2020who} & Anomalous levels of \textbf{coincidental behavior}  \\ [0.5ex]
        \citet{magelinski2022synchronized} & Users [that] take the \textbf{same actions within minutes} of one another \\ [0.5ex]
        \citet{zhang2021vigdet} & Accounts that \textbf{co-appear}, or are \textbf{synchronized} in time \\ [0.5ex]
        \citet{pacheco2021uncovering} & [Users exhibiting a] surprising \textbf{lack of independence} \\ [0.5ex]
        \citet{hristakieva2022spread} & Coordination between users implies a \textbf{shared intent} \\ [0.5ex]
        \citet{keller2020political} & A group of people who want to \textbf{convey specific information} to an audience \\
        \bottomrule
    \end{tabular}
    }
\end{table}
 
\subsection{Concepts and definitions in academic literature}
\label{sec:research_definitions}
As shown in Figure~\ref{fig:publishedpapers}, scientific interest in coordinated online behavior has drastically risen in the last few years. However, akin to social media platforms, scholars have generally refrained from proposing theoretically-grounded and general definitions. The majority of the existing studies adopt Facebook's Definition~\ref{def:cib} of CIB~\cite{mehta2022estimating}. Instead, those who propose their own conceptualization mainly provide \textit{operational} definitions useful for the development of coordination detection methods~\cite{magelinski2022synchronized,nizzoli2021coordinated,pacheco2021uncovering}. Table~\ref{tab:notions} reports some examples of operational definitions proposed recently. As shown, no definition is comprehensive and general enough to adequately describe the multifaceted phenomenon of coordinated online behavior. The existing conceptualizations of the phenomenon appear to be influenced by the specific technique employed for its detection. Consequently, the criteria used to define coordination vary, encompassing aspects ranging from similar behavior~\cite{nizzoli2021coordinated, cinelli2022coordinated, magelinski2020detecting, weber2020who, magelinski2022synchronized, pacheco2021uncovering} to synchronicity~\cite{magelinski2020detecting, magelinski2022synchronized, zhang2021vigdet} and common intent~\cite{hristakieva2022spread, keller2020political}. As practical examples of the above, some definitions and the resulting detection methods revolve around the anomalous use of hashtags or URLs by multiple users~\cite{giglietto2020takes, magelinski2022synchronized}, or repeated screen name swapping~\cite{pacheco2021uncovering}. 

More fundamentally, the field still lacks a formal and statistically grounded definition of coordination, including well-defined null models against which coordinated behavior can be rigorously assessed. As a consequence, many approaches rely on platform- and task-specific heuristics, which complicates cross-domain comparisons, reproducibility, and the estimation of effect sizes. This limitation echoes challenges observed in related areas such as bot detection, where simplistic data collection and labeling practices have been shown to hinder generalization even across datasets from the same platform. While no single definition fully captures the complexity of coordinated online behavior, existing works collectively highlight key properties of the phenomenon. These properties can serve as building blocks toward a more general and theoretically grounded definition of coordination.

\subsection{General definition and fundamental components of coordinated online behavior}
\label{sec:definition-coordinated-behavior}
We propose a new general definition of coordinated online behavior that overcomes the ambiguities and limitations of the existing definitions. The new definition leverages the components of offline coordination introduced in Section~\ref{sec:defining_coordination} and is informed by the various operational definitions proposed by social media platforms and by the academic literature, respectively discussed in Sections~\ref{sec:smp_definitions} and~\ref{sec:research_definitions}.

\begin{ourdefinition}[Coordinated online behavior]
\label{def:cb-definition}
   $\underbrace{\text{A group of users}}_{\textrm{{\normalfont actors}}}$ $\underbrace{\text{who perform synergic actions}}_{\textrm{{\normalfont actions}}}$ $\underbrace{\text{in pursuit of an intent}}_{\textrm{{\normalfont intent}}}$.
\end{ourdefinition}
Definition~\ref{def:cb-definition} delineates coordinated online behavior based on three fundamental components---\textit{actors}, \textit{actions}, and \textit{intent}---that are similar to the components of offline coordination outlined in Definition~\ref{def:coordination-components}. Definition~\ref{def:cb-definition} and its three fundamental components enable the comprehensive mapping of all instances of online coordination, as discussed in the following. 

\subsubsection{Actors}
\label{sec:actors}
Actors refer to the individuals or entities that are engaged in coordinated behavior. The attributes of the actors contribute to characterizing instances of coordination. For example, the way in which the actors represent themselves to those not involved in the coordinated behavior determines whether the coordination is authentic or otherwise. All instances of online coordination where the actors misrepresent themselves, as in the case of social bots~\cite{hristakieva2022spread,graham2020like,nizzoli2021coordinated,pacheco2021uncovering} and state-backed trolls~\cite{ng2023coordinated}, are cases of inauthentic coordination. Conversely, coordination among actors who accurately self-portray is deemed authentic~\cite{graham2020like}. In addition, the relationships between the actors determine whether the coordination is spontaneous, grassroots, or emergent (i.e., bottom-up)~\cite{nizzoli2021coordinated} or whether it is structured and well-organized (i.e., top-down)~\cite{vargas2020detection}. Finally, the number of the involved actors determines the scale of the coordination.

\subsubsection{Actions}
\label{sec:actions}
Actions represent the practical means that allow actors to coordinate. In coordinated behavior the actions are synergic, in that they are mutually reinforcing and potentially capable of producing a larger effect than that obtainable by individual actions alone. While actors and intent can be misrepresented or concealed, actions are typically visible and non-falsifiable. In other words, the actions with which the actors coordinate represent the digital breadcrumbs of the coordination. For this reason, the actions are the component based on which the majority of coordination detection methods are developed. Furthermore, the types and attributes of the actions also provide information towards characterizing instances of coordination. For example, the timing and synchronization of the actions among actors indicates the degree of planning and organization involved~\cite{zhang2021vigdet,magelinski2022synchronized}. The consistency of the actions and the actions performed in response to external events are further characteristics of coordinated behaviors. Finally, the types of actions and their content provides insights into the intent and goals of the actors~\cite{cinelli2022coordinated}.

\subsubsection{Intent}
\label{sec:intent}
The intent is the objective that the actors pursue when they coordinate. When studying coordinated online behavior, the intent of the actors is typically unknown, if not deliberately concealed. For example, actors involved in malicious or harmful coordination conceal their intent to avoid being stopped in their endeavor~\cite{hristakieva2022spread}. However, also actors involved in neutral or benign forms of coordination might not openly state their goals and intent. For this reason, the observer often tries to infer the intent based on the visible actions performed by the actors. Furthermore, intent can be either shared and explicit among the actors, or implicit. For instance, the perpetrators of a disinformation campaign share the explicit goal of disseminating certain pieces of false information~\cite{keller2020political}. Conversely, fans of a sports team or public character may spontaneously engage in coordinated actions to support their idol, without having agreed on a specific objective or course of action~\cite{nizzoli2021coordinated}. The previous examples highlight that the characteristics of the intent are related to the degree of organization of the actors. Importantly, the intent also contributes to determining the harmfulness of the coordinated behavior. However, while there are cases in which it is relatively straightforward to categorize a coordinated behavior as harmful or otherwise---think for example of coordinated hate attacks~\cite{mariconti2019you} or state-backed disinformation campaigns~\cite{cima2024coordinated}---there also exist situations where the harmfulness of the intent is inherently subjective. Coordinated efforts to promote a controversial political ideology may be perceived as harmful by some, while others may view them as legitimate expressions of free speech~\cite{pacheco2020unveiling}. Similarly, coordinated campaigns to boycott a company or criticize a public figure may be seen as harmful by those targeted, but supporters may genuinely view them as justified forms of activism~\cite{ng2022cross}.

\subsection{Defining dimensions of coordinated online behavior}
\label{sec:facets-coordination}
As discussed in the preceding sections, coordinated online behavior constitutes a complex and multifaceted phenomenon whose instances are contingent upon the actions and intent of the involved actors. Here we leverage our discussion about the fundamental components of online coordination to introduce four defining dimensions of this phenomenon: authenticity, harmfulness, orchestration, and time-variance.

\subsubsection{Authenticity}
\label{sec:dim-authenticity}
Authenticity refers to the degree of genuineness and transparency that the actors exhibit in their actions and overall online presence. Coordinated authentic behavior is executed by genuine actors and typically emerges organically within a community of users who share common interests or beliefs. Examples of authentic coordination are activists, social movements, and mutual support groups, which are driven by motivations such as a desire for social or political change or the cultivation of a sense of community and belonging~\cite{steinert2015online,nizzoli2021coordinated}. While authentic forms of coordination are also harmless in the majority of cases, as in the previous examples, there also exist less frequent cases of authentic yet harmful behaviors. For instance, certain coordinated hate groups openly encourage racist, xenophobic, or supremacist ideologies~\cite{ng2022online,ng2022cross,weber2021amplifying}. 
Conversely, coordinated inauthentic behavior entails the use of fake accounts, such as social bots, trolls, and fake personas~\cite{schliebs2021china,cresci2019cashtag}. These are typically employed for purposes such as spreading disinformation, sowing confusion, or eroding trust in democratic institutions. As such, inauthentic coordination is often characterized by its deceptive nature and aim to manipulate unaware users~\cite{starbird2019disinformation}. Nonetheless, there exist cases of inauthentic yet harmless coordination. For example, online participants in the Arab Spring movements concealed their identities to avoid government surveillance and potential reprisals~\cite{steinert2015online}. While their coordinated efforts were inauthentic in terms of individual identity disclosure, they remained largely harmless in intent, aiming to promote democratic ideals, social justice, and human rights. These examples highlight the difference between authenticity and harmfulness, which represent two orthogonal dimensions of coordinated online behavior.

\subsubsection{Harmfulness}
\label{sec:dim-harmfulness}
Harmfulness refers to the negative impact, consequences, or outcomes---both online and offline---resulting from the coordinated actions of the actors. As discussed in Section~\ref{sec:intent}, harmfulness depends both on the shared intent and actions of the actors engaged in coordination, and on the viewpoint of the observer, constituting a much more conceptually intricate dimension of online coordination than authenticity. However, in spite of the inherent subjectivity, there exist many clear cut cases of harmful and harmless coordination. For example, coordinated actors involved in the spread of disinformation, hate speech, and online harassment, represent straightforward cases of harmful coordination~\cite{ng2022coordinated,cima2024coordinated}. In contrast, users who coordinate to collect and share information and other resources, such as in the aftermath of mass emergencies, represent cases of harmless coordination~\cite{magelinski2022synchronized,pacheco2020unveiling,nizzoli2021coordinated}.

\subsubsection{Orchestration}
Orchestration represents the degree of planning and organization between the coordinated actors. This dimension is closely linked to the intent of the actors, in that highly orchestrated campaigns typically imply shared intent and goals between the participants. The orchestration of a coordinated campaign can be centralized or distributed. In centralized orchestration, a single actor or entity exercises control and coordination over the actions of all other actors involved in the coordinated behavior. This centralized authority dictates the timing, content, and strategy of the coordinated actions, allowing for tight coordination and synchronization. An example of strong and centralized orchestration is the coordinated behavior exhibited by social botnets, where large groups of automated accounts quasi-simultaneously perform predefined actions depending on the command of a botmaster entity~\cite{mannocci2022mulbot}.
In decentralized orchestration, coordination and control are distributed among multiple actors within a network, without a single central authority dictating the actions of all participants. Actors may self-organize, collaborate, or communicate autonomously, often guided by shared goals, interests, or ideologies. For instance, in January 2021, retail investors coordinated on Reddit to target short-selling activity by hedge funds on GameStop shares, causing a surge in the share price and triggering significant losses for the funds involved~\cite{lucchini2022reddit}. Instead, non-orchestrated coordinated behavior occurs when the actions of multiple actors spontaneously converge around a given topic, narrative, or activity. Certain viral social media trends are an example of non-orchestrated coordination emerging from the widespread adoption of a particular hashtag or activity that occurs organically as many users observe and emulate others' behavior~\cite{steinert2015online}.

\subsubsection{Time-variance}
\label{sec:time-variance}
Time-variance refers to the temporal characteristics and the dynamic nature of coordinated online behavior. It grasps possible changes in the types, timing, frequency, and intensity of the actions, which in turn may reflect changes in the intent of the actors, as well as adaptations or responses to external stimuli. Examples of largely static coordinated behavior are the activities of some spammers and bots, who repeatedly perform the same actions adhering to a fixed pattern without much adaptation or variation~\cite{pacheco2020unveiling,cresci2019cashtag}. Conversely, many information operations are dynamic and time-varying, presenting different characteristics at different points in time. Among the changing characteristics are the types of actors involved in the coordination (e.g., whether automated or human-operated) or the topics of discussion~\cite{starbird2019disinformation,vargas2020detection}. Time-variance also strongly depends on the duration of the coordinated behavior itself. Actors involved in certain state-sponsored disinformation campaigns operate on online platforms for extended periods, spanning years or even decades~\cite{vargas2020detection}. Over such lengthy time frames, the actors adapt their tactics, narratives, and targets in response to shifts in intent, changes in technology and platforms, or advancements in countermeasures. This extended duration implies a relatively gradual and nuanced evolution of the coordinated behavior. Conversely, other forms of online coordination rely on expendable or disposable accounts created for short-lived and fast-paced activities~\cite{bellutta2023investigating}. These actors are employed for specific tasks and then discarded or deactivated once their purpose is fulfilled or they are detected. As a result, these ephemeral instances of coordination are rapid, intense, and short-lived.

\subsection{Taxonomy and final remarks}
Our conceptual framework of coordinated online behavior encompasses the three fundamental components presented in Section~\ref{sec:definition-coordinated-behavior} and the four defining dimensions discussed in Section~\ref{sec:facets-coordination}, providing a general and flexible scheme for studying, categorizing, and comprehensively mapping the multiple instances of online coordination.

\subsubsection{Taxonomy}
\begin{wrapfigure}{l}{0.4\textwidth}
\includegraphics[width=0.4\textwidth]{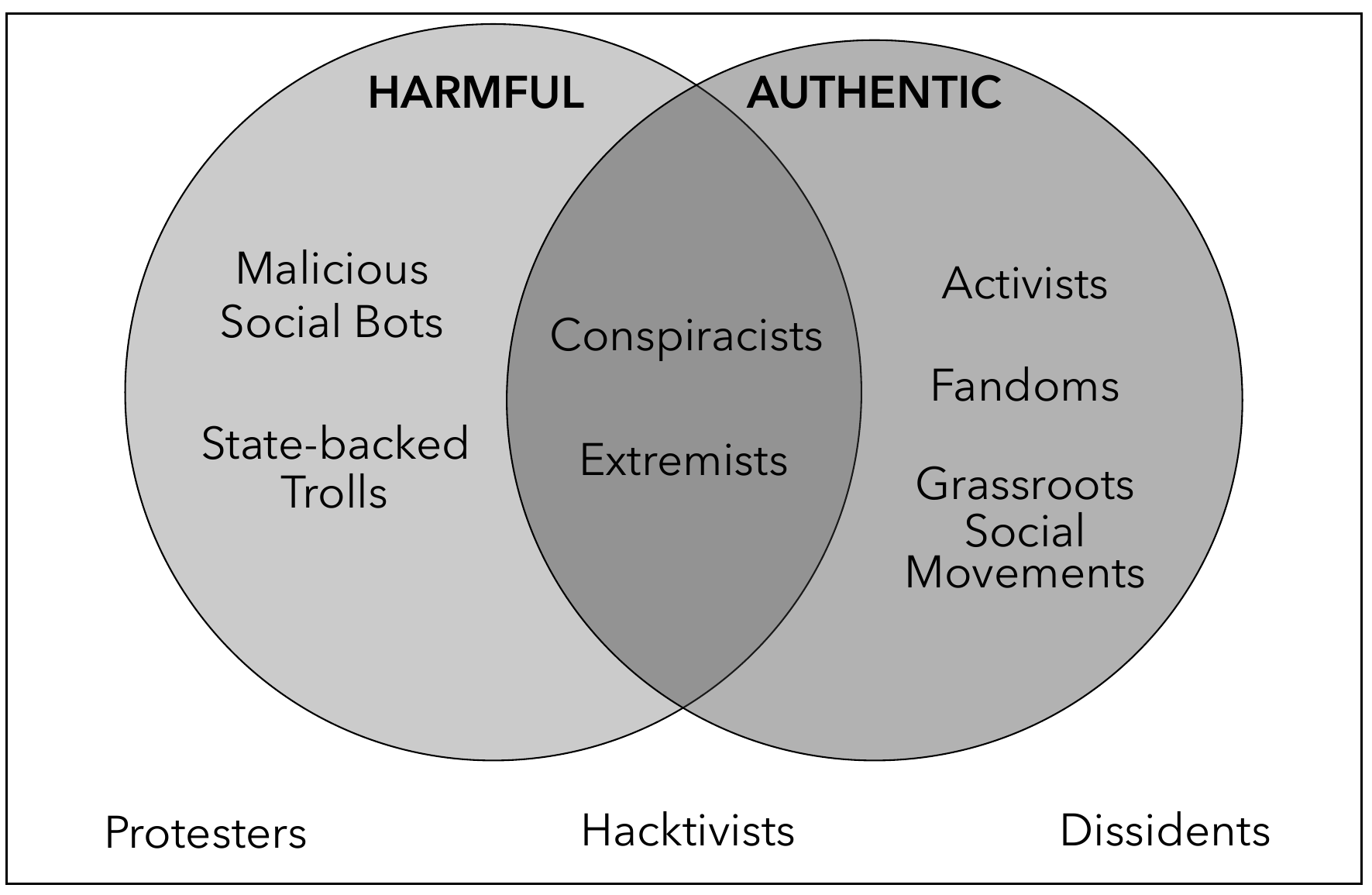}
    \centering
    \caption{Taxonomy of coordinated online behavior obtained by considering the dimensions of \textit{harmfulness} and \textit{authenticity} of our conceptual framework. The framework conveniently allows the mapping of disparate instances of online coordination.}
    \label{fig:venn}
\end{wrapfigure}

Figure~\ref{fig:venn} demonstrates the generality of the conceptual framework by presenting a taxonomy of coordinated online behavior based on the dimensions of authenticity and harmfulness. This analysis takes inspiration from the well-known conceptual framework of \textit{information disorder} that categorizes instances of mis-, mal-, and disinformation based on the falseness and harmfulness of the information~\cite{wardle2017information}. 

As shown in Figure~\ref{fig:venn}, the dimensions of authenticity and harmfulness allow distinguishing between different types of coordinated online behavior, ranging from potentially nefarious to neutral and beneficial ones. Among the most problematic instances of online coordination are harmful and inauthentic phenomena such as information operations and disinformation campaigns, which are often initiated by malicious social bots and state-sponsored trolls~\cite{vargas2020detection,cima2024coordinated,alieva2022investigating}.
Other problematic phenomena are those featuring harmful yet authentic behaviors, such as the activity of conspiracy theorists and hateful extremists~\cite{ng2022cross,ng2022online,mariconti2019you}. Conversely, the activity of coordinated crisis communication by democratic institutions during emergencies~\cite{yoo2024speak}, grassroots social movements, fandoms, and other activists represent instances of harmless and authentic coordination~\cite{steinert2015online}. Finally, harmless yet inauthentic behaviors lay outside of the partially overlapping sets, and are exemplified by hacktivists, dissidents, and other anonymous protesters~\cite{steinert2015online}. Alternative taxonomies can be obtained by leveraging the dimensions of orchestration and time-variance, which would highlight additional phenomena to those shown in Figure~\ref{fig:venn}.

\subsubsection{Final remarks}
The previous instantiation of our conceptual framework in a taxonomy based on the dimensions of harmfulness and authenticity concludes the theoretical part of the survey. The following section bridges the theoretical and methodological parts by presenting the problem definition. Subsequently, we systematically review the proposed methodologies for detecting and characterizing coordinated online behavior, showing how these tasks stem from the conceptual modeling of the phenomenon presented in this section.
 \section{Problem Definition}
\label{sec:problem-definition}

The problem of identifying and investigating different types of coordinated online behavior involves defining two functions $f(\cdot)$ and $g(\cdot)$ that respectively implement the tasks of coordinated behavior \textit{detection} and \textit{characterization}, as outlined in Figure~\ref{fig:problem-definition}. Given a set of users and their actions on one or more online platforms, $f(\cdot)$ identifies possible coordinated groups of users. Instead, $g(\cdot)$ extracts additional information for each detected group, thus contributing to determining the nature, intent, and the overall characteristics of the involved actors (e.g., whether they are inauthentic, harmful, etc.).
The detection and characterization tasks are related to the Definition~\ref{def:cb-definition} of coordinated online behavior and its components in that the function $f(\cdot)$ implementing the detection task does so via the analysis of user \textit{actions}, while the function $g(\cdot)$ implementing the characterization task provides information about the \textit{actors} and their \textit{intent}. third component, i.e.,  \textit{actors} and \textit{intent}, respectively. 
A comprehensive overview of the entire process is depicted in Figure \ref{fig:problem-definition}. The input is represented by the set of users $U$ to analyze and their activities $H$. The detection task differentiates coordinated users from non-coordinated ones. Depending on the detection method, the distinction between the two can be expressed as binary labels assigned to the users, as two or more sets (e.g., clusters) of either coordinated or non-coordinated users, or as two or more coordinated or non-coordinated communities (i.e., nodes and edges) from a network.
These are subsequently scrutinized during the characterization task, which computes a set of indicators for each coordinated user, set, or community.
The indicators are selected so as to provide information about the characteristics of the coordinated actors and their behavior. For example, computing bot scores is a common method to estimate the inauthenticity of coordinated users. 
\def\rectWhalf{4/2}
\def\rectHhalf{8.7/2}
\def\spaceRect{0.6}
\def\opacity{0.05}
\def\fillCol{gray}

\def\titleY{3.6}
\def\spaceY{0.6}
\newcommand{\titleFont}{\LARGE}
\newcommand{\labelFont}{\LARGE}

\def\userY{\spaceY/2} \def\activityY{-6*\spaceY} \def\timeX{1.8} \def\timeY{\activityY+3*\spaceY+\tsY} \def\tsY{0.3} \def\symbolY{0.25} \def\trAUsersY{\userY+3*\spaceY} 

\def\bbY{\titleY-2.8} \def\appY{\bbY-2.5}

\def\trH{0.6} \def\trW{0.6} 

\def\trAX{\trW} \def\trAY{\titleY-2.8*\spaceY} \def\comY{\trAY-1.3*\spaceY} 

\def\trAYbis{\comY-2.8*\spaceY} \def\clusY{\trAYbis-1.3*\spaceY} 

\def\trAYtris{\clusY-2.8*\spaceY} \def\topYline{\trAYtris+\trW+0.2} \def\bottomYline{\topYline-1.7} \def\binaryY{\trAYtris-1.3*\spaceY} 

\def\iconLabelY{\titleY-2}
\def\indicatorY{\binaryY} 

\begin{figure*}[t]
\centering
    \resizebox{0.94\textwidth}{!}{
        \begin{tikzpicture}[inner sep=0pt, outer sep=0pt,
                    mynode/.style={
                    	circle,
                            draw=none,
                    	text opacity=1,
                    	inner sep=0pt,
                    	minimum size=0.2cm,
                    	on grid},
                    circlenode/.style={
                            circle,
                            fill=none, text opacity=1,
                            inner sep=0pt,
                            minimum size=1.1cm,
                            dashed,
                            on grid
                        },
                    labelnode/.style={
                            circle,
                            fill=none, draw=none, text opacity=1,
                            inner sep=0pt,
                            minimum size=0.5cm,
                            on grid
                        }]

\draw (-\rectWhalf,-\rectHhalf) rectangle (\rectWhalf, \rectHhalf);

\draw[fill=\fillCol, fill opacity=\opacity] (\rectWhalf+\spaceRect, -\rectHhalf) rectangle (3*\rectWhalf+\spaceRect, \rectHhalf);

\draw (3*\rectWhalf+2*\spaceRect, -\rectHhalf) rectangle (5*\rectWhalf+2*\spaceRect, \rectHhalf);

\draw[fill=\fillCol, fill opacity=\opacity] (5*\rectWhalf+3*\spaceRect, -\rectHhalf) rectangle (7*\rectWhalf+3*\spaceRect, \rectHhalf);

\draw (7*\rectWhalf+4*\spaceRect, -\rectHhalf) rectangle (9*\rectWhalf+4*\spaceRect, \rectHhalf);

\draw[line width=1mm, ->, >=latex] (\rectWhalf,0) -- (\rectWhalf+\spaceRect,0);
            \draw[line width=1mm, ->, >=latex] (3*\rectWhalf+\spaceRect,0) -- (3*\rectWhalf+2*\spaceRect,0);
            \draw[line width=1mm, ->, >=latex] (5*\rectWhalf+2*\spaceRect,0) -- (5*\rectWhalf+3*\spaceRect,0);
            \draw[line width=1mm, ->, >=latex] (7*\rectWhalf+3*\spaceRect,0) -- (7*\rectWhalf+4*\spaceRect,0);

\begin{scope}[shift={(-\rectWhalf+\rectWhalf+0*\spaceRect,0)}]                        
                \node[draw, labelnode, font=\titleFont] (G) at (0, \titleY) {\textbf{input}};

\node[draw, mynode, fill=black] (A) at (-\trAX, \trAUsersY+\trW) {};
                \node[draw, mynode, fill=black] (B) at (-\trAX+\trW, \trAUsersY+\trW) {};
                \node[draw, mynode, fill=black] (F) at (-\trAX+2*\trW, \trAUsersY+\trW) {};

\node[draw, mynode, fill=black] (C) at (-\trAX+\trW/2, \trAUsersY) {};
                \node[draw, mynode, fill=black] (D) at (-\trAX+3*\trW/2, \trAUsersY) {};
                \node[draw, mynode, fill=black] (E) at (-\trAX+5*\trW/2, \trAUsersY) {};

                \node[draw, labelnode, font=\Large] (H) at (0, \userY+1.5*\spaceY) {$U=\{u_1, \dots, u_N\}$};
                \node[draw, labelnode, font=\labelFont] (I) at (0, \userY) {\textit{users (U)}};

\draw[->, solid] (-\timeX, \timeY) -- (\timeX, \timeY);
                
                \node[draw, labelnode] (K) at (-1.6, -\tsY+\timeY) {$t_1$};
                \node[draw, labelnode] (K1) at (-1.6, \timeY+\symbolY) {\faRetweet};

                \node[draw, labelnode] (L) at (-0.9, -\tsY+\timeY) {$t_2$};
                \node[draw, labelnode] (L1) at (-0.9,  \timeY+\symbolY) {\faReply};
                
                \node[draw, labelnode] (M) at (-0.2, -\tsY+\timeY) {\ldots};
                \node[draw, labelnode] (M) at (-0.2,  \timeY+\symbolY) {\ldots};
                
                \node[draw, labelnode] (N) at (0.5, -\tsY+\timeY) {$t_{m-1}$};
                \node[draw, labelnode] (N2) at (0.5, \timeY+\symbolY) {\faThumbsOUp};
                
                \node[draw, labelnode] (O) at (1.4, -\tsY+\timeY) {$t_m$};
                \node[draw, labelnode] (O1) at (1.4,  \timeY+\symbolY) {\faRetweet};

                 \node[draw, labelnode, font=\Large] (J) at (0, \activityY+1.5*\spaceY) {$H=\{H_1, \dots, H_N\}$};
                
                \node[draw, labelnode, font=\labelFont] (Q) at (0, \activityY) {\textit{activities (H)}};
            \end{scope}

\begin{scope}[shift={(2*\rectWhalf+\spaceRect,0)}]
                \node[draw, labelnode, font=\titleFont] (A) at (0, \titleY) {\textbf{detection task}};
                
                \node (image) at (0, \bbY) {\includegraphics[width=2cm]{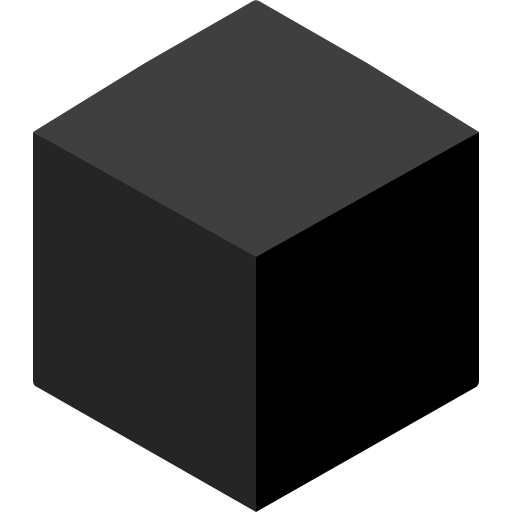}};
                \node[xshift=0cm, yshift=0.5cm, text=white, font=\Large] at (image) {$f(x)$};
                
\node[draw, labelnode, font=\Large] (C) at (0, \appY-\spaceY) {network science};
                \node[draw, labelnode, font=\Large] (D) at (0, \appY-2*\spaceY) {data mining};
                \node[draw, labelnode, font=\Large] (D) at (0, \appY-3*\spaceY) {machine learning};
            
            \end{scope}

\begin{scope}[shift={(4*\rectWhalf+2*\spaceRect,0)}]
                    \node[draw, labelnode, font=\titleFont] (I) at (0, \titleY) {\textbf{detection output}};

                    \node[draw, mynode, fill=color1] (A) at (-\trAX, \trAY+\trW) {};
                    \node[draw, mynode, fill=color1] (B) at (-\trAX+\trW, \trAY+\trW) {};
                    \node[draw, mynode, fill=color1] (C) at (-\trAX+\trW/2, \trAY) {};
                    \node[draw=color1, circlenode] (G) at (-\trAX+\trW/2, \trAY+\trW/2+0.05) {};
    
                    \node[draw, mynode, fill=color2] (D) at (-\trAX+3*\trW/2, \trAY) {};
                    \node[draw, mynode, fill=color2] (E) at (-\trAX+5*\trW/2, \trAY) {};
                    \node[draw, mynode, fill=color2] (F) at (-\trAX+2*\trW, \trAY+\trW) {};
                   \node[draw=color2, circlenode] (H) at (-\trAX+2*\trW, \trAY+\trW/2-0.05) {};
                    
\draw[thick] (A) -- (B);
                    \draw[thick] (A) -- (C);
                    \draw[thick] (B) -- (C);
                    \draw[thick] (D) -- (E);
                    \draw[thick] (D) -- (F);
                    \draw[thick] (E) -- (F);
            
                    \node[draw, labelnode, , font=\labelFont] (K) at (0, \comY) {\textit{communities (P)}};

                    \node[draw, mynode, fill=color1] (A2) at (-\trAX, \trAYbis+\trW) {};
                    \node[draw, mynode, fill=color1] (B2) at (-\trAX+\trW, \trAYbis+\trW) {};
                    \node[draw, mynode, fill=color1] (C2) at (-\trAX+\trW/2, \trAYbis) {};
                    \node[draw=color1, circlenode] (G2) at (-\trAX+\trW/2, \trAYbis+\trW/2+0.05) {};
    
                    \node[draw, mynode, fill=color2] (D2) at (-\trAX+3*\trW/2, \trAYbis) {};
                    \node[draw, mynode, fill=color2] (E2) at (-\trAX+5*\trW/2, \trAYbis) {};
                    \node[draw, mynode, fill=color2] (F2) at (-\trAX+2*\trW, \trAYbis+\trW) {};
                   \node[draw=color2, circlenode] (H2) at (-\trAX+2*\trW, \trAYbis+\trW/2-0.05) {};
                    
                    \node[draw, labelnode, font=\labelFont] (I2) at (0, \clusY) {\textit{clusters (C)}};

                    \node[draw, mynode, fill=color1] (A3) at (-\trAX, \trAYtris+\trW) {};
                    \node[draw, mynode, fill=color1] (B3) at (-\trAX+\trW, \trAYtris+\trW) {};
                    \node[draw, mynode, fill=color1] (C3) at (-\trAX+\trW/2, \trAYtris) {};

                    \node[draw, mynode, fill=color2] (D3) at (-\trAX+3*\trW/2, \trAYtris) {};
                    \node[draw, mynode, fill=color2] (E3) at (-\trAX+5*\trW/2, \trAYtris) {};
                    \node[draw, mynode, fill=color2] (F3) at (-\trAX+2*\trW, \trAYtris+\trW) {};
            
\draw[dashed] (-\trAX+3*\trW/2, \trAYtris+3*\trW/2) -- (-\trAX+\trW, \trAYtris-\trW/2);

                    \node[draw, labelnode, font=\LARGE] (I3) at (0, \binaryY) {\textit{labels (B)}};
            \end{scope}
            
\begin{scope}[shift={(6*\rectWhalf+3*\spaceRect,0)}]
                \node[draw, labelnode, font=\titleFont, align=center] (A) at (0, \titleY) {\textbf{characterization}\\\textbf{task}};
            
                \node (image) at (0, \bbY) {\includegraphics[width=2cm]{renamed_images/figure_6_cube.png}};
                \node[xshift=0cm, yshift=0.5cm, text=white, font=\Large] at (image) {$g(x)$};

\node[draw, labelnode, font=\Large] (C) at (0, \appY-0*\spaceY) {inauthenticity};
                \node[draw, labelnode, font=\Large] (D) at (0,  \appY-1*\spaceY) {harmfullnes};
                \node[draw, labelnode, font=\Large] (E) at (0,  \appY-2*\spaceY) {orchestration};
                \node[draw, labelnode, font=\Large] (E) at (0,  \appY-3*\spaceY) {time-variance};
\end{scope}

\begin{scope}[shift={(8*\rectWhalf+4*\spaceRect,0)}]
                \node[draw, labelnode, font=\titleFont, align=center] (A) at (0, \titleY) {\textbf{characterization}\\\textbf{output}};
            
                \node[draw, labelnode, font=\Large] (B) at (0, \iconLabelY+\spaceY) {\faWarning};
                \node[draw, labelnode, font=\Large] (C) at (0, \iconLabelY) {toxicity score};
                
                \node[draw, labelnode, font=\Large] (D) at (0, \iconLabelY-2*\spaceY) {\faSmileO \hspace{0.3em} \faMehO \hspace{0.3em} \faFrownO};
                \node[draw, labelnode, font=\Large] (B) at (0, \iconLabelY-3*\spaceY) {sentiment score};
            
                \node[draw, labelnode, font=\Large] (B) at (0, \iconLabelY-5*\spaceY) {\faGears}; \node[draw, labelnode, font=\Large] (B) at (0, \iconLabelY-6*\spaceY) {bot score};

                \node[draw, labelnode] (G) at (0, \iconLabelY-7*\spaceY) {\vdots};
            
                \node[draw, labelnode, font=\labelFont] (B) at (0, \indicatorY) {\textit{indicators (M)}};
            \end{scope}
        \end{tikzpicture}
    }
    \caption{The analytical process of studying coordinated online behavior, involving the \textit{detection} and \textit{characterization} tasks. The input to the overall process is a set of users $U$ and their activities $H$ on one or more platforms. The output of the detection task is either a set of binary labels $B$, clusters $C$, or network communities $G$ that differentiate coordinated and non-coordinated users. The characterization task receives these in input and outputs a set of indicators $M$.}
    \label{fig:problem-definition}
\end{figure*}
 
\subsection{Detection of coordinated online behavior}
\subsubsection{Input}
Let $I=\langle U, H \rangle$ be the problem input, where $U=\{u_1,\ldots, u_N\}$ denotes the set of users and $H=[H^{u_1}, \ldots, H^{u_N}]$ represents an ordered vector of activities performed by those users. We define the activity of a user $u_j$ as $H^{u_j}=[h_1^{u_j}, \ldots, h_T^{u_j}]$ representing the vector of chronologically ordered actions  performed by $u_j$. An action is defined by the quadruple $h=\langle type, target, content, timestamp \rangle$ which describes the \textit{type} of action executed by a user on a specific \textit{target} or \textit{content}, at a given \textit{timestamp}. Users can execute actions of different \textit{type} such as posting, resharing, befriending, and more. A \textit{target} is another user of the platform who is affected by the action. For example, in the case of a retweet action on the platform Twitter/X, the target is the author of the retweeted tweet. For some actions the target is undefined, as in the case of the posting action.
The \textit{content} of an action is a post (e.g., a tweet, comment, submission, and more, depending on the platform). Posts contain one or more elements of content, such as text, image, URL, mention, hashtag, and more. In case the content contains multiple elements, the corresponding action is called compound action~\cite{magelinski2022synchronized}. Similarly to the target, also the content can be undefined depending on the type of action, as in the case of a befriending or following action. To wrap up, the type of action and its timestamp are always defined, while one of content and target might be optional, depending on the type of action.

\subsubsection{Methods and output}
As presented in Section~\ref{sec:detection}, most of the existing literature on coordinated behavior detection analyzes both the set of users and their actions. Some studies only leverage the content of the actions, without considering their type~\cite{mariconti2019you,suresh2023tracking,danaditya2022curious}. Furthermore, certain works do not take into consideration the timings of the actions~\cite{mariconti2019you,cao2015organic,chomel2023manipulation,cinelli2022coordinated}, while some others only consider the timings~\cite{bellutta2023investigating}. The task of detecting coordinated online behavior is modelled by the function $f(U,H)$, which can provide three different outputs depending on the adopted method, corresponding to different levels of detail and information on the coordinated users:
\begin{equation}
\label{eq:output-detection}
    f(U,H)=
    \begin{cases}
        \begin{aligned}
            & P=\{P_1, \ldots, P_i, \ldots P_k\}, && P_i=(V_i,E_i), V_i \subseteq U && \text{communities}\\
            & C=\{C_1, \ldots, C_i, \ldots C_k\}, &&C_i \subseteq U && \text{clusters}\\
            & B=\{B_c, B_u\}, && B_c  \cup B_u = U && \text{binary labels} \\
        \end{aligned}
    \end{cases}    
\end{equation}
In the most general case, the output of $f(\cdot)$ is a set $P$ of \textit{communities} of coordinated users. Coordinated communities $P_i$ are sub-networks where the nodes are users from $U$, and the edges---with their weights---encode the level of coordination among the users. Communities are typically outputted by those methods that adopt an internal network representation, which is then analyzed with community detection algorithms. Coordinated communities are an information rich representation, given that the presence and weight of links between the coordinated users facilitates subsequent analyses, such as those needed for the characterization task. 
Another possible output consists of a set of \textit{clusters} of users. The clusters $C_i$ are produced by methods that adopt tabular representations of the users, which are then analyzed with clustering algorithms. These methods typically ignore the relationships between the users but are able to identify multiple groups of coordinated users. 
Finally, the least informative output is given by those methods based on classification algorithms. These methods assign \textit{binary labels}, partitioning the initial set of users $U$ in two labeled groups of coordinated ($B_c$) and non-coordinated ($B_u$) users. These labelled groups do not provide information about neither the relationships between the users nor the existence of multiple coordinated groups of users in $U$.

\subsection{Characterization of coordinated online behavior}
\subsubsection{Input}
The characterization task is modelled by the function $g(Y,H) = M$ whose inputs are the groups of coordinated users resulting from the detection task $f(U,H) = Y \in \{P, C, B\}$, defined in Eq.~\eqref{eq:output-detection}, with their activities $H$.

\subsubsection{Methods and output}
As discussed in Section~\ref{sec:characterization}, the characterization task aims at computing a set of quantitative indicators $M$ to measure distinctive properties of the detected coordinated behaviors in terms of the defining dimensions that we presented in Section~\ref{sec:facets-coordination}: authenticity, harmfulness, orchestration, and time-variance. The indicators that can be used in the characterization task partly depend on the methods and outputs of the detection task. For example, assortativity measures the extent to which nodes with a high degree in a network are connected to other nodes with a high degree, and vice versa. This indicator was used to gain insights into the inner structure and organization of certain coordinated communities~\cite{nizzoli2021coordinated}. However, assortativity can be computed only if the coordination detection method outputs communities, rather than clusters or binary labels. On the contrary, other indicators can be computed independently of the detection method, such as the aforementioned bot scores that are commonly used as an estimator of the inauthenticity of the coordinated users~\cite{hristakieva2022spread,graham2020like,nizzoli2021coordinated,pacheco2021uncovering}. The utility of the characterization task is not limited to shedding light on the nature of the detected coordinated behaviors nor to distinguishing between different instances of the phenomenon. In fact, the output of characterization task can also be leveraged to validate the output of the detection task, as in those frequent cases when a ground-truth of coordinated users is unavailable. 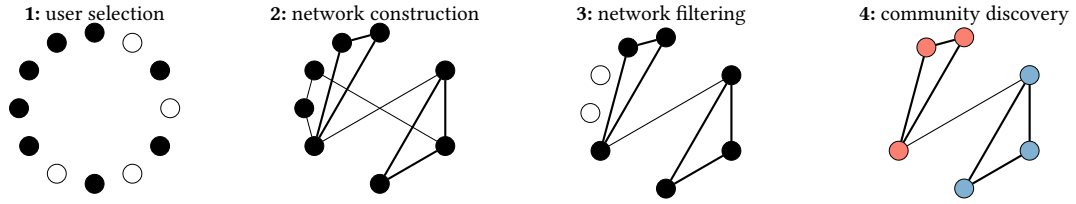
\begin{figure}[t]
    \centering
    \begin{minipage}[t]{0.22\textwidth}
        \centering
        \small{\textbf{1:} user selection}
        \scalebox{0.5}{
\begin{tikzpicture}[
                    mynode/.style={
                    	circle,
                    	draw=black,
                    	text opacity=1,
                    	inner sep=0pt,
                    	minimum size=0.5cm,
                    	on grid}]

\foreach \i/\col in {1/black, 2/white, 3/black, 4/black, 5/black, 6/black, 7/black, 8/white, 9/black, 10/white, 11/black, 12/white} {
                \pgfmathsetmacro{\angle}{30 + 30 * (\i - 1)}
                \node[mynode] (\i) at (\angle:2) [fill=\col] {};
            }
        \end{tikzpicture}
        }
    \end{minipage}\hspace{0.02\textwidth}
\begin{minipage}[t]{0.22\textwidth}
        \centering
        \small{\textbf{2:} network construction} \scalebox{0.5}{
\begin{tikzpicture}[
                        mynode/.style={
                        	circle,
                        	draw=black,
                        	text opacity=1,
                        	inner sep=0pt,
                        	minimum size=0.5cm,
                        	on grid}]

\foreach \i/\col in {1/black, 3/black, 4/black, 5/black, 6/black, 7/black, 9/black, 11/black} {
                    \pgfmathsetmacro{\angle}{30 + 30 * (\i - 1)}
                    \node[mynode] (\i) at (\angle:2) [fill=\col] {};
                }
                
\draw[ultra thick] (11) -- (1);
                \draw[ultra thick] (9) -- (1);
                \draw[ultra thick] (9) -- (11);
                \draw (11) -- (5);
                \draw[thick] (7) -- (1);
                
                \draw[ultra thick] (7) -- (3);
                \draw[ultra thick] (7) -- (4);
                \draw[ultra thick] (3) -- (4);
                \draw (5) -- (6);
                \draw (6) -- (7);
            \end{tikzpicture}
        }
    \end{minipage}
    \hspace{0.02\textwidth}
\begin{minipage}[t]{0.22\textwidth}
        \centering
        \small{\textbf{3:} network filtering}
        \scalebox{0.5}{
\begin{tikzpicture}[
                        mynode/.style={
                        	circle,
                        	draw=black,
                        	text opacity=1,
                        	inner sep=0pt,
                        	minimum size=0.5cm,
                        	on grid}]

\foreach \i/\col in {1/black, 3/black, 4/black, 5/white, 6/white, 7/black, 9/black, 11/black} {
                    \pgfmathsetmacro{\angle}{30 + 30 * (\i - 1)}
                    \node[mynode] (\i) at (\angle:2) [fill=\col] {};
                }
                
\draw[ultra thick] (11) -- (1);
                \draw[ultra thick] (9) -- (1);
                \draw[ultra thick] (9) -- (11);
            
                \draw[thick] (7) -- (1);
                \draw[ultra thick] (7) -- (3);
                \draw[ultra thick] (7) -- (4);
                \draw[ultra thick] (3) -- (4);
            
            \end{tikzpicture}
        }
    \end{minipage}
\hspace{0.02\textwidth}
    \begin{minipage}[t]{0.25\textwidth}
        \centering
        \small{\textbf{4:} community discovery}
        \scalebox{0.5}{
\begin{tikzpicture}[
                        mynode/.style={
                        	circle,
                        	draw=black,
                        	text opacity=1,
                        	inner sep=0pt,
                        	minimum size=0.5cm,
                        	on grid}]

\foreach \i/\col in {1/color1, 3/color2, 4/color2, 7/color2, 9/color1, 11/color1} {
                    \pgfmathsetmacro{\angle}{30 + 30 * (\i - 1)}
                    \node[mynode] (\i) at (\angle:2) [fill=\col] {};
                }
                
\draw[ultra thick] (11) -- (1);
                \draw[ultra thick] (9) -- (1);
                \draw[ultra thick] (9) -- (11);
            
                \draw[thick] (7) -- (1);
                \draw[ultra thick] (7) -- (3);
                \draw[ultra thick] (7) -- (4);
                \draw[ultra thick] (3) -- (4);
            \end{tikzpicture}
        }
    \end{minipage}
    \caption{Main steps of the network science methods for the detection of coordinated online behavior. \textbf{1:} The selected users become nodes in a network. \textbf{2:} User similarities are computed with a similarity function and assigned to the edge weights of the network. \textbf{3:} The network is filtered so as to retain only similarities with given properties. \textbf{4:} Community discovery is performed to detect groups of strongly coordinated users.}
\label{fig:tikz_networks}
\end{figure}

\section{Detection of coordinated behavior}
\label{sec:detection}
Coordination detection methods can be classified into two main categories depending on their underlying approach: network science or machine learning. The following sections discuss the existing solutions in each category.

\subsection{Network science methods}
\label{sec:network-science-approaches}
Network science coordination detection methods build a network of users or posts, where the links between the nodes in the network represent the presence, and possibly also the extent~\cite{nizzoli2021coordinated}, of coordination. In spite of the existing differences, all methods in this category carry out the sequence of steps shown in Figure~\ref{fig:tikz_networks}, namely: (\textit{i}) user selection, (\textit{ii}) coordination network construction, (\textit{iii}) network filtering, and (\textit{iv}) community discovery. We now discuss the objective and the implementation options available for each step.

\subsubsection{User selection} This step selects an initial subset $U' \subseteq U$ of users according to some criteria that depend on the purpose of the analysis. This initial selection is motivated by the observation that a small fraction of users accounts for the majority of actions on a social network, especially those associated with harmful behaviors~\cite{nizzoli2021coordinated}. Selecting a subset of users also has the positive consequence of reducing the computational cost of the subsequent steps, which can easily become time- and computation-intensive for large networks. \vspace{0.1cm}

\textit{Implementation.} Multiple choices can be made to select a subset of relevant users. A common choice is to select the most active users as they produce the largest share of actions and content. Most active users can be defined as those who publish a large number of original posts (\textit{super-producers})~\cite{lee2013campaign,magelinski2020detecting,pacheco2021uncovering}, or as those having a large number of re-shares (\textit{super-spreaders})~\cite{nizzoli2021coordinated,cinelli2022coordinated,hristakieva2022spread,tardelli2024multifaceted,loru2024influence,di2025post,neha2024understanding,loru2025compression}. Other than activity, network centrality or influence, suspicious behavior, location, and timings are used for user selection~\cite{luceri2024unmasking}. Furthermore, many other general criteria can also be adopted, such as selecting all users who posted certain keywords, or all followers of a given user. Combining multiple criteria allows for even more fine-grained user selections.

\subsubsection{Coordination network construction.} 
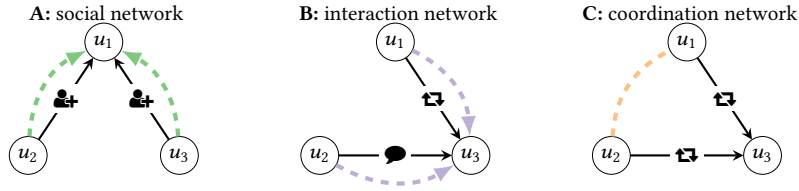
\begin{figure}
    \centering
    \begin{subfigure}{0.25\textwidth}
        \centering
        \small{\textbf{A:} social network}
        \begin{tikzpicture}[
            mynode/.style={
            circle,
            draw=black,
            text opacity=1,
            inner sep=0pt,
            minimum size=0.5cm,
            on grid}]
            
\path[] (0,0) rectangle (2cm,2cm);

\node[mynode,ultra thin,fill=white] (u1) at (1,2) {$u_1$};
            \node[mynode,ultra thin,fill=white,below left=1.5cm and 1cm of u1] (u2) {$u_2$};
\node[mynode,ultra thin,fill=white,below right=1.5cm and 1cm of u1] (u3) {$u_3$};

            \draw[-stealth,thick] (u3) -- (u1)
                node[midway, fill=white, inner sep=2pt] {\faUserPlus};
            \draw[-stealth,thick] (u2) -- (u1)
                node[midway, fill=white, inner sep=2pt] {\faUserPlus};

\draw[ultra thick,dashed,bend left=30, color3, ->, >=latex]
            (u2) to (u1);
            \draw[ultra thick,dashed,bend right=30, color3, ->, >=latex]
            (u3) to (u1);
        \end{tikzpicture} 
        \label{fig:social-network}
    \end{subfigure}
\begin{subfigure}{0.25\textwidth}
        \centering
        \small{\textbf{B:} interaction network}
        \begin{tikzpicture}[
            mynode/.style={
            	circle,
            	draw=black,
            	text opacity=1,
            	inner sep=0pt,
            	minimum size=0.5cm,
            	on grid}]
            	
\path[] (0,0) rectangle (2cm,2cm);

\node[mynode,ultra thin,fill=white] (u1) at (1,2) {$u_1$};
            	\node[mynode,ultra thin,fill=white,below left=1.5cm and 1cm of u1] (u2) {$u_2$};
            	\node[mynode,ultra thin,fill=white,below right=1.5cm and 1cm of u1] (u3) {$u_3$};
                
\draw[-stealth,thick] (u1) -- (u3)
                        node[midway, fill=white, inner sep=2pt] {\faRetweet};
                    \draw[-stealth,thick] (u2) -- (u3)
                        node[midway, fill=white, inner sep=2pt] {\faComment};

\draw[ultra thick,dashed,bend left=30, color4, ->, >=latex]
                    (u1) to (u3);
                    \draw[ultra thick,dashed,bend right=30, color4, ->, >=latex]
                    (u2) to (u3);
        \end{tikzpicture}
        \label{fig:interaction-network}
    \end{subfigure}
\begin{subfigure}{0.25\textwidth}
        \centering
        \small{\textbf{C:} coordination network}
        \begin{tikzpicture}[
            mynode/.style={
            	circle,
            	draw=black,
            	text opacity=1,
            	inner sep=0pt,
            	minimum size=0.5cm,
            	on grid}]
            	
\path[] (0,0) rectangle (2cm,2cm);

\node[mynode,ultra thin,fill=white] (u1) at (1,2) {$u_1$};
            	\node[mynode,ultra thin,fill=white,below left=1.5cm and 1cm of u1] (u2) {$u_2$};
            	\node[mynode,ultra thin,fill=white,below right=1.5cm and 1cm of u1] (u3) {$u_3$};

                    \draw[-stealth,thick] (u1) -- (u3)
                        node[midway, fill=white, inner sep=2pt] {\faRetweet};
                    \draw[-stealth,thick] (u2) -- (u3)
                        node[midway, fill=white, inner sep=2pt] {\faRetweet};
                    
\draw[dashed,ultra thick,bend left=30, color5]
                    (u2) to (u1);
        \end{tikzpicture}
        \label{fig:coordination-network}
    \end{subfigure}
    \caption{Differences between social (\textbf{A}), interaction (\textbf{B}), and coordination (\textbf{C}) networks. Solid black edges represent actions on the online platform, while dashed colored edges show how actions are translated into edges in the corresponding type of network. Coordination networks are typically undirected and link users performing similar actions at around the same time. Differently to social and interaction networks, coordination networks allow connecting users even if they never directly interact with one another.}
\label{fig:network-difference}
\end{figure}

This step builds a coordination network between the users $U' \subseteq U$ previously selected.\footnote{A few works follow the general approach of network science methods, but build networks of \textit{content} rather than \textit{users}. These are discussed in Section~\ref{sec:content-networks}.} A coordination network is a type of network where links exist only between coordinated nodes. As per Definition~\ref{def:cb-definition}, coordination implies synergic actions between users. In network science methods, this concept is operationalized with \textit{co-actions}. A co-action represents two users performing the same action on the same target or content. For instance, two users who comment, like, or re-share the same post are generating a co-action. As shown in Figure~\ref{fig:network-difference}, the two coordinated users need not be directly connected in the social or interaction network. This characteristic makes coordination networks particularly suitable for surfacing coordination between seemingly unrelated users, such as those involved in inauthentic or harmful behavior, so much so that some scholars specifically refer to \textit{latent} coordination networks~\cite{weber2021amplifying}. In the most general case, a coordination network is a multiplex network $G(V, E, W, \mathcal{L})$. In order to build $G$, one must define the types of co-actions $C_a$ to consider (e.g., re-shares, mentions, follows, etc.). When multiple co-actions are used, $G$ is a multiplex network with $L$ layers, where $\mathcal{L}=\{1,\dots,L\}$ is the set of layers, each corresponding to a co-action $i \in C_a$. However, the majority of existing works leverage a single co-action. In this case, the number of layers is $L=1$ and $G$ is a single layer network. Each layer in $G$ is an undirected weighted graph $G^i(V^i, E^i, W^i )$, where $V^i \subseteq U'$, $E^i$ and $W^i$ respectively denote the set of nodes, edges, and weights of layer $i$. We highlight that $V= \bigcup_{i \in \mathcal{L}} V^i$, $E= \bigcup_{i \in \mathcal{L}} E^i$, and $W= \bigcup_{i \in \mathcal{L}} W^i$. Given a layer $i$, an edge $e^i_{jk}$ is created if there exists a co-action of type $i$ between two users $(u_j, u_k)$. The edge weight $w^i_{jk}=sim_i(u_j, u_k)$ is obtained via a similarity function that computes pairwise user similarities in $U'$. Different similarity functions can be used for different co-actions. \vspace{0.1cm}

\begin{table}
    \caption{Network science methods for detecting coordinated behavior based on \textit{single layer} user networks. For each group of works we report the considered co-actions, similarity functions, filtering criteria, and community detection methods.} \label{tab:single-network}
    \centering
    \footnotesize
    \setlength{\tabcolsep}{3pt}
\scalebox{0.915}{
    \begin{tabular}{L{2.3cm}L{3.8cm}L{2.9cm}L{2.2cm}L{2.7cm}}
        \toprule
        \textbf{reference} & \textbf{action} & \textbf{similarity} & \textbf{filters$^\dagger$} & \textbf{community detection}\\
        \midrule
\cite{cima2024coordinated} & retweet & cardinality & threshold, ADJ & modularity clustering \\ [0.5ex]
        \cite{graham2020like} & retweet & cardinality & EDO & Louvain \\ [0.5ex]
        \cite{kulichkina2025protest,kulichkina2026connective} & retweet & cardinality & threshold, ADO & Louvain \\ [0.5ex]
        \cite{linhares2022uncovering} & retweet & cardinality & backbone, ADJ & Louvain \\ [0.5ex]
        \cite{schliebs2021china} & retweet & cardinality & EDO & \\ [0.5ex]
        \cite{nizzoli2021coordinated,cinelli2022coordinated,hristakieva2022spread,loru2024influence,di2025post,tardelli2024multifaceted} & retweet & cosine similarity TF-IDF & backbone & Louvain \\ [0.5ex]
        \cite{tardelli2024temporal,sartori2025insights} & retweet & cosine similarity TF-IDF & backbone, EDO & Leiden \\ [0.5ex]
        \cite{weber2021temporal} & retweet & cardinality & threshold, ADJ & \\ [0.5ex]
        \cite{axelrod2025structure} & retweet & cosine similarity & kNN graph, correlation  & HDBSCAN \\ [0.5ex]

        \cite{chomel2023manipulation} & retweet, tweet & cardinality & threshold, EDO & Louvain, connected components \\ [0.5ex]
        \cite{keller2020political} & retweet, tweet & cardinality & threshold, EDO & Louvain \\ [0.5ex]
        \cite{schoch2022coordination} & retweet, tweet & cardinality & threshold, EDO & \\ [0.5ex]

        \cite{blas2025large} & tweet & cosine similarity & threshold, EDO & \\ [0.5ex] 
        \cite{de2024detecting} & tweet & cardinality & ADO & \\ [0.5ex]
        \cite{lee2013campaign} & tweet & cardinality & threshold & cohesive campaign \\ [0.5ex]
        \cite{pacheco2020unveiling} & tweet & text similarity & threshold, ADO & \\ [0.5ex]
        \cite{ng2023coordinating} & parley & cardinality & threshold, kNN graph & Leiden \\ [0.5ex]
        \cite{venancio2024unraveling} & text & cardinality & backbone & Louvain \\ [0.5ex]
        \cite{wu2025unmasking} & tweet &  & threshold, kNN graph, ADO & \\ [0.5ex]
        \cite{kansaon2025fake} & message & cardinality & threhsold, ADO & Louvain \\ [0.5ex]
        \cite{ng2023you} & tweet, URL  & cardinality,cosine similarity & threshold, EDO & Louvain \\ [0.5ex]
        \cite{ng2022cross} & tweet, parley, URL, username & text similarity, cardinality, cosine similarity & threshold, kNN graph & Louvain \\ [0.5ex]

        \cite{luceri2024unmasking} & retweet, tweet, URL, hashtag  & cosine similarity TF-IDF, text similarity  & threshold, ADO & \\ [0.5ex]
        \cite{pante2025beyond} & retweet, tweet, URL, hashtag, fast retweet & cosine similarity TF-IDF & ADO (fast retweet) & connected components \\ [0.5ex]
        \cite{weber2020who,weber2021amplifying} & retweet, URL, hashtag, mention, reply & cardinality & ADJ & focal structures \\ [0.5ex]
        \cite{vargas2020detection} & retweet, tweet, URL, hashtag, mention & cosine similarity TF-IDF & ADJ & Leiden \\ [0.5ex]
        \cite{pacheco2021uncovering} & retweet, hashtag, image, handle change, synchronization & Jaccard coefficient, cardinality, cosine similarity & threshold, EDO & \\ [0.5ex]
        \cite{dash2024decoding} & retweet, tweet, image, synchronization & cosine similarity TF-IDF & threshold & Louvain \\ [0.5ex]
        \cite{loru2025compression} & interaction, text, synchronization & Normalized Compression Distance & kNN graph & \\ [0.5ex]
        \cite{luceri2026coordinated} & text, synchronization, hashtag, URL, duet, stitch, reply & cosine similarity TF-IDF & threshold & connected components \\ [0.5ex]
        \cite{wohlert2025detecting} & hashtag, URL, video description, music, audio & cardinality & kNN graph& connected components \\ [0.5ex]

        \cite{cinus2025exposing} & URL & cosine similarity TF-IDF & threshold, EDO & connected components \\ [0.5ex]
        \cite{minici2024uncovering} & URL & cosine similarity TF-IDF & threshold & \\ [0.5ex] 
        \cite{cao2015organic} & URL & cardinality & kNN graph & Louvain \\ [0.5ex]
        \cite{giglietto2020coordinated,giglietto2020takes,broniatowski2021towards,gruzd2022coordinated,ghasiya2022rapid,righetti2022political,rogers2025coordinated,song2025spread} & URL & cardinality & threshold, ADO & connected components \\ [0.5ex]
        \cite{yang2025coordinated} & URL & cardinality & & \\ [0.5ex]
        \cite{alieva2022investigating} & URL, hashtag, mention & unweighted & EDO & Leiden \\ [0.5ex]
        \cite{ng2022online} & URL, hashtag, mention & cardinality & threshold, EDO & Louvain \\ [0.5ex]
        \cite{colizzi2025investigating} & URL, text & cosine similarity TF-IDF & threshold, ADJ & \\ [0.5ex]
        \cite{giglietto2023workflow,soares2023sharing} & URL, text-image  & cardinality & threshold, ADO & connected components \\ [0.5ex]
        \cite{ng2022coordinated} & image & cardinality & threshold, kNN graph & \\ [0.5ex]
        \cite{yu2022framework} & image, video & cardinality & threshold, ADO & connected components \\ [0.5ex]

        \cite{burghardt2024socio} & hashtag & cardinality & threshold & connected components \\ [0.5ex]
        \cite{ng2023coordinated} & hashtag & cardinality & threshold, backbone & Louvain \\ [0.5ex]
        \cite{wang2023evidence} & hashtag & cardinality & & \\ [0.5ex]

        \cite{jahn2023towards} & like & cardinality & threshold & \\ [0.5ex]
        
        \cite{kirdemir2022towards} & comment & cardinality & threshold & k-means, hierarchical clustering \\ [0.5ex]
        \cite{piercey2023coordinated} & comment & cardinality & threshold, ADO & connected components \\ [0.5ex]
        
        \cite{neha2024understanding} & mention & cosine similarity TF-IDF & threshold & Louvain \\ [0.5ex]

        \cite{zhao2025unveiling} & report & cardinality & threshold & connected components \\ [0.5ex]

        \cite{zouzou2024unsupervised} & follow & cosine similarity & & Louvain \\ [0.5ex]

        \bottomrule 
        \multicolumn{5}{l}{{\footnotesize $\dagger$ ADJ: adjacent time window,  EDO: evenly distributed overlapping time window, ADO: action-driven overlapping time window}}
    \end{tabular}
    }
\end{table}

\textit{Implementation.} Building the coordination network $G$ requires defining the types of co-actions and the corresponding similarity functions. The vast majority of works in literature rely on a single co-action and the resulting networks are single-layered. Table~\ref{tab:single-network} reports the main implementation details for the works that built single layer coordination networks. In table, when multiple co-actions are listed for the same author or work, this means that multiple single layer networks were built, rather than a multiplex network resulting from the simultaneous analysis of multiple co-actions. The few existing works that built multiplex coordination networks are instead described in Table~\ref{tab:multiplex-network}.

As shown in Table~\ref{tab:single-network}, the most common type of co-action is \textit{co-sharing} (e.g., the \textit{co-retweet} action on Twitter/X)~\cite{chomel2023manipulation,graham2020like,keller2017manipulate,linhares2022uncovering,luceri2024unmasking,nizzoli2021coordinated,cinelli2022coordinated,hristakieva2022spread,loru2024influence,di2025post,pacheco2021uncovering,schliebs2021china,schoch2022coordination,tardelli2024temporal,vargas2020detection,weber2020who,weber2021amplifying,emeric2023interpretable,weber2021temporal,tardelli2024multifaceted,cima2024coordinated,kulichkina2025protest,dey2024coordinated,axelrod2025structure,pante2025beyond,luceri2026coordinated,mannocci2026multimodal,sartori2025insights,kulichkina2026connective,elmas2026israel}. Other frequently used co-actions are \textit{co-reply}/\textit{co-comment}~\cite{kirdemir2022towards,weber2021amplifying,weber2020who,piercey2023coordinated,luceri2026coordinated,mannocci2026multimodal} and \textit{co-like}~\cite{jahn2023towards}, which occur when two users comment or leave a reaction to the same post. The previous co-actions are based on the type of interaction between users and content in an online platform. Other co-actions are instead based on the content of user posts. For example, \textit{co-post} (e.g., \textit{co-tweet} on Twitter/X, \textit{co-parley} on Parler, and \textit{co-message} on WhatsApp)~\cite{keller2020political,schoch2022coordination,vargas2020detection,chomel2023manipulation,emeric2023interpretable,suresh2023tracking,de2024detecting,venancio2024unraveling,graham2024coordination,wu2025unmasking,blas2025large,pante2025beyond,loru2025compression,luceri2026coordinated,colizzi2025investigating,kansaon2025fake}, \textit{co-image}~\cite{elmas2026israel,ng2022coordinated,yu2022framework,pacheco2021uncovering}, and \textit{co-video}~\cite{yu2022framework} represent the publishing of posts with the same text, image, or video by multiple users. More specific co-actions are also possible, such as \textit{co-text-image} that occurs when multiple users post images that contain the same text~\cite{soares2023sharing}. Similarly, \textit{co-mention}~\cite{alieva2022investigating,vargas2020detection,weber2020who,weber2021amplifying,magelinski2022synchronized,ng2022combined,ng2022online,neha2024understanding,luceri2026coordinated,mannocci2026multimodal,ng2024tiny}, \textit{co-URL}~\cite{alieva2022investigating,cao2015organic,luceri2024unmasking,ng2022cross,ng2023you,vargas2020detection,weber2020who,weber2021amplifying,magelinski2020detecting,magelinski2022synchronized,ng2022combined,ng2022online,emeric2023interpretable,giglietto2020coordinated,giglietto2020takes,broniatowski2021towards,gruzd2022coordinated,ghasiya2022rapid,giglietto2023workflow,dey2024coordinated,luceri2026coordinated,wohlert2025detecting,yang2025coordinated,colizzi2025investigating,luceri2026coordinated,mannocci2026multimodal,ng2024tiny,rogers2025coordinated,song2025spread,minici2024uncovering,elmas2026israel}, and \textit{co-hashtag}~\cite{alieva2022investigating,burghardt2024socio,luceri2024unmasking,ng2023coordinated,pacheco2021uncovering,vargas2020detection,weber2020who,weber2021amplifying,magelinski2020detecting,magelinski2022synchronized,ng2022combined,ng2022online,emeric2023interpretable,wang2023evidence,dey2024coordinated,luceri2026coordinated,wohlert2025detecting,luceri2026coordinated,mannocci2026multimodal,ng2024tiny,elmas2026israel} represent two users publishing a post with the same user mention, URL, or hashtag. Regarding the latter, the majority of works consider publishing a post with a \textit{single} common hashtag as a valid co-action~\cite{alieva2022investigating,ng2023coordinated,vargas2020detection,weber2021amplifying,weber2020who,magelinski2020detecting,magelinski2022synchronized,ng2022combined,ng2022online,emeric2023interpretable,wang2023evidence}. However, others argued that using the same \textit{set} or \textit{sequence} of hashtags represents a stronger and more reliable signal of coordination~\cite{burghardt2024socio,luceri2024unmasking,dey2024coordinated,pacheco2021uncovering}.
More recently, on platforms such as TikTok, coordinated behaviors have also been studied through co-stitch~\cite{luceri2026coordinated} and co-duet~\cite{luceri2026coordinated}, which correspond to users jointly engaging with the same source content by embedding or responding to it within their own videos, as well as through the sharing of videos that reuse the same~\cite{wohlert2025detecting} or similar audio tracks~\cite{wohlert2025detecting}. Additional forms of coordination include co-follow~\cite{zouzou2024unsupervised} and co-report~\cite{zhao2025unveiling}, where users collectively follow or report the same account.
When a co-action is defined in such a way that multiple atomic actions are required for two users to be considered as coordinated, that co-action is a \textit{compound} action~\cite{magelinski2022synchronized}. More generally, compound actions refer to actions that involve the simultaneous occurrence or combination of multiple individual actions or sub-events.
Therefore, a co-action requiring the posting of the same set or sequence of hashtags is a compound action composed of multiple homogeneous elements: $\langle$\textit{hashtag, hashtag, }$\ldots\rangle$. However, compound co-actions can also be defined with heterogeneous elements, such as $\langle$\textit{hashtag, mention}$\rangle$ or $\langle$\textit{URL, mention}$\rangle$~\cite{magelinski2022synchronized}. When building coordination networks, the use of compound actions increases the confidence of labelling a group of users as coordinated, since compound actions are less likely to occur by chance. However, this approach risks neglecting simpler, milder, or looser forms of coordination. 
Lastly, \textit{co-handle changes} refer to multiple users using the same handle or username at different points in time~\cite{pacheco2021uncovering}. 
After selecting one or more co-actions to identify similar user behaviors, it is necessary to define the corresponding similarity functions. A similarity function computes the weight $w_{jk}$ of the edges connecting two coordinated users $u_j$ and $u_k$ based on how similar their behaviors are according to the chosen co-action. Independently of the type of co-action, the majority of works use similarity functions based on the \textit{cardinality} of the co-action between the two users~\cite{burghardt2024socio,cao2015organic,chomel2023manipulation,graham2020like,jahn2023detecting,keller2020political,kirdemir2022towards,linhares2022uncovering,ng2022cross,ng2022coordinated,ng2023coordinating,ng2023you,schliebs2021china,schoch2022coordination,weber2021amplifying,weber2020who,emeric2023interpretable,wang2023evidence,giglietto2020coordinated,giglietto2020takes,broniatowski2021towards,gruzd2022coordinated,ghasiya2022rapid,giglietto2023workflow,soares2023sharing,yu2022framework,weber2021temporal,tardelli2024multifaceted,cima2024coordinated,de2024detecting,venancio2024unraveling,righetti2022political,graham2024coordination,kulichkina2025protest,wohlert2025detecting,rogers2025coordinated,song2025spread,yang2025coordinated,piercey2023coordinated,zhao2025unveiling,luceri2026coordinated,ng2024tiny,ng2022combined,kulichkina2026connective,kansaon2025fake}. For example, when using \textit{co-hashtags}, the similarity function may simply count the number of times the two users used the same hashtags. 
Another widely adopted function is the \textit{cosine similarity} between the two user vectors~\cite{dey2024coordinated,luceri2024unmasking,ng2022cross,ng2023you,nizzoli2021coordinated,cinelli2022coordinated,hristakieva2022spread,loru2024influence,di2025post,pacheco2021uncovering,tardelli2024temporal,vargas2020detection,dey2024coordinated,neha2024understanding,axelrod2025structure,sartori2025insights,blas2025large,pante2025beyond,luceri2026coordinated,cinus2025exposing,colizzi2025investigating,zouzou2024unsupervised,mannocci2026multimodal,antonakaki2026coordinated,minici2024uncovering,elmas2026israel}. These can be binary vectors, frequency vectors, or TF-IDF weighted vectors. The latter allows discounting the importance of popular or viral content, boosting instead the relevance of unpopular items~\cite{luceri2024unmasking,nizzoli2021coordinated,cinelli2022coordinated,hristakieva2022spread,loru2024influence,di2025post,tardelli2024temporal,vargas2020detection,dey2024coordinated,neha2024understanding,pante2025beyond,luceri2026coordinated,cinus2025exposing,colizzi2025investigating,mannocci2026multimodal,antonakaki2026coordinated,minici2024uncovering,elmas2026israel}.
Depending on the choice of co-action, the previous similarity functions must be preceded by additional processing steps. For example, the use of co-actions such as \textit{co-post}, \textit{co-image}, and \textit{co-video} involve counting the number of times two users shared the same content, which severely limits the possibility to detect certain coordinated behaviors. For this reason, some scholars relaxed this requirement by also considering the posting of \textit{similar}---as opposed to \textit{equal}---content. This solution requires defining additional similarity functions for texts, images, and videos. In literature, text similarity was computed via correlation~\cite{lee2013campaign}, cosine similarity of document embeddings~\cite{ng2023coordinating,ng2023you,emeric2023interpretable,luceri2024unmasking,dey2024coordinated}, Ratcliff/Obershelp algorithm~\cite{pacheco2020unveiling}, Jaccard similarity~\cite{graham2024coordination}, or hashing~\cite{kansaon2025fake}. All these works first computed similarities between the text of two users' posts. Then, they set a threshold to select highly similar texts. Finally, they computed user similarities based on the number of similar texts~\cite{ng2023coordinating,ng2023you,emeric2023interpretable,graham2024coordination,kansaon2025fake}, or as the average of the similarities between the highly similar texts~\cite{luceri2024unmasking,dey2024coordinated}. 
Analogously, \textit{co-image} involves a pre-processing step for representing the images with their embeddings~\cite{ng2022coordinated,elmas2026israel} or with RGB color histograms~\cite{pacheco2021uncovering}. Then, image similarity is computed via measures such as the Euclidean distance~\cite{ng2022coordinated,elmas2026israel}. Finally, a similarity threshold is applied and user similarity is computed as the cardinality~\cite{ng2022coordinated} or the Jaccard coefficient~\cite{pacheco2021uncovering} of the sets of similar images.

\subsubsection{Network filtering.} This optional, yet crucial, step allows to filter out nodes and edges from the coordination network so as to only retain network structures that convey meaningful and reliable coordination. Independently of the method used to the achieve this objective, performing network filtering also has the advantage of reducing the size of the final coordination network, which speeds-up subsequent analyses. \vspace{0.1cm}

\textit{Implementation.} There are three main approaches for coordination network filtering: (\textit{i}) fixed thresholds, (\textit{ii}) statistical validation, and (\textit{iii}) the timings of the co-actions. Methods based on fixed similarity thresholds discard all edges in the network whose weight $w < w_{\text{th}}$, and all the resulting disconnected nodes~\cite{burghardt2024socio,chomel2023manipulation,jahn2023towards,keller2020political,kirdemir2022towards,luceri2024unmasking,ng2023coordinating,ng2022coordinated,ng2023coordinated,ng2022online,ng2023you,pacheco2020unveiling,pacheco2021uncovering,schoch2022coordination,cresci2019cashtag,suresh2023tracking,emeric2023interpretable,giglietto2020coordinated,giglietto2020takes,broniatowski2021towards,gruzd2022coordinated,ghasiya2022rapid,giglietto2023workflow,soares2023sharing,yu2022framework,weber2021temporal,cima2024coordinated,kulichkina2025protest,neha2024understanding,blas2025large,wu2025unmasking,luceri2026coordinated,cinus2025exposing,colizzi2025investigating,piercey2023coordinated,zhao2025unveiling,ng2024tiny,ng2022combined,mannocci2026multimodal,kulichkina2026connective,kansaon2025fake,elmas2026israel}. The threshold $w_{\text{th}}$ is chosen in such a way to retain only strongly coordinated users, as they are implicitly considered to be more relevant. Relevant nodes can also be identified via eigenvector centrality, by pruning those nodes that do not exceed a certain threshold~\cite{luceri2024unmasking,dey2024coordinated,minici2024uncovering}. The similarity and centrality thresholds are typically selected arbitrarily, without a strong underlying theoretical motivation~\cite{nizzoli2021coordinated}. Moreover, coordination networks resulting from the analysis of different datasets, or from the use of different types of co-actions, inevitably result in different edge weight and node centrality distributions. This variability makes the repeated use of ``standard'' thresholds unsuitable and mandates in-depth case-by-case analyses.

To alleviate this burden some works leverage $k$-nearest neighbors graphs ($k$-NNG), where two users $u_j$ and $u_k$ are connected only if $u_j$ is among the $k$-nearest neighbors of $u_k$~\cite{cao2015organic,ng2023coordinating,ng2022coordinated,axelrod2025structure,wu2025unmasking,loru2025compression,wohlert2025detecting}. This filtering operation retains only the $k$ strongest neighborhoods in the coordination network, thus reducing the emphasis on the edge weight. However, determining the best value for $k$ is also challenging since $k$ strongly depends on the characteristics of the different networks and their layers. Other methods for identifying relevant network structures are those that retain statistically-meaningful edges, independently of their weight. This approach, used by several recent works~\cite{linhares2022uncovering,ng2023coordinated,nizzoli2021coordinated,cinelli2022coordinated,hristakieva2022spread,loru2024influence,di2025post,tardelli2024temporal,tardelli2024multifaceted,venancio2024unraveling}, is not biased towards fixed arbitrary levels of similarity or coordination, but instead erases network structures that convey limited information, allowing to focus on meaningful expressions of coordination. 

\begin{table}[!t]
    \caption{Types and characteristics of the time windows and the coordination networks used by network science methods.}
\label{tab:time-window}
    \centering
    \small
    \setlength{\tabcolsep}{6pt}
    \scalebox{0.75}{
    \begin{tabular}{L{2.5cm}L{3.5cm}L{3.5cm}L{1cm}L{2.5cm}}
        \toprule
        & \multicolumn{2}{c}{\textit{time window}} & \multicolumn{2}{c}{\textit{network}} \\ 
        \cmidrule(lr){2-3} \cmidrule(l){4-5}
        \textbf{reference} & \textbf{type} & \textbf{size} & \textbf{type} & \textbf{layer(s)}\\
        \midrule
        \cite{weber2020who,weber2021amplifying,weber2021temporal} & adjacent & 15 min, 1 hour, 6 hour, 1 day &  user & single \\
        \cite{vargas2020detection}  & adjacent &  1 day, 1 week & user & single \\
        \cite{linhares2022uncovering,cima2024coordinated,colizzi2025investigating} &  adjacent & 1 week & user & single \\  
        \cite{al2025coordinated} & adjacent & 1 hour, 1 day, & content & single \\

\cite{graham2020like} & evenly distributed overlapping  & 1 sec & user & single \\
        \cite{chomel2023manipulation} &  evenly distributed overlapping & from 1 sec to 250 sec & content & single\\
        \cite{keller2020political,schliebs2021china,schoch2022coordination} & evenly distributed overlapping & 1 min &  user & single \\
        \cite{ng2022online} & evenly distributed overlapping & from 1 min to 30 min & user & single \\
        \cite{alieva2022investigating} & evenly distributed overlapping & 5 min & user & single \\
        \cite{magelinski2020detecting,ng2023you}  & evenly distributed overlapping & 5 min & user & multiple (L=2) \\
        \cite{magelinski2022synchronized,ng2022combined,graham2024coordination} & evenly distributed overlapping  & 5 min & user & multiple (L=3)\\
        \cite{pacheco2021uncovering} & evenly distributed overlapping & 30 min &user & single \\
        \cite{cinus2025exposing,luceri2026coordinated} & evenly distributed overlapping & 1 day & user & single \\
        \cite{blas2025large} & evenly distributed overlapping & 2 day & user & single \\ 
        \cite{mannocci2026multimodal}  & evenly distributed overlapping &  6 hour, 1 week & user & multiple (L=5) \\
        \cite{tardelli2024temporal} & evenly distributed overlapping &  1 week & user & single \\

        \cite{kulichkina2025protest} & action-driven overlapping & from 1 sec to 1 hour & user & single \\
        \cite{de2024detecting} & action-driven overlapping & from 1 sec to 11 day & user & single \\
        \cite{pacheco2020unveiling,pante2025beyond} & action-driven overlapping & 10 sec &  user & single \\
        \cite{luceri2024unmasking,dey2024coordinated} & action-driven overlapping &  10 sec &  user & single, multiple (L=4) \\ \cite{gruzd2022coordinated,ghasiya2022rapid,giglietto2020coordinated,giglietto2020takes,giglietto2023workflow,soares2023sharing,righetti2022political} & action-driven overlapping & from 10 sec to 1 min &  user & single \\
        \cite{broniatowski2021towards} & action-driven overlapping & 25 sec &  user & single \\
        \cite{yu2022framework,kansaon2025fake} &  action-driven overlapping & 1 min & user & single \\
        \cite{kulichkina2026connective} & action-driven overlapping & 2 min & user & single \\
        \cite{emeric2023interpretable} & action-driven overlapping & 1 min, 1 hour, 1 day &  user & multiple (L=4) \\
        \cite{wu2025unmasking} &  action-driven overlapping & 1 hour & user & single \\
        \cite{suresh2023tracking} & action-driven overlapping & 10 tweets & content & single \\
        
        \bottomrule
\end{tabular}
    }
\end{table}
 
Filtering methods based on the timings of the co-actions use a sequence of time windows of equal size $Z = t_{\text{end}} - t_{\text{start}}$.
The filtering typically occurs when computing user similarities, by only retaining the co-actions that occur within the same time window. In other words, this filter corresponds to adding a further constraint to the actions that determine the coordination, in that such actions must be temporally close to one another. Table~\ref{tab:time-window} displays the type and length of the time windows used in literature, along with the type of coordination network built. As shown in Tables~\ref{tab:time-window} and~\ref{fig:time-window}, time windows can be either \textit{adjacent}~\cite{alieva2022investigating,linhares2022uncovering,vargas2020detection,weber2020who,weber2021amplifying,cima2024coordinated,al2025coordinated} or \textit{overlapping}. Furthermore, overlapping time windows can be \textit{evenly distributed} in time~\cite{graham2020like,chomel2023manipulation,keller2020political,schliebs2021china,schoch2022coordination,ng2022online,alieva2022investigating,magelinski2020detecting,ng2023you,graham2024coordination,magelinski2022synchronized,ng2022combined,pacheco2021uncovering,tardelli2024temporal,cinus2025exposing,luceri2026coordinated,blas2025large,mannocci2026multimodal}, or \textit{action-driven}---that is, positioned based on the timings of the actions~\cite{kulichkina2025protest,de2024detecting,pacheco2020unveiling,dey2024coordinated,luceri2024unmasking,ghasiya2022rapid,gruzd2022coordinated,ghasiya2022rapid,giglietto2020coordinated,giglietto2020takes,giglietto2023workflow,soares2023sharing,righetti2022political,broniatowski2021towards,yu2022framework,emeric2023interpretable,suresh2023tracking,pante2025beyond,wu2025unmasking,kulichkina2026connective,kansaon2025fake}. As shown in Table~\ref{fig:time-window}, in the latter case each time window starts when a relevant action occurs. In the former case instead, the additional parameter step $\delta < Z$ defines the temporal offset between two consecutive time windows. The amount of overlap $O$ between two consecutive overlapping time windows is thus $O = Z - \delta$. Given a sequence of actions by two or more users, the choice of time windows and their parameters influence the number of such actions that are valid co-actions. These are indicated with the {\small\faLock} lock icon in the example shown in the rightmost column of Table~\ref{fig:time-window}. Let $d = t_2 - t_1$ be the delay with which user $u_2$ performs an action at time $t_2$, with respect to the same action that $u_1$ performed at time $t_1$. Independently of the type of time window, actions whose $d > Z$ are never considered as co-actions. Then, actions with $d \le Z$ are always co-actions when using action-driven overlapping time windows. Conversely, actions with $d \le O$ are always co-actions when using evenly distributed overlapping time windows, and can possibly be co-actions when $O < d \le Z$. Instead, with adjacent time windows, also actions that are close in time (i.e., $d \ll Z$) are occasionally not considered valid co-actions, when they occur across the boundary between two windows, as in the case of the actions occurred at $t_3$ and $t_4$ in the example of Table~\ref{fig:time-window}.
Therefore, using overlapping rather than adjacent time windows ensures that no actions close in time are missed. However, it increases the number of windows required to cover the same time frame, leading to more computational demands. Another critical consideration is the length $Z$ of the time window. A larger $Z$ includes more actions as co-actions, while a shorter one restricts valid co-actions to highly synchronized events, while also increasing the number of windows and the resulting computations. The choice of $Z$ also depends on the goal of the analysis. Literature indicates that inauthentic or harmful coordinated behaviors are highly synchronized and can often be detected with short windows~\cite{pacheco2021uncovering,magelinski2022synchronized}, whereas emergent human behaviors that are typically less orchestrated and loosely synchronized, require longer windows to capture their relaxed temporal dynamics~\cite{nizzoli2021coordinated}.

\begin{table}[t]
    \caption{Time window types, their parameters, and their effect on valid co-actions. To the right, a sequence of actions occurs at times $t_1, \ldots, t_6$. The same sequence results in different valid co-actions, marked by the {\small\faLock} lock icon, depending on the time window type.}
    \label{fig:time-window}
    \centering
\setlength{\tabcolsep}{6pt}
    \scalebox{0.75}{
\begin{tabular}{lR{3.1cm}C{8cm}}
            \toprule
            \textbf{type} & \textbf{parameters} & \textbf{action sequence and valid co-actions}\\
            \midrule
            \multirow{3}{*}{adjacent} & \multicolumn{2}{l}{\multirow{9}{*}{\scalebox{0.75}{\def\Xsingletw{-3.5}
\def\spaceYIcon{0.05}
\def\stickH{0.2}
\def\stickW{0.4}
\def\paddingW{0.02}
\def\overW{1}

\def\livETW{1}
\def\livOTW{3}
\def\livATW{5}

\def\maxHor{10}
\def\minHor{0}
\def\maxVer{6}
\def\minVer{0.5}
\def\bottomVer{0.2}
\def\rectH{0.5}
\def\rectW{1.5}
\def\stepOTW{2*\rectW}

\definecolor{lightgray}{gray}{0.80}
\definecolor{darkgray}{gray}{0.5}
\colorlet{highlightcol}{black}

\begin{tikzpicture}
\centering

\draw[thin, fill=white] (\Xsingletw, \livATW) rectangle (\Xsingletw + \rectW, \livATW + \rectH);
\draw[thin, fill=white] (\Xsingletw + \rectW, \livATW) rectangle (\Xsingletw + \rectW + \rectW, \livATW + \rectH);

\node[above, black] at (\Xsingletw + \rectW/2, \livATW + \rectH- \spaceYIcon) {\small Z};
\draw[darkgray] (\Xsingletw + \paddingW, \livATW + \rectH + \spaceYIcon) -- (\Xsingletw  + \paddingW,  \livATW + \rectH + \spaceYIcon + \stickH); \draw[darkgray] (\Xsingletw + \rectW  - \paddingW, \livATW + \rectH + + \spaceYIcon) -- (\Xsingletw + \rectW - \paddingW,  \livATW + \rectH + \spaceYIcon  + \stickH); \draw[darkgray] (\Xsingletw + \paddingW, \livATW + \rectH + \spaceYIcon  + \stickH/2) -- (\Xsingletw + \stickW  + \paddingW,  \livATW + \rectH + + \spaceYIcon  + \stickH/2); \draw[darkgray] (\Xsingletw + \rectW - \stickW  - \paddingW, \livATW + \rectH + \spaceYIcon  + \stickH/2) -- (\Xsingletw + \rectW  - \paddingW, \livATW + \rectH + \spaceYIcon  + \stickH/2);

\draw[thin, fill=white] (\Xsingletw, \livOTW) rectangle (\Xsingletw + \rectW, \livOTW + \rectH);
\draw[thin, fill=white] (\Xsingletw + 1, \livOTW + \rectH) rectangle (1 + \Xsingletw + \rectW, \livOTW + 2*\rectH);

\node[above, black] at (\Xsingletw + 7*\rectW/6, \livOTW + 2*\rectH - \spaceYIcon) {\small Z};
\draw[darkgray] (\Xsingletw + 2*\rectW/3 + \paddingW, \livOTW + 2*\rectH + \spaceYIcon) -- (\Xsingletw + 2*\rectW/3 + \paddingW,  \livOTW + 2*\rectH + \spaceYIcon + \stickH); \draw[darkgray] (\Xsingletw + 5*\rectW/3 - \paddingW, \livOTW + 2*\rectH + + \spaceYIcon) -- (\Xsingletw + 5*\rectW/3 - \paddingW,  \livOTW + 2*\rectH + \spaceYIcon  + \stickH); \draw[darkgray] (\Xsingletw + 2*\rectW/3 + \paddingW, \livOTW + 2*\rectH + \spaceYIcon  + \stickH/2) -- (\Xsingletw + 2*\rectW/3 + \stickW + \paddingW,  \livOTW + 2*\rectH + \spaceYIcon  + \stickH/2); \draw[darkgray] (\Xsingletw + 5*\rectW/3 - \paddingW - \stickW, \livOTW + 2*\rectH + \spaceYIcon  + \stickH/2) -- (\Xsingletw + 5*\rectW/3 - \paddingW, \livOTW + 2*\rectH + \spaceYIcon  + \stickH/2);

\node[above, black] at (\Xsingletw + \rectW/3, \livOTW - 3*\rectH/4) {\small$\delta$};
\draw[darkgray] (\Xsingletw + \paddingW, \livOTW - \rectH/2) -- (\Xsingletw  + \paddingW,  \livOTW - \rectH/2 + \stickH); \draw[darkgray] (\Xsingletw + 2*\rectW/3  - \paddingW, \livOTW - \rectH/2) -- (\Xsingletw + 2*\rectW/3 - \paddingW,  \livOTW - \rectH/2 + \stickH); 

\draw[darkgray] (\Xsingletw + \paddingW,  \livOTW - \rectH/2 + \stickH/2) -- (\Xsingletw + \stickW-0.2 + \paddingW,   \livOTW - \rectH/2 + \stickH/2); \draw[darkgray] (\Xsingletw + 2*\rectW/3 - \stickW + 0.2 - \paddingW,  \livOTW - \rectH/2 + \stickH/2) -- (\Xsingletw + 2*\rectW/3  - \paddingW,  \livOTW - \rectH/2 + \stickH/2);

\node[above, black] at (\Xsingletw + 5*\rectW/6, \livOTW - 3*\rectH/4) {\small O};
\draw[darkgray] (\Xsingletw + 2*\rectW/3  + \paddingW, \livOTW - \rectH/2) -- (\Xsingletw + 2*\rectW/3 + \paddingW,  \livOTW - \rectH/2 + \stickH); \draw[darkgray] (\Xsingletw + \rectW - \paddingW, \livOTW - \rectH/2) -- (\Xsingletw + \rectW - \paddingW,  \livOTW - \rectH/2 + \stickH); 

\draw[darkgray] (\Xsingletw + 2*\rectW/3  + \paddingW,  \livOTW - \rectH/2 + \stickH/2) -- (\Xsingletw + 2*\rectW/3 + \paddingW + \stickW - 0.32, \livOTW - \rectH/2 + \stickH/2); \draw[darkgray] (\Xsingletw + \rectW - \paddingW - \stickW + 0.32 - \paddingW,  \livOTW - \rectH/2 + \stickH/2) -- (\Xsingletw + \rectW - \paddingW,  \livOTW - \rectH/2 + \stickH/2);

\node[below, black] at (\Xsingletw + 0.5, \bottomVer) {$t_i$};
\draw[dashed, darkgray] (\Xsingletw + 0.5, \bottomVer) -- (\Xsingletw + 0.5,  \livETW);
\draw[thin, fill=white] (\Xsingletw + 0.5, \livETW) rectangle (\Xsingletw + 0.5 + \rectW, \livETW + \rectH);

\node[below, black] at (\Xsingletw + 2*\rectW/3, \bottomVer) {$t_{i+1}$};
\draw[dashed, darkgray] (\Xsingletw + 2*\rectW/3, \bottomVer) -- (\Xsingletw + 1, \livETW + \rectH);
\draw[thin] (\Xsingletw + 2*\rectW/3, \livETW + \rectH) rectangle (\Xsingletw + 1 + \rectW, \livETW + 2*\rectH);

\node[above, black] at (\Xsingletw + 7*\rectW/6, \livETW + 2*\rectH - \spaceYIcon) {\small Z};

\draw[darkgray] (\Xsingletw + 2*\rectW/3 + \paddingW, \livETW + 2*\rectH + \spaceYIcon) -- (\Xsingletw + 2*\rectW/3 + \paddingW,  \livETW + 2*\rectH + \spaceYIcon + \stickH); \draw[darkgray] (\Xsingletw + 5*\rectW/3 - \paddingW, \livETW + 2*\rectH + + \spaceYIcon) -- (\Xsingletw + 5*\rectW/3 - \paddingW,  \livETW + 2*\rectH + \spaceYIcon  + \stickH); \draw[darkgray] (\Xsingletw + 2*\rectW/3 + \paddingW, \livETW + 2*\rectH + \spaceYIcon  + \stickH/2) -- (\Xsingletw + 2*\rectW/3 + \stickW + \paddingW,  \livETW + 2*\rectH + \spaceYIcon  + \stickH/2); \draw[darkgray] (\Xsingletw + 5*\rectW/3 - \paddingW - \stickW, \livETW + 2*\rectH + \spaceYIcon  + \stickH/2) -- (\Xsingletw + 5*\rectW/3 - \paddingW, \livETW + 2*\rectH + \spaceYIcon  + \stickH/2);

\draw[-, thick] (\minHor, \minVer) -- (\minHor, \maxVer);
\draw[->, thick] (\minHor, \minVer) -- (\maxHor, \minVer);
    
\draw[thick] (\minHor, \livETW - 0.01) -- (\maxHor, \livETW - 0.01);
\draw[thick] (\minHor, \livOTW - 0.01) -- (\maxHor, \livOTW - 0.01);
\draw[thick] (\minHor, \livATW - 0.01) -- (\maxHor, \livATW - 0.01);

\foreach \x/\col in {0/lightgray, 1/white, 2/white, 3/white, 4/white, 5/white} {
\draw[thin, fill=\col] (\x*\rectW, \livATW) rectangle (\x*\rectW + \rectW, \livATW + \rectH);
}
    
\foreach \x/\col in {0/lightgray, 1/white, 2/white} {
    \draw[thin, fill=\col] (\x*\stepOTW, \livOTW) rectangle (\x*\stepOTW + \rectW, \livOTW + \rectH);
}
\foreach \x/\col in {0/white, 1/lightgray, 2/white} {
    \draw[thin, fill=\col] (1+\x*\stepOTW, \livOTW + \rectH) rectangle (1+\x*\stepOTW + \rectW, \livOTW + 2*\rectH);
}
\foreach \x in {0, 1, 2} {
    \draw[thin] (2+\x*\stepOTW, \livOTW + 2*\rectH) rectangle (2+\x*\stepOTW + \rectW, \livOTW + 3*\rectH);
}
    
\node[below] at (0, \bottomVer) {$t_0$};
\draw[solid, thick] (0, \bottomVer) -- (0, \maxVer);
    
\node[below] at (0.5, \bottomVer) {$t_1$};
\draw[dashed, highlightcol] (0.5, \bottomVer) -- (0.5, \maxVer);
\draw[thin, fill=lightgray] (0.5, \livETW) rectangle (0.5 + \rectW, \livETW + \rectH);
    
\node[below] at (1, \bottomVer) {$t_2$};
\draw[dashed, highlightcol] (1, \bottomVer) -- (1, \maxVer);
\draw[thin] (1, \livETW + \rectH) rectangle (1 + \rectW, \livETW + 2*\rectH);
    
\node[below] at (4.2, \bottomVer) {$t_3$};
\draw[dashed, highlightcol] (4.2, \bottomVer) -- (4.2, \maxVer);
\draw[thin, fill=lightgray] (4.2, \livETW) rectangle (4.2 + \rectW, \livETW + \rectH);
    
\node[below] at (4.7, \bottomVer) {$t_4$};
\draw[dashed, highlightcol] (4.7, \bottomVer) -- (4.7, \maxVer);
\draw[thin] (4.7, \livETW + \rectH) rectangle (4.7 + \rectW, \livETW + 2*\rectH);
    
\node[below] at (6.9, \bottomVer) {$t_5$};
\draw[dashed, highlightcol] (6.9, \bottomVer) -- (6.9, \maxVer);
\draw[thin, fill=lightgray] (6.9, \livETW) rectangle (6.9 + \rectW, \livETW + \rectH);
    
\node[below] at (8.2, \bottomVer) {$t_6$};
\draw[dashed, highlightcol] (8.2, \bottomVer) -- (8.2, \maxVer);
\draw[thin] (8.2, \livETW + \rectH) rectangle (8.2 + \rectW, \livETW + 2 * \rectH);

\node[above, yshift=-0.1cm, highlightcol] at (0.75, \livETW) {\faArrowsH};
\node[above, yshift=+0.1cm, highlightcol] at (0.75, \livETW) {\small\faLock};
\node[above, yshift=-0.1cm, highlightcol] at (0.75, \livOTW) {\faArrowsH};
\node[above, yshift=+0.1cm, highlightcol] at (0.75, \livOTW) {\small\faLock};
\node[above, yshift=-0.1cm, highlightcol] at (0.75, \livATW) {\faArrowsH};
\node[above, yshift=+0.1cm, highlightcol] at (0.75, \livATW) {\small\faLock};

\node[above, yshift=-0.1cm, highlightcol] at (4.45, \livETW) {\faArrowsH};
\node[above, yshift=+0.1cm, highlightcol] at (4.45, \livETW) {\small\faLock};
\node[above, yshift=-0.1cm, highlightcol] at (4.45, \livOTW + \rectH) {\faArrowsH};
\node[above, yshift=+0.1cm, highlightcol] at (4.45, \livOTW + \rectH) {\small\faLock};

\node[above, highlightcol] at (7.15, \livETW) {\faLongArrowLeft};
\node[above, highlightcol] at (7.55, \livETW) {\small\faLock};
\node[above, highlightcol] at (7.95, \livETW) {\faLongArrowRight};
    
\node[right] at (9.5, 0) {time};

\end{tikzpicture}
 }}} \\[10ex]
            \multirow{1}{*}{evenly distributed overlapping} & & \\  [7ex]
            \multirow{1}{*}{action driven overlapping} & & \\ [9ex]

\bottomrule
        \end{tabular}
    }
\end{table}

A few works departed from the traditional time windows filters presented above. For example,~\cite{weber2020who,weber2021amplifying,tardelli2024temporal} built a distinct coordination network for each time window. Then,~\cite{weber2020who,weber2021amplifying} aggregated all networks, each corresponding to a different adjacent time window, by computing the pairwise sums of the network weights. The summation includes a temporal decay weighting scheme that emphasizes the contribution of recent time windows over older ones. Instead,~\citet{tardelli2024temporal} built a multiplex temporal network where the layers are obtained from a sequence of evenly distributed overlapping time windows. Differently from all other approaches, the multiplex temporal network is then analyzed as a whole.
A further innovation is introduced in~\cite{ng2022online}, which solves an optimization task to select the best window size $Z$ in an overlapping time windows setting. 
Finally, we remark that the filtering methods discussed in this section can be, and oftentimes are, used in combination for greater efficiency and to further reduce the network size when analyzing very large datasets.

\subsubsection{Community discovery.}\label{sec:CDA} This step aims to detect a set of coordinated communities $P=\{P_1,\dots,P_n\}$ via community discovery on the coordination network. Multiplex networks are either flattened before performing community discovery~\cite{weber2020who,weber2021amplifying} or an algorithm suitable for multiplex networks is used~\cite{tardelli2024temporal,magnani2021community}.\vspace{0.1cm}

\textit{Implementation.} Most of the works dealing with single layer networks carry out community discovery with \textsc{Louvain}~\cite{cao2015organic,chomel2023manipulation, graham2020like,keller2020political,linhares2022uncovering,ng2022cross,ng2023coordinated,ng2022online,ng2023you,danaditya2022curious,ng2022coordinated,venancio2024unraveling,kulichkina2025protest,neha2024understanding,zouzou2024unsupervised,kulichkina2026connective,kansaon2025fake}, or more broadly via modularity clustering~\cite{cima2024coordinated}. Other recent works relied on \textsc{Leiden}~\cite{alieva2022investigating,ng2023coordinating,tardelli2024temporal,vargas2020detection,sartori2025insights}. Among the advantages of these approaches is their scalability, which makes them suitable for the analysis of large networks. In addition, modularity clustering provides a hierarchical community structure, allowing for analyses at different levels of granularity. The authors of~\cite{nizzoli2021coordinated,tardelli2024multifaceted} combined community discovery via \textsc{Louvain} with the iterative application of a progressively increasing edge weight filtering threshold on the coordination network. As a result of the moving threshold, they studied how the structure and the properties of the coordinated communities change across the whole spectrum of coordination. Moreover, the moving threshold implicitly defines a measure for the extent of coordination observed at each iteration, for each coordinated community, thus providing a continuous score of coordination rather than a binary label. 
Others proposed a variation of focal structures analysis to identify influential sets of nodes in a coordination network~\cite{weber2020who,weber2021amplifying}. 

An alternative to community discovery is the application of a particularly restrictive set of filters that results in splitting the coordination network in multiple connected components~\cite{burghardt2024socio,giglietto2020coordinated,giglietto2020takes,broniatowski2021towards,gruzd2022coordinated,ghasiya2022rapid,giglietto2023workflow,soares2023sharing,yu2022framework,righetti2022political,graham2024coordination,pante2025beyond,wohlert2025detecting,cinus2025exposing,rogers2025coordinated,song2025spread,zhao2025unveiling,piercey2023coordinated,elmas2026israel}. Since each component is disconnected from the others, this process essentially produces an output that is similar to that of community discovery, in that it identifies groups of connected and highly coordinated users. However, the application of very restrictive filters also discards much of the coordination network~\cite{nizzoli2021coordinated}.

\begin{table}[t]
    \caption{Network science methods for detecting coordinated behavior based on \textit{multiplex} user networks, where each layer corresponds to a different co-action. For each group of works we report the considered co-actions, similarity functions, filtering criteria, and the optional flattening step applied before the community detection method.}
\label{tab:multiplex-network}
    \centering
    \small
    \setlength{\tabcolsep}{3pt}
    \scalebox{0.75}{
    \begin{tabular}{L{1.4cm}L{3.5cm}L{3.3cm}L{2.2cm}L{2.8cm}L{3cm}}
        \toprule
        \textbf{reference} & \textbf{action} & \textbf{similarity} & \textbf{filters$^\dagger$} & \textbf{flattening} & \textbf{community detection} \\
        \midrule
        \cite{emeric2023interpretable} & share, message, URL, hashtag & cardinality & threshold, ADO & & Louvain, IPVC$^\ddagger$ \\ [0.5ex]
        \cite{luceri2024unmasking,dey2024coordinated} & retweet, tweet, URL, hashtag  & cosine similarity TF-IDF & threshold, ADO & unweighted edge union & \\ [0.5ex]
        \cite{elmas2026israel} & retweet, URL, hashtag, image, token  & cosine similarity TF-IDF & threshold & unweighted edge union & connected components \\ [0.5ex]
        \cite{iannucci2026detecting} & retweet, URL, hashtag, mention & temporal weighted cardinality & & & Generalized Leiden \\ [0.5ex]
        \cite{graham2024coordination} & retweet, text, URL, reply & cardinality & EDO & multigraph & connected components \\ [0.5ex]
        \cite{magelinski2020detecting} & URL, hashtag & cardinality & EDO & & multi-view clustering \\ [0.5ex]
        \cite{magelinski2022synchronized} & URL, hashtag, mention & cardinality & EDO & & multi-view clustering  \\ [0.5ex]
        \cite{ng2022combined,ng2024tiny} & URL, hashtag, mention & cardinality & EDO & sum cardinality & \\ [0.5ex]
        \cite{mannocci2026multimodal} & URL, hashtag, mention, reply, retweet & cosine similarity TF-IDF & threshold, EDO & & Generalized Louvain, Generalized Infomap \\ [0.5ex]
        \cite{ng2023you} & URL, tweet & cardinality, cosine similarity & threshold, EDO & unweighted edge union & Louvain \\ [0.5ex]
        \bottomrule
        \multicolumn{6}{l}{{\small $\dagger$ EDO: evenly distributed overlapping time window, ADO: action-driven overlapping time window. $\ddagger$ IPVC: iterative probabilistic voting consensus}}
    \end{tabular}
    }
\end{table}
 
The works presented in Table~\ref{tab:multiplex-network} performed community discovery on a multiplex coordination network. Some authors reduced the complexity of dealing with multiplex networks by flattening them into single layer networks where nodes and edges are the union of the nodes and edges of each layer. Flattened networks can be either  weighted~\cite{ng2022combined} or unweighted~\cite{luceri2024unmasking,ng2023you,dey2024coordinated,elmas2026israel}, with the latter resulting in the loss of some information. Alternative approaches involve representing the multiplex network as a single-layer multigraph with labeled parallel edges~\cite{graham2024coordination}, leveraging multi-view clustering to combine different network layers~\cite{magelinski2020detecting,magelinski2022synchronized}, or applying \textsc{Louvain} on each layer and an iterative probabilistic voting consensus algorithm to achieve consensus clustering~\cite{emeric2023interpretable}. Only a couple of works exploit the multiplex version of \textsc{Leiden}~\cite{luceri2026coordinated}, \textsc{Infomap}~\cite{mannocci2026multimodal}, and \textsc{Louvain}~\cite{mannocci2026multimodal}, fully exploiting the multimodal structure of the coordination network.

\subsubsection{Content networks}
\label{sec:content-networks}
A few works built and analyzed networks of \textit{content} rather than \textit{users}. After selecting the initial set of users, these methods build a fully connected network where the nodes are contents posted by the users (e.g., posts) and edge weights are proportional to the similarity between the linked contents. The community discovery process on a content network yields clusters of highly similar contents, regarded as proxies for coordination. User and content networks are interchangeable because each node in a user network can be mapped to the content they published, and each node in a content network can be mapped to its respective author, allowing for a dual representation of the same underlying actions. The main characteristics of the works based on content networks are reported in Table~\ref{tab:content-networks}.
Some works built and analyzed content networks where the nodes were Twitter cashtags~\cite{cresci2019cashtag}, or hashtags~\cite{danaditya2022curious,weber2021amplifying}. Edge weights were given by the number of co-occurrences of the cashtags and hashtags in the same tweets. The analyses revealed suspicious clusters of contents that were later linked to online manipulations by coordinated actors. 
Other works created a network of similar texts~\cite{lee2013campaign,antonakaki2026coordinated} or comments~\cite{al2025coordinated} published by multiple users.~\citet{lee2013campaign} identify clusters through connected components and maximal clique analysis. Similarly, \cite{suresh2023tracking} constructed two content networks: one for copypasta tweets, clustering coordinated actors via hierarchical clustering, and another to analyze the temporal evolution of co-occurring hashtags. \cite{ng2022coordinated} built an image similarity network by computing image embeddings and using Euclidean distance for comparison, then applied \textsc{Louvain} to detect groups of similar images.

\begin{table}[t]
    \caption{Network science methods for detecting coordinated behavior based on \textit{content} networks, where nodes are posted contents and edge weights encode the similarity between the linked contents. For each group of works we report the types of content, similarity functions, filtering criteria, and the community detection methods.}
\label{tab:content-networks}
    \centering
    \small
    \setlength{\tabcolsep}{6pt}
    \scalebox{0.75}{
    \begin{tabular}{L{1.5cm}L{2cm}L{3.5cm}L{2cm}L{5cm}}
        \toprule
        \textbf{reference} & \textbf{nodes} & \textbf{similarity} & \textbf{filters$^\dagger$} & \textbf{community detection}\\
        \midrule

        \cite{antonakaki2026coordinated} & texts & cosine similarity TF-DID & & \\ [0.5ex]
        \cite{lee2013campaign} & texts & text similarity scores & threshold & loose strict campaign, cohesive campaign \\ [0.5ex]
        \cite{suresh2023tracking} & texts, hashtags & cosine similarity, cardinality & EDO, threshold & hierarchical clustering \\ [0.5ex]
        \cite{danaditya2022curious,weber2021amplifying} & hashtags & cardinality & threshold & Louvain \\ [0.5ex]
        \cite{cresci2018fake,cresci2019cashtag} & cashtags & cardinality & threshold & \\ [0.5ex]
        \cite{ng2022coordinated} & images & Euclidean distance & kNN graph & Louvain \\ [0.5ex]
        \cite{al2025coordinated} & comments & message similarity & Louvain \\ [0.5ex]
        \bottomrule
        \multicolumn{5}{l}{{\small $\dagger$ EDO: evenly distributed overlapping time window}}
    \end{tabular}
    }
\end{table}

\subsection{Data mining and machine learning methods}
\label{sec:machine-learning-approaches}
Methods in this category follow the typical knowledge discovery in databases (KDD) analytical process, involving data collection and preparation, application of data mining and machine learning techniques, validation of the discovered patterns, and extraction of insights by interpretation of the results. The types of analyzed data include texts~\cite{assenmacher2020two,assenmacher2020towards,pohl2022artificial,stampe2023towards,saeed2024tuberaider,barbero2023multi,stampe2024benchmarking}, images~\cite{stampe2023towards}, audio~\cite{mariconti2019you}, interactions~\cite{sharma2021identifying,zhang2021vigdet,steinert2015online,zhang2023capturing,uyheng2022mapping,manchanayaka2024identifying,manchanayaka2025using,kalenkova2025discovering,zareie2025identifying}, and temporal data~\cite{bellutta2023investigating,keller2017manipulate,francois2021measuring}. In terms of machine learning techniques, some works leverage unsupervised approaches (see Table~\ref{tab:machine-learning-unsupervised-approach}) and focus on identifying groups of users exhibiting similar behaviors, while others rely on the availability of labeled data and apply supervised techniques (see Table~\ref{tab:machine-learning-supervised-approach}). These supervised approaches can be further distinguished by their \textit{prediction target}, namely \textit{user} (individual account classification), \textit{community} (groups of users), \textit{network} (entire interaction graphs), \textit{target} (entities targeted by coordinated actions).

\subsubsection{Unsupervised} Unsupervised methods work with unlabelled data and, therefore, do not exploit any knowledge on the membership of users to coordinated groups. 
Multiple unsupervised methods~\cite{assenmacher2020towards,assenmacher2020two,pohl2022artificial,stampe2023towards,stampe2024benchmarking} are based on stream clustering algorithms, designed to group similar textual documents into micro-clusters that represent the topics recently discussed in the stream. 
The detection of rapidly growing clusters of documents is used to identify inorganic or orchestrated campaigns by coordinated users. This approach is conceptually similar to the analysis of a content network of similar documents that includes a time-based filter. A similar approach is also used in~\cite{barbero2023multi}, where text and node embeddings are clustered via \textsc{DBSCAN}. In \cite{smith2025unsupervised}, hashtags used by different users are leveraged to perform clustering via a Bayesian model. In contrast, \cite{uyheng2022mapping,kuznetsova2025amplifying} apply clustering on interaction networks, such as the mention network~\cite{kuznetsova2025amplifying}, or more comprehensive networks incorporating mentions, replies, retweets, textual content, and hashtags~\cite{uyheng2022mapping}.
\citet{keller2017manipulate} modeled user daily tweeting activity with a binary matrix whose cells $x_{jt}=1$ represent a user $u_j$ who tweeted at least once on day $t$, while cells $x_{jt}=0$ indicate no tweeting on that day. Then, groups of users with similar tweeting behaviors are found via expectation-maximization. Here, highly similar tweeting behaviors are considered as a proxy for coordination. 
Others modeled the sequence of user activities as temporal point processes. 
\citet{sharma2021identifying} proposed the \textsc{AMDN-HAGE} generative model to jointly account for user activities and hidden group behaviours. The \textsc{Attentive Mixture Density Network} (\textsc{AMDN}) component models observed activity traces as a temporal point process, while the \textsc{Hidden Account Group Estimation} (\textsc{HAGE}) component models user groups as mixtures of multiple distributions. The synchronized groups detected by \textsc{AMDN-HAGE} are deemed coordinated and assumed to be malicious. Similarly,~\citet{zhang2021vigdet} jointly learned a distribution of user-group assignments based on how consistent each assignment is to the user embedding space and to some prior knowledge such as temporal logic. Finally, they used expectation-maximization to cluster the users according to the different distributions.

\begin{table}[t]
    \caption{Data mining and machine learning methods, based on \textit{unsupervised} learning for detecting coordinated behavior. For each group of works we report the input types, the machine learning approach, and whether the method takes time into account.}

    \label{tab:machine-learning-unsupervised-approach}
    \centering
    \small
    \setlength{\tabcolsep}{4pt}
    \scalebox{0.75}{
    \begin{tabular}{L{1.6cm}L{4.8cm}L{7.5cm}C{1cm}}
        \toprule
        \textbf{reference} & \textbf{input} & \textbf{machine learning approach}& \textbf{time}\\
        \midrule
        \cite{assenmacher2020towards,assenmacher2020two,pohl2022artificial,stampe2024benchmarking} & text streams & text stream clustering & \faCheck \\ [0.5ex]
        \cite{stampe2023towards} & text streams, image captions & text stream clustering & \faCheck\\ [0.5ex]
        \cite{barbero2023multi} & text, user-content network & clustering of text and node embeddings & \faCheck\\ [0.5ex]
        \cite{keller2017manipulate} & daily tweeting activity & expectation-maximization & \faCheck\\ [0.5ex]
        \cite{steinert2015online} &  hashtags & peak detection & \faCheck \\ [0.5ex]
        \cite{bellutta2023investigating} & account creation timestamps & burst detection & \faCheck\\ [0.5ex]
        \cite{manchanayaka2024identifying} & user activities & contrast pattern mining & \faCheck\\ [0.5ex]
         \cite{manchanayaka2025using} & user activities & convergent cross mapping & \faCheck \\ [0.5ex]
         \cite{erhardt2023detecting,erhardt2024hidden} & URL, hashtag, image, mention & networked Markov chains & \faCheck \\ [0.5ex]
        \cite{sharma2021identifying} & user activities & temporal point processes, gaussian mixture models & \faCheck \\ [0.5ex]
        \cite{zhang2021vigdet} & user activities & temporal point processes, expectation-maximization & \faCheck \\ [0.5ex]
        \cite{kalenkova2025discovering} & user activities & Petri net  & \faCheck \\ [0.5ex]
        \cite{zareie2025identifying,rivero2026detecting} &  user activities & outlier detection & \faCheck \\ [0.5ex]
        \cite{jakesch2021trend} & text, posting time & text similarity, timings of campaign launch and posting tweets & \faCheck \\ [0.5ex]
        
        \cite{smith2025unsupervised} & hashtags & bayesan model & \\ [0.5ex]
        \cite{uyheng2022mapping} & user activities & cluster interactions networks  & \\ [0.5ex]
        \cite{kuznetsova2025amplifying} & mention & cluster interactions networks  & \\ [0.5ex]

        \cite{stringhini2024understanding} & URLs & exploratory analysis & \\ [0.5ex]
        
        \bottomrule
\end{tabular}
    }
\end{table}

\begin{table}[t]
    \caption{Data mining and machine learning methods, based on \textit{supervised} learning, for detecting coordinated behavior. For each group of works we report the input types, the machine learning approach, whether the method takes time into account, and the prediction target.}
    \label{tab:machine-learning-supervised-approach}
    \centering
    \small
    \setlength{\tabcolsep}{4pt}
    \scalebox{0.75}{
    \begin{tabular}{L{1.6cm}L{4.8cm}L{7.5cm}C{1cm}L{1.6cm}}
        \toprule
        \textbf{reference} & \textbf{input} & \textbf{machine learning approach} & \textbf{time} & \textbf{target}\\
        \midrule
        
        \cite{zhang2023capturing} & user activities & representation learning, conditional embedding, neural encoding  & \faCheck & user \\ [0.5ex]
        \cite{francois2021measuring} & network, temporal, semantic features & outlier detection  & \faCheck & network \\ [0.5ex]
        \cite{saeed2024tuberaider} & text & peak detection, multiclass classification  & \faCheck & community \\
        \cite{pote2025coordinated} & user activities & classifier  & \faCheck & user, target \\ [0.5ex]

        \cite{mariconti2019you} & metadata, audio transcripts, thumbnails & ensemble classification & & target \\ [0.5ex]
        \cite{anand2025density} & user activities & random weighted walk  & & network \\ [0.5ex]
        \cite{matsuzaki2025edcoc} & co-retweet network & graph neural network & & community \\ [0.5ex]
        \cite{kanakaris2025network} & retweet & retrieval-augmented generation & & network \\ [0.5ex]
        \cite{minici2025iohunter} & network, text & graph neural network & & user\\ [0.5ex]

        \bottomrule
\end{tabular}
    }
\end{table}
 
Other works proposed simpler methods or indicators as signals of possible coordinated behavior. \citet{bellutta2023investigating} analyze account creation times under the hypothesis that accounts involved in coordinated malicious activities tend to be created in bursts.
They computed the daily histogram of account creations to which they applied a burst detection algorithm, identifying spikes in account creations by comparing the number of accounts created in a given day against the average number of daily accounts created in a reference time window.
Another simple unsupervised technique is proposed in~\cite{steinert2015online}, where coordination is measured as the extent to which users converge---spontaneously or in an organized fashion---on the use of certain hashtags. The Gini coefficient, an indicator of inequality, is applied to the distribution of used hashtags to create time series where saddles close to 0---indicative of low inequality---correspond to no coordination whereas peaks close to 1---indicative of a situation where all users use the same few hashtags---correspond to strong coordination.
Two alternative techniques are proposed in~\cite{manchanayaka2024identifying} and~\cite{manchanayaka2025using}. The former leverages contrast pattern mining to extract anomalous behavior, while the latter uses convergent cross mapping to discover cause-and-effect relationship and to construct an influence network. Similarly,~\cite{erhardt2023detecting,erhardt2024hidden} use a discrete-time stochastic model to analyze coordinated activity, representing the user behaviors as interacting Markov chains. In~\cite{kalenkova2025discovering} social media interactions are modeled as Petri nets, while in~\cite{zareie2025identifying,rivero2026detecting} coordination is detected by identifying anomalies in account sharing behavior. In~\cite{stringhini2024understanding} an exploratory analysis is conducted on YouTube links shared on 4chan.

\subsubsection{Supervised}
Supervised methods can be categorized based on their \textit{prediction target}, leading to approaches that operate at the \textit{user}, \textit{community}, \textit{network}, and \textit{target} levels.

\textit{User.} ~\citet{zhang2023capturing} tackles a binary classification task to distinguish between coordinated and non-coordinated users that participate in cross-platform campaigns. They leverage information from an \textit{aid platform} where coordinated users are known, to detect unknown coordinated users on a \textit{target platform}. Input data consists of a known coordinated activity set on the aid platform, plus unknown user activities on the target platform. The relationship between the two activities are modeled with neural time series encoders before being fed to a multi-layer perceptron for prediction. In~\cite{pote2025coordinated} traditional classifiers (e.g., Random Forest) are employed on user- and content-level engagement features to address to detect both coordinated users and content targeted by coordination. In~\cite{minici2025iohunter}, a similarity network is built following~\cite{luceri2024unmasking}, and multimodal node embeddings---combining one-hot features and text---are fed into a GNN to classify users as information operation drivers or legitimate accounts.\vspace{0.1cm}

\textit{Community.}
\citet{saeed2024tuberaider} focused on attributing coordinated hate attacks to the communities that organized them. A peak detector identifies an abrupt rise in the comment activity of a YouTube video, a signal of a coordinated attack. Then, it leverages a trained classifier based on linguistic features from the comments to the video and a set of Reddit and 4chan communities, identifying the community responsible for each attack based on linguistic patterns and similarities. In~\cite{matsuzaki2025edcoc}, a graph neural network model is exploited to recognize normal and abnormal communities.\vspace{0.1cm}

\textit{Network.}
A simpler approach is proposed in~\cite{francois2021measuring}, where a small expert-labeled ground truth of organic and inorganic campaigns is used to characterize campaigns along network, temporal, and semantic dimensions. Unknown campaigns are then labeled as coordinated if their indicators fall within the 95\% confidence interval of inorganic campaigns and outside that of organic ones. In~\cite{anand2025density} a graph classification method based on random weighted walks leverages the density of local network structures to distinguish coordinated from non-coordinated networks. \citet{kanakaris2025network} model retweet networks as propagation trees, encode them into text via graph prompting, and combine them with content and similarity-based examples in a RAG framework, enabling an LLM to classify coordinated disinformation campaigns.\vspace{0.1cm}

\textit{Target.} In~\cite{mariconti2019you}, authors propose a system to predict YouTube videos likely to be targeted by coordinated hate attacks, using features from metadata, transcripts, and thumbnails, and an ensemble classifier trained on previously raided videos. As previously discussed, the approach in~\cite{pote2025coordinated} operates at both the \textit{user} and \textit{target} levels.

\subsection{Discussion}
\label{sec:detection-final-remarks}

\subsubsection{Modeling complex coordination}
In Section~\ref{sec:conceptual-framework} we highlighted that coordinated online behavior is complex and multifaceted, often involving various actions across multiple platforms. Consequently, methods that analyze only a single type of action---particularly single-layer network approaches---risk missing significant coordination activities. To address this limitation, using \textit{multiplex} networks and \textit{compound} actions can offer more comprehensive insights. The network science framework presented in Section~\ref{sec:network-science-approaches} supports both single- and multi-layer (i.e., multiplex) networks. However, few studies considered multiple co-actions and built multiplex networks, as summarized in Table~\ref{tab:multiplex-network}. Moreover, to fully leverage the benefits of multiplex networks, these should not be flattened before running the community detection algorithm, which must be specifically designed for multiplex networks so as to utilize the multiple layers effectively~\cite{magnani2021community}. Unfortunately, only a few studies possess these characteristics~\cite{emeric2023interpretable,luceri2024unmasking,ng2022combined,dey2024coordinated,magelinski2020detecting}, making the analysis of coordinated behavior across multiple actions a largely unexplored area of research. Compound co-actions, combining simpler actions (e.g., a post with both a hashtag and user mention), can enhance detection confidence because they are less likely to occur by chance~\cite{magelinski2022synchronized}. However, this approach is also almost completely unexplored. In conclusion, too little research has been conducted so far on using multiple co-actions for detecting coordinated online behavior, leaving the actual advantages of these methods over single-action analysis unclear. Moreover, any potential benefits must be weighed against the increased complexity and computational costs that they introduce.

\subsubsection{Network science vs. machine learning}
Most methods for detecting coordinated behavior rely on network science, offering greater generality and expressiveness than machine learning approaches. They effectively model complex interactions and relationships, capturing varying degrees of coordination~\cite{nizzoli2021coordinated}, tracking temporal changes~\cite{tardelli2024temporal}, and providing detailed characterizations of coordinated groups~\cite{tardelli2024multifaceted}. The disadvantage is the computational cost of analyzing large networks and the requirement for human analysts to interpret results, making them less suitable for being directly used in automatic decision making systems. A further drawback is their sensitivity to specific parameters, such as those defining the way in which the timings of user actions are modeled~\cite{weber2021temporal}, which can significantly impact the obtained results and for which few guidelines currently exist~\cite{ng2022online,tardelli2024temporal}. Instead, data mining and machine learning methods often rely on oversimplified assumptions, such as the existence of a sharp binary distinction between coordinated and non-coordinated actors, which overlooks the nuanced nature of online coordination. This binary approach can undermine the theoretical and practical reliability of results. Furthermore, these methods face limitations due to the lack of comprehensive labeled datasets needed for training effective models. Despite these challenges, machine learning methods excel in encoding diverse actions and characteristics of users, and offer outputs that are directly applicable for automatic decision-making systems, unlike the more complex network science methods. In conclusion, network science methods are particularly suitable when the goal is an in depth understanding of the studied phenomenon, while machine learning methods can be used---with due caution---for quick decisions and when scalability is a concern.

\subsection{Validation}
\label{sec:validation}
The scarcity of labeled data limits the development of coordinated behavior detection methods. This constraint hampers the adoption of supervised approaches, contributing to the predominance of unsupervised techniques, and complicates both model tuning and evaluation. As a result, many methods are still assessed through post-hoc analyses rather than standardized quantitative benchmarks. To address this challenge, existing works adopt three main validation strategies. \vspace{0.1cm}

\textit{Characterization-based validation.}
A common approach consists in validating methods through qualitative or descriptive analyses of the detected behaviors. In this setting, outputs are assessed a posteriori by examining account activity, temporal patterns, content and socio-linguistic signals, and network structure, to verify their consistency with known characteristics of coordinated behavior. While this strategy enables exploratory insights, it lacks objective ground truth and limits reproducibility and comparability across methods. Section~\ref{sec:characterization} is dedicated to the task of characterization.\vspace{0.1cm}

\textit{Public ground-truth datasets.}
Publicly available datasets can act as ground truth proxies, with some studies constructing balanced datasets from authoritative repositories\footnote{\url{https://transparency.x.com/en/reports/moderation-research.html} (accessed: 28/07/2026)} enabling both training~\cite{luceri2024unmasking} and evaluation~\cite{vargas2020detection,ng2023coordinated,ng2022coordinated,cima2024coordinated,mannocci2026multimodal,luceri2026coordinated,loru2025compression,pante2025beyond,minici2025iohunter,pote2025coordinated,smith2025unsupervised,kalenkova2025discovering,zareie2025identifying,manchanayaka2024identifying,manchanayaka2025using,sharma2021identifying,zhang2021vigdet} of detection methods. However, these datasets are often limited to specific platforms and coordination types, restricting model generalization. \vspace{0.1cm}

\textit{Simulation-based validation.}
An alternative strategy consists in evaluating detection methods in controlled simulated environments, where coordinated behaviors can be explicitly modeled. Early approaches rely on agent-based models grounded in empirical observations. For instance, \citet{mehta2022estimating} develop a multi-agent simulation of Reddit to analyze the impact of coordinated activity on recommendation systems, while \citet{jahn2023detecting} design an agent-based model with heterogeneous agents (e.g., authentic users and coordinated boosters) to generate synthetic data. Other approaches generate synthetic datasets by injecting controlled anomalous patterns to enable systematic evaluation across coordination scenarios~\cite{zouzou2024unsupervised,pohl2022artificial}. Recent work explores large language models (LLMs) to simulate more realistic and adaptive behaviors. For instance, in \cite{orlando2026emergent} LLM-driven agents are used to mimic both organic users and coordinated actors under different regimes, from implicit alignment to explicit collaboration. Overall, simulation-based validation offers a flexible framework, though current efforts remain limited in scale and standardization.\vspace{0.1cm}
 \begin{table}[t]
    \centering
    \caption{Indicators used for characterizing coordinated behavior. For each group of indicators we report the works that used them, the high-level concept implemented, and the defining dimensions of coordination for which the indicators provide {\small\fulldot}, or could provide {\small\emptydot}, information. Table rows are grouped based on the type of information (i.e., user, content, network) leveraged by the indicator.}
\label{tab:characterization-measures}
    \small
    \setlength{\tabcolsep}{3pt}
\scalebox{0.75}{
    \begin{tabular}{cL{4.85cm}L{2.15cm}L{4.65cm}ccccc}
        \toprule
        &&&& \multicolumn{4}{c}{\textbf{defining dimensions}} & \\
        \cmidrule{5-8}
        & \textbf{reference} & \textbf{concept} & \textbf{indicators} & \textit{auth.} & \textit{harm.} & \textit{orch.} & \textit{time} & \textit{other} \\
        \midrule
        \multirow{5}{*}{\rotatebox[origin=c]{90}{\textit{user}}} & \cite{assenmacher2020two,bellutta2023investigating,chomel2023manipulation,danaditya2022curious,graham2020like,hristakieva2022spread,ng2022combined,ng2022online,ng2023you,nizzoli2021coordinated,pacheco2020unveiling,pacheco2021uncovering,weber2021amplifying,tardelli2024multifaceted,de2024detecting,neha2024understanding,ng2024tiny,zouzou2024unsupervised}  & automation & bot scores &\fulldot & & & \emptydot & \\
        & \cite{hristakieva2022spread,linhares2022uncovering,lee2013campaign,ng2022online,nizzoli2021coordinated,schliebs2021china,tardelli2024multifaceted,gruzd2022coordinated,yu2022framework,de2024detecting,sharma2021identifying,zhang2021vigdet,neha2024understanding}  & moderation &  suspended users & & \fulldot & & \emptydot & \\
        & \cite{ng2022online} & username diversity & entropy & \fulldot &  &  & \emptydot & \\
        & \cite{schliebs2021china,blas2025large,matsuzaki2025edcoc,kulichkina2026connective} & activity & account creation burstiness & \fulldot &  &  & \fulldot & \\
        \midrule
        & \cite{burghardt2024socio,cresci2019cashtag,danaditya2022curious,keller2020political,lee2013campaign,schliebs2021china,schoch2022coordination,vargas2020detection,wang2023evidence,weber2021amplifying,ghasiya2022rapid,righetti2022political,venancio2024unraveling,steinert2015online,keller2017manipulate,jakesch2021trend} & activity & number of posts, retweets, \ldots &  & &  & \fulldot & \fulldot \\
        & \cite{righetti2022political,venancio2024unraveling,uyheng2022mapping} & engagement & number of views, likes, \ldots &  & &  & \fulldot & \fulldot \\
        & \cite{chomel2023manipulation,pacheco2020unveiling,broniatowski2021towards,jakesch2021trend,antonakaki2026coordinated,yang2025coordinated,kulichkina2026connective} & timings & action time interval & \emptydot && \emptydot & \fulldot & \fulldot \\
        \multirow{5}{*}{\rotatebox[origin=c]{90}{\textit{content}}} & \cite{burghardt2024socio,bellutta2023investigating,danaditya2022curious,ng2022cross,ng2022online,ng2023coordinated,venancio2024unraveling,zhang2021vigdet,wang2023evidence,weber2020who,weber2021amplifying,saeed2024tuberaider,cresci2019cashtag,nizzoli2021coordinated,keller2017manipulate,sharma2021identifying,zhang2021vigdet,dey2024coordinated,neha2024understanding,manchanayaka2025using,blas2025large,cinus2025exposing,song2025spread,wu2025unmasking,uyheng2022mapping,kulichkina2026connective,elmas2026israel,kansaon2025fake} & socio-linguistics & attitudes, concerns, emotions, stances, topics, words &  & \fulldot &  & \emptydot & \fulldot \\
        & \cite{weber2020who,weber2021amplifying,ng2023coordinated,jakesch2021trend} & repetitiveness & text similarity scores &  & \fulldot & \emptydot & \emptydot & \fulldot \\
        & \cite{suresh2023tracking,tardelli2024temporal} & socio-linguistics & topic and hashtag evolution & & \fulldot & & \fulldot & \fulldot \\    
        & \cite{cao2015organic,giglietto2020takes} & news diversity & entropy, Gini coefficient &  &  & \emptydot & \emptydot & \fulldot \\
        & \cite{wang2023evidence,giglietto2020coordinated,bellutta2023investigating,cinus2025exposing} & news reliability & suspended and blacklisted URLs && \fulldot && \emptydot & \\
        & \cite{righetti2022political,tardelli2024multifaceted,barbero2023multi,bellutta2023investigating,cinus2025exposing,song2025spread,yang2025coordinated} & news reliability & NewsGuard and MBFC$^\dagger$ scores && \fulldot && \emptydot &  \\
        & \cite{hristakieva2022spread,matsuzaki2025edcoc} & manipulation & propaganda scores &  & \fulldot & & \emptydot &  \\
        & \cite{loru2024influence,saeed2024tuberaider,elmas2026israel} & offensiveness & toxicity scores &  & \fulldot & & \emptydot &  \\
        & \cite{loru2024influence,nizzoli2021coordinated,kulichkina2025protest,tardelli2024multifaceted,tardelli2024temporal} & political bias & MBFC$^\dagger$ scores, hashtag bias &&&& \fulldot & \fulldot \\ 
        & \cite{cinus2025exposing} & fake content & AI-generated && \fulldot &&& \\ 
        \midrule    
        & \cite{chomel2023manipulation,hristakieva2022spread,loru2024influence,ng2022combined,ng2022online,ng2023coordinated,nizzoli2021coordinated,tardelli2024multifaceted,cima2024coordinated,kulichkina2025protest,neha2024understanding,matsuzaki2025edcoc} & coordination & edge weights & & & \fulldot & \emptydot & \fulldot \\
        \multirow{3}{*}{\rotatebox[origin=c]{90}{\textit{network}}}& \cite{ng2023you,giglietto2020takes,graham2024coordination,kulichkina2025protest,cinelli2022coordinated,di2025post,pacheco2020unveiling,kirdemir2022towards,ng2023coordinated,de2024detecting,venancio2024unraveling,neha2024understanding,kuznetsova2025amplifying} & influence & centrality scores & & & \fulldot & \emptydot & \\    
        & \cite{cresci2019cashtag,nizzoli2021coordinated,cima2024coordinated,neha2024understanding,wu2025unmasking} & homophily & assortativity & & & \fulldot & \emptydot & \\ 
        & \cite{kirdemir2022towards,ng2022coordinated,nizzoli2021coordinated,giglietto2020takes,graham2024coordination,ng2023you,kulichkina2025protest,ng2023coordinated,cima2024coordinated,de2024detecting,neha2024understanding,song2025spread} & cohesiveness & clustering coefficient, modularity, density & & & \fulldot & \emptydot & \\    
        & \cite{tardelli2024temporal,linhares2022uncovering} & stationarity & user influx and outflux, user evolution  & & & & \fulldot & \\
        \bottomrule 
\multicolumn{9}{l}{$\dagger$: Media Bias/Fact Check} \\
\end{tabular}
    }
\end{table}
 
\section{Characterization of coordinated behavior}
\label{sec:characterization}
When the coordination detection method is not integrated into an automated decision-making system, its output may initiate the characterization task. The aim of this task is to describe each coordinated user, group, or community along one or more of the defining dimensions outlined in Section~\ref{sec:facets-coordination}. Characterizing the detected instances of coordination can also provide valuable information for validating detection results, especially in the absence of ground truth data. 
Characterizing coordinated online behavior is a semi-automatic task, as it involves some degree of manual analyses and observations by human analysts supported by automatically computed indicators that provide information on the coordinated actors. Table~\ref{tab:characterization-measures} summarizes the main indicators used for characterizing coordinated actors, whose discussion is presented in the remainder of this section. Indicators in the table are grouped based on the information they leverage, which can be related to (\textit{i}) users, (\textit{ii}) content, or (\textit{iii}) networks, a distinction that has also been adopted in related analyses of bots~\cite{ng2025global}. The table also highlights the dimensions of coordination that each indicator addresses. Full dots indicate dimensions where the indicator has already been applied, while empty dots denote dimensions where it has potential use that has not yet been explored.

\subsection{Authenticity}
Authenticity refers to the extent to which users, groups, or communities correctly represent themselves to others on a platform. As a consequence, inauthenticity is found when an actor misrepresents itself, such as to mislead others on who they are or what they do. In other words, authenticity is a property of the actors. In literature, the most common proxy for inauthenticity is an account's degree of \textit{automation}, obtained via a \textit{bot score}~\cite{assenmacher2020two,bellutta2023investigating,chomel2023manipulation,danaditya2022curious,graham2020like,hristakieva2022spread,ng2022combined,ng2022online,ng2023you,nizzoli2021coordinated,pacheco2020unveiling,pacheco2021uncovering,weber2021amplifying,tardelli2024multifaceted,de2024detecting,neha2024understanding,ng2024tiny,zouzou2024unsupervised}. By definition, social bots are accounts that make use of some or full automation. Thus, accounts with a high bot score are likely to be mostly automated. Unfortunately however, while useful, the use of bot scores as indicators of inauthenticity faces some limitations. First, computing bot scores is a challenging task per se, and it is prone to errors~\cite{ng2025global}. Second, not all bots are inauthentic, as there exist some types of self-declared bots that operate for neutral or benign purposes. Finally and most importantly, there exist multiple types of users that are inauthentic but that do not make use of automation (e.g., trolls, dissidents). Therefore, using bot scores as indicators of inauthenticity can cause both false positive and false negative errors. In addition to bot scores, some also used username similarity~\cite{ng2022online} and bursts of account creations~\cite{schliebs2021china,blas2025large,matsuzaki2025edcoc,kulichkina2026connective} as proxies for inauthenticity. 
Table~\ref{tab:characterization-measures} also shows that the few existing indicators of inauthenticity are all based on user information. While this is expected since authenticity is a property of the actors rather than their actions, we also note that extremely short time intervals between user actions~\cite{chomel2023manipulation,pacheco2020unveiling,broniatowski2021towards,jakesch2021trend,antonakaki2026coordinated,yang2025coordinated,kulichkina2026connective} could be used as a red flag of automation and, by extension, also of inauthenticity. 
\subsection{Harmfulness}
Harmfulness measures the extent to which the actions of the coordinated actors can potentially cause negative consequences. Hence, the analysis of the actions can provide valuable information about the intent, or potential, for harm of the coordinated actors. The production or re-sharing of content (e.g., text, images, links) is the most information-rich type of action and, in fact, nearly all indicators of harmfulness in Table~\ref{tab:characterization-measures} are derived from the analysis of content. 
Traditional NLP methods are used to analyze attitudes, concerns, emotions, stances, topics, hashtags, and words, as these provide useful insights into the intent and aims of coordinated actors~\cite{burghardt2024socio,bellutta2023investigating,danaditya2022curious,ng2022cross,ng2022online,ng2023coordinated,venancio2024unraveling,zhang2021vigdet,wang2023evidence,weber2020who,weber2021amplifying,saeed2024tuberaider,cresci2019cashtag,nizzoli2021coordinated,keller2017manipulate,sharma2021identifying,zhang2021vigdet,dey2024coordinated,neha2024understanding,manchanayaka2025using,blas2025large,cinus2025exposing,song2025spread,wu2025unmasking,uyheng2022mapping,elmas2026israel,kansaon2025fake}. For instance, certain words, negative emotions, and stances can amplify social discord and polarize public opinion. The prevalence of these indicators is largely due to the abundance of readily available tools and their ease of application. However, these indicators are often quite generic and consequently lack substantial power and informativeness. Text similarity scores were also used to identify harmful campaigns that aim to create false consensus through repeated sharing of similar posts~\cite{weber2020who,weber2021amplifying,ng2023coordinated,jakesch2021trend}. 
Other straightforward indicators of harmfulness are those based on the presence of toxic (i.e., hateful or offensive)~\cite{loru2024influence,saeed2024tuberaider,elmas2026israel} or propagandistic~\cite{hristakieva2022spread,matsuzaki2025edcoc} content. 
A different line of work measured harmfulness in terms of the unreliability of the news shared by coordinated actors, which was measured via NewsGuard's\footnote{\url{https://www.newsguardtech.com/} (accessed: 28/07/2026)} reliability ratings and Media Bias/Fact Check\footnote{\url{https://mediabiasfactcheck.com/} (accessed: 28/07/2026)} factual reporting and credibility ratings~\cite{righetti2022political,tardelli2024multifaceted,barbero2023multi,bellutta2023investigating,cinus2025exposing,song2025spread,yang2025coordinated}. Others measured unreliability in terms of suspended and blacklisted URLs~\cite{wang2023evidence,giglietto2020coordinated,bellutta2023investigating,cinus2025exposing}. 
The fraction of coordinated users that were suspended or banned by a social platform has been extensively used as an indicator of harmfulness~\cite{hristakieva2022spread,linhares2022uncovering,lee2013campaign,ng2022online,nizzoli2021coordinated,schliebs2021china,tardelli2024multifaceted,gruzd2022coordinated,yu2022framework,de2024detecting,sharma2021identifying,zhang2021vigdet,neha2024understanding}. Finally, the use of AI generated content  may be a signal of potentially harmful activity~\cite{cinus2025exposing}. This stands out as the only indicator of harmfulness based on user, rather than content, information. However, we note that harmfulness indicators based on content moderation lack predictive capabilities and can only be used retrospectively, since moderation actions typically occur some time after the account have engaged in the harmful activities.

\subsection{Orchestration}
\label{sec:beyond-authenticity-harmfulness}
Orchestration expresses the extent to which the coordination is well-organized, rather than spontaneous and emergent. Since orchestration reflects the level of organization among coordinated actors, its indicators are derived from coordination networks (Table~\ref{tab:characterization-measures}) and typically computed after network-based detection. High coordination scores are used as proxies for orchestration, under the assumption that tightly synchronized groups are more organized~\cite{chomel2023manipulation,hristakieva2022spread,loru2024influence,ng2022combined,ng2022online,ng2023coordinated,nizzoli2021coordinated,tardelli2024multifaceted,cima2024coordinated,kulichkina2025protest,neha2024understanding,matsuzaki2025edcoc}. Centrality scores measure node importance within a network or community. Highly central nodes have a greater influence on the dissemination of information, and their presence in coordinated communities may indicate a structured hierarchy~\cite{ng2023you,giglietto2020takes,graham2024coordination,kulichkina2025protest,cinelli2022coordinated,di2025post,pacheco2020unveiling,kirdemir2022towards,ng2023coordinated,de2024detecting,venancio2024unraveling,neha2024understanding}. Conversely, assortativity is a measure of homophily in a network. Measuring assortativity can help determine whether coordinated behavior is orchestrated or spontaneous by assessing the extent to which highly connected nodes tend to connect with other highly connected nodes~\cite{cima2024coordinated,cresci2019cashtag,nizzoli2021coordinated,neha2024understanding,wu2025unmasking}. For example, high assortativity between nodes with large degree might indicate a centralized orchestrated structure. Instead, high assortativity between nodes with low degree might indicate decentralized orchestration, while low assortativity or disassortativity might indicate spontaneous behaviors. Likewise, low modularity, high density, or high clustering coefficient indicates a tendency for nodes to form tightly-knit groups, suggesting well-organized and potentially orchestrated behavior~\cite{kirdemir2022towards,ng2022coordinated,nizzoli2021coordinated,giglietto2020takes,graham2024coordination,ng2023you,kulichkina2025protest,ng2023coordinated,cima2024coordinated,de2024detecting,neha2024understanding,song2025spread}. Finally, the time intervals between the actions of coordinated actors can provide information on whether the coordinated behavior is orchestrated or spontaneous because tightly synchronized actions suggest a higher level of premeditated organization, whereas more irregular intervals might indicate spontaneous, less structured coordination.
The study of the internal structure and organization of coordinated communities is also relevant for investigating the diverse strategies of online manipulation. Organized influence campaigns often unfold in distinct ways: some aim to mobilize pre-existing organic communities, leveraging existing social ties and trust, while others seek to establish new communities controlled by a central entity~\cite{starbird2019disinformation}. Additionally, organic communities can emerge from grassroots efforts, where users are genuinely motivated by a common goal without external orchestration. These different strategies leave ``network footprints'' that can be captured by orchestration indicators~\cite{cima2024coordinated}. 
\subsection{Time-variance}
\label{sec:time-variance-analysis}
Time-variance indicators aim at quantifying temporal changes in a wide array of characteristics of the coordinated actors. These include the number, intent, and behavior of the actors, which may result in changes in the types, timing, frequency, and intensity of their actions. Many measures can serve as time-variance indicators, including the indicators previously discussed for the dimensions of authenticity, harmfulness, and orchestration. Indeed, as shown in Table~\ref{tab:characterization-measures}, all indicators for authenticity, harmfulness, and orchestration can potentially be analyzed over time, even if they have not been already studied in this way yet. However, utilizing a measure as a time-variance indicator requires repeated and extended monitoring to detect possible changes over time. This requirement for prolonged monitoring poses an additional challenge compared to other types of indicators, which explains the scarcity of studies that employed time-variance indicators or that investigated the temporal dynamics of coordinated behavior~\cite{tardelli2024temporal}. The few existing detailed temporal analyses examined user flows between coordinated communities or the variation of users between different time windows~\cite{tardelli2024temporal,linhares2022uncovering}. Other temporal analyses of user activity can involve examining bursts of account creations~\cite{schliebs2021china}, abnormal post or retweet volumes~\cite{burghardt2024socio,cresci2019cashtag,danaditya2022curious,keller2020political,lee2013campaign,schliebs2021china,schoch2022coordination,vargas2020detection,wang2023evidence,weber2021amplifying,ghasiya2022rapid,righetti2022political,venancio2024unraveling,steinert2015online,keller2017manipulate,jakesch2021trend}, or inflated engagement metrics, such as the number of views~\cite{righetti2022political,venancio2024unraveling}. The analysis of topics and hashtags over time can reveal the evolution of the narratives promoted or discussed by the coordinated actors~\cite{suresh2023tracking,tardelli2024temporal}. The timing of the actions, such as the intervals between pairs of actions that result in a co-action, can provide insights into the type of coordination and strategies employed, which may evolve over time~\cite{broniatowski2021towards,chomel2023manipulation,pacheco2020unveiling}. 

\subsection{Other general indicators}
\label{sec:characterization-others}
In addition to the indicators that provide information about the four defining dimensions of coordinated behavior, other general-purpose indicators have also been used. These come in handy to provide additional context on the coordinated behavior and may be particularly relevant depending on the application context, as in the case of the indicators of political polarization or bias that are used to characterize coordinated communities involved in online electoral debates~\cite{loru2024influence,nizzoli2021coordinated,kulichkina2025protest,tardelli2024multifaceted,tardelli2024temporal}. Political bias was computed based on MBFC scores or by leveraging hashtag polarity. Analyses of the level of activity of coordinated actors on a platform~\cite{burghardt2024socio,cresci2019cashtag,danaditya2022curious,keller2020political,lee2013campaign,schliebs2021china,schoch2022coordination,vargas2020detection,wang2023evidence,weber2021amplifying,ghasiya2022rapid,righetti2022political,venancio2024unraveling,steinert2015online,keller2017manipulate,jakesch2021trend}, of the engagement they obtain~\cite{venancio2024unraveling,righetti2022political}, or of the content they produce, can provide additional contextual information. Regarding the latter, standard socio-linguistic analyses were used to draw insights into the narratives and stances of coordinated actors~\cite{burghardt2024socio,bellutta2023investigating,danaditya2022curious,ng2022cross,ng2022online,ng2023coordinated,venancio2024unraveling,zhang2021vigdet,wang2023evidence,weber2020who,weber2021amplifying,saeed2024tuberaider,cresci2019cashtag,nizzoli2021coordinated,keller2017manipulate,sharma2021identifying,zhang2021vigdet,suresh2023tracking,tardelli2024temporal,neha2024understanding,song2025spread,kulichkina2026connective}. Similarly, others considered the diversity of the news shared by the coordinated actors~\cite{cao2015organic,giglietto2020takes} or assessed the presence of patterns in the timings of their actions~\cite{broniatowski2021towards,chomel2023manipulation,pacheco2020unveiling,jakesch2021trend}. Finally, acknowledging that coordination is a nuanced and non-binary concept, some studies computed coordination scores, for example based on the edge weights of the coordination network~\cite{chomel2023manipulation,hristakieva2022spread,loru2024influence,ng2022combined,ng2022online,ng2023coordinated,nizzoli2021coordinated,tardelli2024multifaceted,cima2024coordinated,kulichkina2025protest,neha2024understanding,kuznetsova2025amplifying}.Analyzing the degree of coordination helps reveal the level of collaboration, influence, and effectiveness of collective actions within a group.

\subsection{Compound indicators}
\label{sec:characterization-compound}
While the previous discussion considered indicators individually, combining multiple indicators can provide deeper insights by capturing complex interactions. For instance, studies~\cite{hristakieva2022spread,nizzoli2021coordinated,cima2024coordinated,tardelli2024multifaceted,neha2024understanding} analyze trends and correlations across dimensions such as coordination, propaganda, moderation, cohesiveness, and automation, showing that coordinated communities can exhibit distinct behaviors. They also highlight that relying on a single indicator may lead to unreliable assessments, emphasizing the importance of compound indicator approaches~\cite{hristakieva2022spread}. Similarly, others considered the interplay between coordination, toxicity, and political bias~\cite{loru2024influence}. In \cite{tardelli2024multifaceted}, a comprehensive characterization of coordinated communities is proposed using multiple indicators, including coordination, automation, suspensions, news unreliability, and political bias, visualized via radar charts. Higher scores—and thus larger chart areas—indicate more suspicious behavior, making the radar area a compound indicator of suspiciousness. This approach highlights the potential of combining multiple indicators, an area that remains largely underexplored.

\subsection{Discussion}
\label{sec:characterization-final-remarks}
Table~\ref{tab:characterization-measures} shows that many and diverse indicators are used to measure harmfulness. These include indicators of hate speech, toxicity, and propaganda, as well as indicators based on platform moderation decisions. 
Conversely, inauthenticity has only been examined through the lens of automation, despite being the focus of many works. From a practical standpoint, understanding the authenticity of an account is inherently more challenging than assessing its harmfulness, as authenticity pertains to the nature of the actors rather than their actions. Many online platforms afford users a significant degree of anonymity, making it difficult to verify their true identities. In contrast, many actions performed online leave unforgeable traces, which can be more easily analyzed to determine harmful behavior. Despite the challenges however, future work should focus on developing additional indicators of authenticity. Among the untapped sources of information are the completeness and credibility of the user profile, the regularity of activity patterns, the duplication or repetitiveness of posted content, the lack of consistency in language or interests, the authenticity of posted multimedia content, the cross-platform consistency of profile and behavior, the similarity of profile and behavior to that of previously moderated accounts, and the verification status, which however should be evaluated differently depending on whether such status can be purchased on the analyzed platform.\footnote{\url{link.gale.com/apps/doc/A746844400/AONE} (accessed: 28/07/2026)}

 \section{Open challenges and future research directions}
\label{sec:discussion}

\begin{figure}\begin{minipage}[b]{0.5\textwidth}\centering
    \includegraphics[width=1\linewidth]{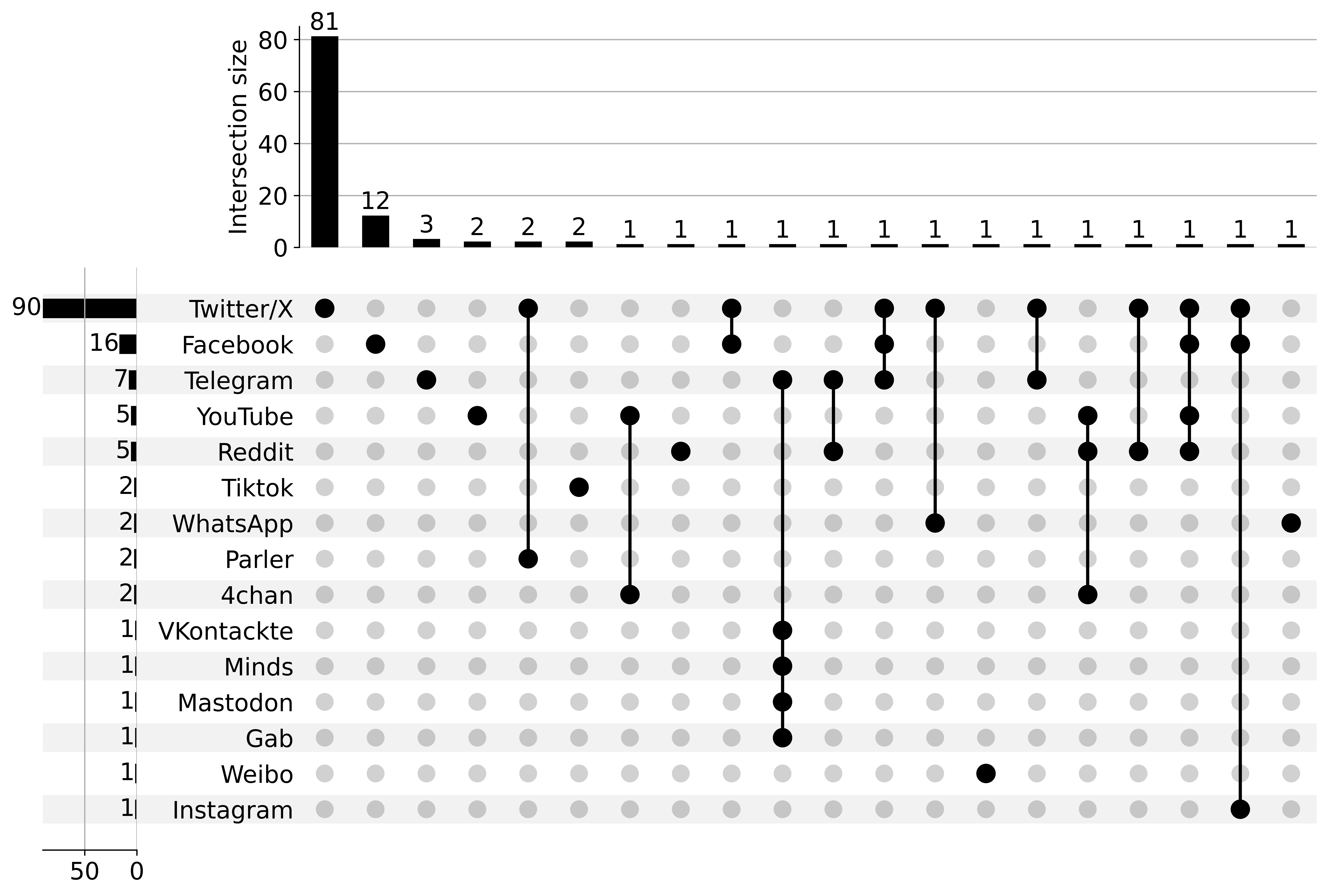}\caption{Distribution of works based on the combination of analyzed platforms. The vast majority of works analyzed a single platform, mainly Twitter/X. Only a few works performed cross-platform analyses, represented by multiple connected dots.}\label{fig:upset-platforms}\end{minipage}\hspace{0.02\textwidth}\begin{minipage}[b]{0.44\textwidth}\centering
    \includegraphics[width=1\linewidth]{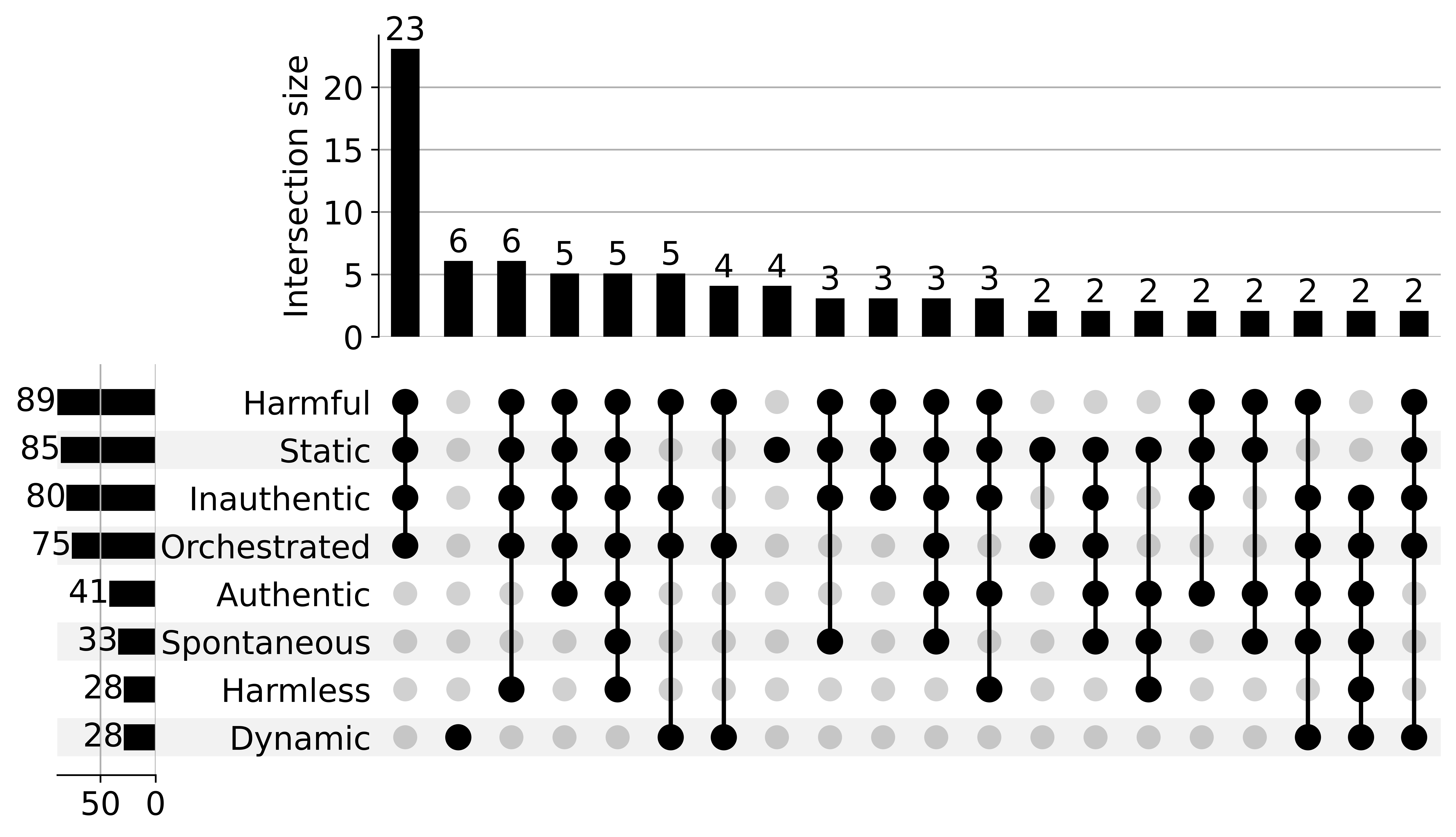}\caption{Distribution of works focused on different types of coordinated behavior, categorized based on the defining dimensions of authenticity, harmfulness, orchestration, and time-variance. Vertical bars show the number of works for each combination of dimensions, while horizontal bars show the number of works for each individual dimension. Only combinations covered by 2+ works are shown.}\label{fig:upset-dimensions}\end{minipage}\end{figure}

\textit{Multiplatform-ness.} Some coordinated campaigns unfold simultaneously on multiple platforms, resulting in \textit{multiplatform} coordinated behavior~\cite{wilson2020cross}. These campaigns exhibit overall similar behaviors, such as the use of campaign-specific hashtags, with minor variations between platforms. In contrast, \textit{cross-platform} campaigns involve different platforms playing distinct roles, with coordinated behavior varying significantly across the involved platforms. For example, organizers of a targeted hate attack might use private groups on a messaging platform to plan their actions before moving to the target platform to flood it with hateful comments~\cite{jakesch2021trend}. Detecting coordination on the planning platform might involve analyzing group joinings, while on the target platform it could involve examining synchronized commenting. The example highlights the challenges of multi- and cross-platform analyses, which require extensive data collection, careful selection of actions or machine learning features for each platform, and identification of the same groups of users across multiple platforms. While the latter can be mitigated via the analysis of \textit{content} rather than \textit{user} networks---as discussed in Section~\ref{sec:content-networks}---the remaining outstanding challenges result in a scarcity of multi- and cross-platform analyses. Figure~\ref{fig:upset-platforms} shows that most studies analyzed coordinated behavior on Twitter/X, with only a few multiplatform analyses~\cite{ng2022cross,ng2023coordinating,emeric2023interpretable,righetti2022political,barbero2023multi,zhang2023capturing,saeed2024tuberaider,jakesch2021trend,mannocci2026multimodal,iannucci2026detecting}, highlighting that this area still necessitates much investigation. \vspace{0.1cm}

\textit{Temporal variability.} Most studies on coordinated behavior incorporate time to some degree, as shown in Tables~\ref{tab:time-window},~\ref{tab:machine-learning-unsupervised-approach},~\ref{tab:machine-learning-supervised-approach}, and~\ref{tab:characterization-measures}. However, the majority of such works only perform superficial temporal analyses. Figure~\ref{fig:upset-dimensions} highlights that only a few works have deeply investigated the \textit{temporal dynamics} of coordinated behavior, using methods like multiplex temporal networks~\cite{tardelli2024temporal} and temporal point processes~\cite{sharma2021identifying,zhang2021vigdet}. Thus, analyzing temporal dynamics remains a promising yet relatively unexplored research avenue. Temporal analyses offer numerous advantages, including examining (\textit{i}) the temporal stability of coordinated groups, (\textit{ii}) changes in membership, structure, and actions, and (\textit{iii}) adaptation to countermeasures or other stimuli~\cite{giglietto2020adapting}. Further, dynamic analyses seem to yield more accurate results compared to static ones~\cite{tardelli2024temporal}. However, this approach faces several challenges, such as setting appropriate temporal parameters and addressing the computational demands resulting from the time-intensive nature of dynamic analyses.\vspace{0.1cm}

\textit{Heterogeneity.} The great degree of heterogeneity inherent of coordinated behavior does not only manifests through temporal variations, but also through the multitude of characteristics that the different instances of coordinated behaviors exhibit. Figure~\ref{fig:upset-dimensions} shows the distribution of studies that addressed specific types of online coordination, based on the defining dimensions introduced in Section~\ref{sec:facets-coordination}. As shown by the vertical bars, coordinated \textit{harmful inauthentic} and \textit{orchestrated} behavior is the type of coordination that was studied the most. As discussed in Section~\ref{sec:conceptual-framework}, this is due to both practical reasons related to the urgency of contrasting online manipulations, as well as to the large body of work that adopted Facebook's initial Definition~\ref{def:cib} of coordinated inauthentic behavior. Almost all the other remaining combinations involve \textit{harmful} behavior and \textit{static} analyses. Figure~\ref{fig:upset-dimensions} also shows that \textit{authentic spontaneous} and \textit{harmless} coordinated behavior is completely unstudied. While harmful, inauthentic, and orchestrated behaviors are of particular practical relevance for content moderation purposes, the study of harmless, authentic, and spontaneous coordination should not be overlooked as it provides valuable insights into fundamental aspects of online human interactions. Therefore, the analysis of the less problematic forms of online coordination should be increased in the near future.

\textit{Multimodality.} Online coordination can involve multiple content modalities, such as text, images, and videos. This diversity of content types adds complexity to detecting and analyzing coordinated behavior, as each modality has unique characteristics requiring different analytical techniques. For instance, text-based coordination might involve analyzing text similarities or hashtags use~\cite{venancio2024unraveling,de2024detecting,wang2023evidence}. Image-based coordination could require image recognition tasks or computing image similarities~\cite{yu2022framework,giglietto2023workflow,pacheco2021uncovering}, while video-based coordination demands techniques like video content analysis and transcript examination. As shown in Tables~\ref{tab:single-network},~\ref{tab:content-networks},~\ref{tab:machine-learning-unsupervised-approach},~\ref{tab:machine-learning-supervised-approach} however, only a handful of works investigated modalities other than text~\cite{mariconti2019you,pacheco2021uncovering,ng2022coordinated,yu2022framework,soares2023sharing,giglietto2023workflow}. Accounting for and integrating different modalities offers the potential to capture a more comprehensive picture of coordinated behavior, but also poses new challenges. These include developing features and methodologies to assess similarities based on multimedia content and addressing the computational demands of multimodal analyses. Nonetheless, given the rising trend of online platforms centered around multimedia content, multimodal analyses appear promising and well motivated.\vspace{0.1cm}

\textit{Generative AI} The rise of generative AI presents both challenges and opportunities. One potential challenge is the increased difficulty in detecting coordinated actors that use such techniques to mask their activities. AI is already capable of producing human-like text with given properties, and increasingly also images, audio, and video~\cite{cinus2025exposing}. Moreover, it has already been used to simulate authentic online human behaviors in such a way as to evade detection by existing systems~\cite{ng2025global}. Progress in generative AI could thus make it harder to detect future instances of coordinated behavior or to assess the authenticity of coordinated actors. However, it is still unknown what the effect of these techniques will be on the landscape of online coordination. 
At the same time, as discussed in Section~\ref{sec:validation}, recent work has begun to explore LLM-driven agents to simulate coordinated behavior
in controlled environments~\cite{orlando2026emergent}. Looking ahead, these approaches could enable more realistic and scalable simulations, supporting the development and stress-testing of detection methods. However, the use of LLMs in this context is still in its early stages, and a systematic understanding of their capabilities and limitations for modeling coordinated behavior is currently lacking.\vspace{0.1cm}

\textit{Scalability.} Large-scale coordinated campaigns can involve tens of thousands of users and millions of interactions, posing significant computational challenges, especially for methods based on pairwise user similarities. As a result, some studies analyze only a small fraction of data—sometimes as little as 1\%~\cite{nizzoli2021coordinated,di2025post}---limiting reliability. Future efforts should focus on increasing the scalability of current detection methods to handle the vast amounts of data generated by large-scale campaigns. Possible solutions to this issue can include the use of \textit{approximation algorithms} to obtain near-accurate results with reduced computational complexity, or \textit{sampling techniques} to select representative data subsets. Additionally, \textit{online algorithms} can update results incrementally as new data arrives, while \textit{graph summarization} techniques can condense large networks while preserving key structural properties.\vspace{0.1cm}

\textit{Validation and formalization.} As discussed in Section~\ref{sec:validation}, current validation strategies only partially alleviate the lack of labeled data. The challenge of data availability is expected to intensify as access to platform data continues to decline in the post-API era~\cite{tromble2021have}. Although some works leverage publicly available repositories to construct ground-truth datasets~\cite{luceri2024unmasking,mannocci2026multimodal,cima2024coordinated}, these resources remain limited in scope, typically focusing on a single platform or a specific type of coordinated behavior. This restricts the ability of models to generalize across the diverse manifestations of coordination, such as those illustrated in Figure~\ref{fig:venn}. At the same time, simulation-based solutions are still underexplored in the coordinated behavior literature~\cite{mehta2022estimating,jahn2023detecting,orlando2026emergent}. As discussed in Section~\ref{sec:research_definitions}, the field lacks a unified and statistically grounded definition of coordinated behavior, which limits principled validation. A key challenge is to translate existing operational definitions into formal, testable frameworks, for instance through appropriate null models capturing baseline behavior. Such models would enable more robust and comparable assessments across datasets and domains. In this direction, theory-driven approaches offer a promising path, drawing from socio-psychological models of collective action and the statistical physics of complex networks to develop more general and interpretable frameworks.\vspace{0.1cm}

\textit{Subjectivity.} The existence of many indicators of harmfulness---reported in Table~\ref{tab:characterization-measures}---is advantageous from a practical standpoint. However, this multitude of indicators reflects great variability in how harmfulness has been practically defined. More broadly, the subjective nature of harmfulness---discussed in Sections~\ref{sec:intent} and~\ref{sec:dim-harmfulness}---presents several challenges to its assessment. The first of such challenges is the variability with which different stakeholders define what constitutes harmful behavior, which hinder reaching a consensus on definitions and indicators~\cite{douek2020does}. Additionally, perceptions of harmfulness are influenced by cultural and social norms, which differ across individuals, regions, and communities. Consequently, what is seen as harmful by some might not be perceived the same way by others. Finally, assessing harmfulness involves evaluating both the intent behind the actions and their impact. While the impact can sometimes be measured---as with misinformation that leads to real-world violence~\cite{suresh2023tracking} or that reduces vaccine uptake~\cite{song2025spread}---the intent is often hidden or ambiguous, making it difficult to measure accurately. Given that a certain degree of subjectivity in the assessment of online coordination is inevitable and bound to persist, future works must prioritize transparency in defining and implementing it, favoring nuanced analyses based on multiple indicators.\vspace{0.1cm}

\textit{Attribution.} Detecting coordination, especially in cases where malicious actors aim to remain hidden, is already challenging. Attributing these instances to the responsible entities is even more difficult. Attribution does not merely involve identifying the accounts that take part in the coordination, but most importantly uncovering the entities behind them---such as movements, organizations, or states---and understanding their goals. Several obstacles hinder this process, including the anonymity afforded by online platforms and the sophisticated techniques used to obfuscate identities, activities, and intentions. Potential solutions include strengthening the collaboration with platforms, which have access to more data than what is publicly visible or available via APIs or scraping. Additionally, advanced forensic techniques and multidisciplinary approaches that combine technical, social, and behavioral analysis are essential. Despite these solutions, the task will remain daunting and require continuous research and experimentation.\vspace{0.1cm}

\textit{Multidisciplinarity.} The study of coordinated behavior is intrinsically multidisciplinary due to the complex interplay of technical, social, and regulatory factors involved. This multidisciplinary nature is evident in the diversity of the studies on the subject, which employ a wide array of methods from multiple scientific communities. For instance, computer scientists and complex systems scholars develop computational methods to detect instances of coordinated behavior~\cite{tardelli2024multifaceted,magelinski2022synchronized,weber2021amplifying}. Social scientists study the emergent behaviors and interactions that lead to massive online coordination~\cite{danaditya2022curious,kulichkina2025protest}, while political scientists analyze the impact of coordinated campaigns on real-world events, such as elections~\cite{nizzoli2021coordinated,chomel2023manipulation,yu2022framework}. The involvement of media and communication experts in understanding the dissemination of (mis)information further highlights the multidisciplinary scope~\cite{soares2023sharing,giglietto2020takes}. In this context, multidisciplinarity is both a challenge and an opportunity. Challenges arise from the need to coordinate efforts among various scientific communities and to organize and keep track of the extensive body of work on the subject---a task for which this survey aims to provide a contribution. In spite of the challenges however, the comprehensive understanding of the phenomenon and the effective mitigation of its nefarious instances can only be achieved through the collaboration between diverse stakeholders, such as scholars, platform operators, policymakers, and civil society organizations. \vspace{0.1cm}

\textit{Ethical dilemmas.} Research on coordinated online behavior also faces some ethical dilemmas. The process of collecting and analyzing online data involves monitoring user interactions and behaviors, which can raise concerns about surveillance and consent. The inherent subjectivity of coordinated behavior leads to additional challenges, as certain coordinated actions may be punished on some platforms while being tolerated on others~\cite{douek2020does}, resulting in imbalanced interventions and potentially unfair decisions. Characterizing coordinated actors further complicates matters, as this requires assessing the authenticity and harmfulness of actors who might have strong motivations---such as social and political reasons~\cite{steinert2015online,kulichkina2026connective}---for remaining anonymous, raising privacy concerns. Additionally, developing methods to detect coordinated actors presents risks as these tools could be misused to silence certain groups or minorities, thereby selectively threatening freedom of expression. Ultimately, these issues demand that great attention to ethical and normative considerations should be devoted in future research.\vspace{0.1cm}
 \section{Conclusions}
\label{sec:conclusions}
Coordination is a fundamental dynamic of human behavior and its study holds great theoretical significance and numerous practical implications. Theoretically, understanding coordinated online behavior contributes to shedding light on social dynamics, group formation, and collective action. Practically, this research is pivotal in combating online manipulations, such as disinformation campaigns and orchestrated hate attacks, and fostering more inclusive online spaces that promote genuine cooperation and constructive discourse. However, the field faces several significant challenges, including the complexity of multimodal and cross-platform analyses, the opportunities and perils of generative AI, the scarcity of labeled data, the subjectivity and ethical dilemmas inherent to the evaluation of coordinated behavior. These diverse challenges, coupled with the broad interest in the topic, highlight the need for multidisciplinary efforts from diverse research communities. In the coming years, these efforts should aim to develop a comprehensive array of datasets, methods, and studies to better understand and address this complex and dynamic phenomenon.
 \begin{acks}
    This work was supported in part by the European Union under the scheme HORIZON-INFRA-2025-01-DEV-02 - Early phase implementation of ESFRI Projects that entered the ESFRI Roadmap in 2021, Grant Agreement n.101292277, “SoBigData IP: SoBigData Implementation Phase”; in part by the HORIZON Europe projects TANGO - Grant Agreement n. 101120763; in part by the European Union – Next Generation EU within the ERC project DEDUCE (Data-driven and User-centered Content Moderation) under grant \#101113826.
\end{acks}

\section*{Author Contributions}
Lorenzo Mannocci: Conceptualization, Methodology, Investigation, Formal Analysis, Data Curation, Writing – Original Draft, Writing – Review \& Editing, Visualization. Michele Mazza: Conceptualization. Anna Monreale: Conceptualization, Supervision, Writing – Review \& Editing. Maurizio Tesconi: Conceptualization, Supervision, Funding Acquisition, Writing – Review \& Editing. Stefano Cresci: Conceptualization, Supervision, Writing – Original Draft, Review \& Editing. 
\bibliographystyle{ACM-Reference-Format}
\bibliography{mybib}

\end{document}